\documentclass[aps,prx,preprintnumbers,nofootinbib,twocolumn,amsmath,amssymb,floatfix,superscriptaddress]{revtex4-2}
\usepackage[utf8]{inputenc}
\usepackage[T1]{fontenc}
\usepackage[german,english]{babel}
\usepackage{tensor}
\usepackage{amsmath}
\usepackage{mathtools}
\usepackage{amssymb}
\usepackage{amsthm}
\usepackage{bbm}
\usepackage{microtype} 
\usepackage{quoting} 
\quotingsetup{font=small}

\usepackage{graphicx}
\usepackage[usenames]{color}
\usepackage[usenames,dvipsnames]{xcolor}
\usepackage{comment}
\usepackage{braket}
\usepackage{svg}
\usepackage{float} 
\usepackage{tikz}
\usetikzlibrary{arrows.meta, positioning}
\usetikzlibrary{calc}

\usepackage{commath}

\usepackage{color}

\definecolor{Blue}{rgb}{0.00, 0.00, 1.00}
\definecolor{Green}{rgb}{0.00, 1.00, 0.00}

\DeclareUnicodeCharacter{2212}{-}

\usepackage[bookmarks=true,colorlinks,linkcolor=blue,urlcolor=blue,citecolor=blue]{hyperref}

\begin{document}

\title{Nonanaliticities and ergodicity breaking in noninteracting many-body dynamics via stochastic resetting and global measurements}

\author{David Soldner}
\affiliation{Institut für Theoretische Physik, Eberhard Karls Universität Tübingen, Auf der Morgenstelle 14, 72076 Tübingen, Germany}

\author{Igor Lesanovsky}
\affiliation{Institut für Theoretische Physik and Center for Integrated Quantum Science and Technology, Eberhard Karls Universität Tübingen, Auf der Morgenstelle 14, 72076 Tübingen, Germany}
\affiliation{School of Physics and Astronomy and center for the Mathematics and Theoretical Physics of Quantum Non-Equilibrium Systems, The University of Nottingham, Nottingham, NG7 2RD, United Kingdom}

\author{Gabriele Perfetto}
\affiliation{Institut für Theoretische Physik, Eberhard Karls Universität Tübingen, Auf der Morgenstelle 14, 72076 Tübingen, Germany}

\date{\today}

\begin{abstract}
Stochastic resetting generates nonequilibrium steady states by interspersing unitary quantum dynamics with resets at random times. When the state to which the system is reset is chosen conditionally on the outcome of a global and spatially resolved measurement, the steady state can feature collective behavior similar to what is typically observed at phase transitions. Here we investigate such conditional reset protocol in a system of noninteracting spins, where the reset state is chosen as a magnetization eigenstate, that is selected (conditioned) on the outcome of a previous magnetization measurement. The stationary states that emerge from this protocol are characterized by the density of spins in a given magnetization eigenstate, which is the analogue of the order parameter. The resulting stationary phase diagram features multiple nonanalytic points. They are of first-order type for half-integer spin, while multicritical behavior, signalled by both first and second-order discontinuities, is found for integer spin. We also show that the associated dynamics is nonergodic, i.e., which stationary state the system ultimately assumes is determined be the initial state. Interestingly, the mechanism underlying these phenomena does not rely on interactions, but the emergent nonlinear behavior is solely a consequence of correlations induced by the measurement. 
\end{abstract}

\maketitle

\section{Introduction} 
Stochastic resetting is an intriguing phenomenon that is relevant in various fields of science, such as biology, ecology, computer science and physics. In ecology \cite{foraging1,foraging2}, stochastic resetting models animals foraging for food, where the agent, after a period of unsuccessful search, returns back at random times to the position where food was initially located. In biology \cite{DNA2004,DNA2009,DNA2018}, similarly, resetting describes proteins searching for a binding site on a DNA molecule. A common aspect of all these different processes is that stochastic resetting can make the search process more efficient by minimizing the time needed to locate the target. This follows by alternating local search periods with long-range relocation moves, see, e.g., Ref.~\cite{IntermittentSearch}. In physics, these aspects have been first shown in Ref.~\cite{evans2011resetting}, for the model of a Brownian particle whose position is reset to its initial value, c.f. Refs.~\cite{evans2020review,nagar2023review} for comprehensive reviews on the subject. Here, the mean first passage time through a given point in space can be minimized as a function of the resetting rate \cite{renewal1,renewal2,renewal3,Campos2015,renewal5,Sandev_1,Sandev_book}. In addition, stochastic resetting represents a simple model to generate nonequilibrium stationary states (NESS), since the resetting position is a sink of probability and thereby violates detailed balance \cite{evans2011resetlong,evans2014diffarbitraryd,DPTreset2015,eule2016non,power_law_reset_ness,ness_reset_random_walk,renewal8,Biroli_extreme}.     

In quantum systems, resetting, in its simplest implementation, amounts to reinitializing the state of the system to a predefined (pure or mixed) state. As such, resetting can be considered as an external monitoring of the quantum dynamics, which presents similarities to measurements. It is, indeed, by now understood, see, e.g., Refs.~\cite{Hartmann2006,Linden2010,Rose2018spectral,carollo2019unravelling,Armin2020,riera2020measurement,quantum_reset_ldev,carollo24universality}, that stochastic resetting superimposed to unitary evolution leads to an effective open-system dynamics, even in the absence of an actual environment being present. Such dissipative evolution leads at long times to a NESS \cite{Mukherjee2018,perfetto2021designing,turkeshireset,Magoni_PRA_quantum,dattagupta2022stochastic,sevilla2023dynamics,kulkarni2025dynamically,wald2025stochastic}. The combination of both resets and measurements has also been addressed. In the field of quantum walks \cite{dhar2015,dhar2015b,grunbaum2013recurrence,Barkai2016,Barkai2017,Barkai2018,tornow2023,majumdar2023,Walter25,yin2025resonance}, resetting leads to a faster mean hitting time of a target site for a single-particle subject to projective measurements at stroboscopic times, as shown in Refs.~\cite{restart_Yin,Modak_resetting,roy2025causality}. For many-body spin systems, the stationary state obtained upon interspersing resets and global projective measurements has been studied in Ref.~\cite{perfetto2021designing}. In the specific case of noninteracting spin $1/2$, it has been shown in Ref.~\cite{Magoni_PRA_quantum} that the stationary state features nonanalytic behavior in the thermodynamic limit. Due to the simplicity of the spin $1/2$ algebra, the kind of nonanalyticity observed is strongly limited to a discontinuity of the density of excited spins, akin to a first-order phase transition. Continuous nonanalytic points are not permitted. 

It is therefore interesting to develop a systematic classification of this collective behavior for generic spin $S$ systems and to understand how this links to microscopic symmetries. First of all, higher spin values might yield a wealth of different stationary phenomena with different kind of nonanalytic behavior. This is reminiscent of the $q$-state Potts model, where, for example in two spatial dimensions \cite{potts1,potts2,potts3,potts4}, the order of the equilibrium phase transitions changes from second ($1 \leq q\leq 4$) to first ($q>4$) order as the number $q$ of states on each lattice site is poised. Second, the ensuing collective behavior might be exploited for sensing protocols and metrology applications \cite{sensing_1,sensing_2,sensing_3,sensing_4,sensing_5}, where the different spin $S$ values could be exploited and distinguished on the basis of the different nonanalyic behavior displayed as the system parameters are tuned.

\begin{figure}[t]
    \centering
    \includegraphics[]{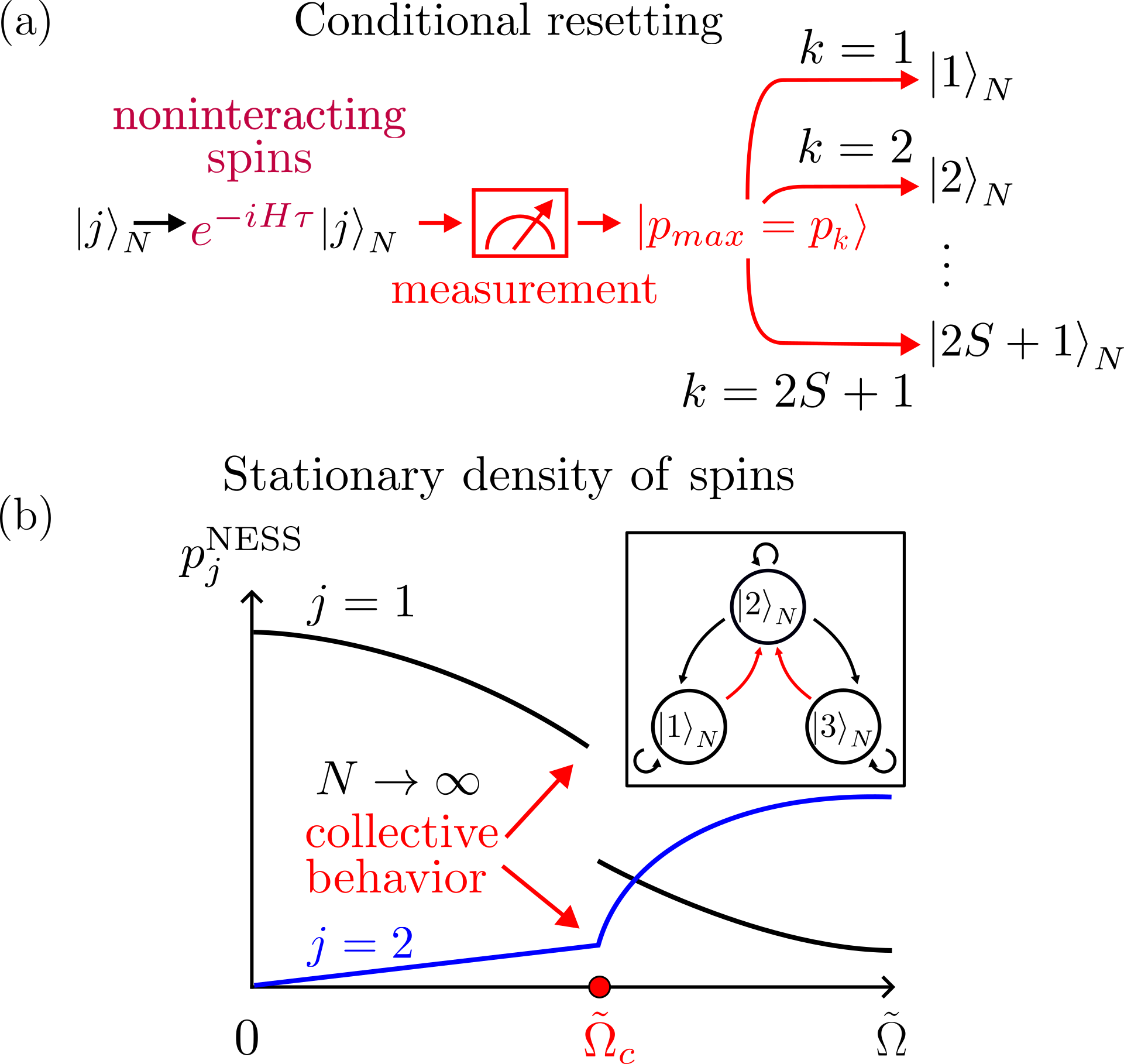}
    \caption{\textbf{Collective behavior in the conditional resetting protocol of spins}.
    (a) In the conditional reset protocol for a number $N$ of spin $S$ spins, there are $j=1,2,\dots 2S+1$ reset states $\ket{j}_N$ \eqref{eq:reset_states}. The system undergoes unitary time evolution according to the Hamiltonian \eqref{eq:Hamiltonian} from the reset state $\ket{j}_N$ for a random time $\tau$. The next reset state $\ket{k}_N$ is chosen conditionally on the outcome of a projective measurement of the total magnetization $S^z$. The majority rule is adopted, so that one chooses the state $\ket{j}_N$ corresponding to the magnetization value $j$ measured for the majority of spins. (b) A Markov chain between reset states $\ket{j}_N$ and $\ket{k}_N$ can then be associated with the dynamics. As $N\to \infty$, the Markov chain model (in the inset of the figure) is reducible into smaller Markov chains containing absorbing states. In the cartoon plot, we exemplify this mechanism for $S=1$, with three reset states. For $\tilde{\Omega}<\tilde{\Omega}_c$, the Markov chain contains two absorbing reset states, namely $\ket{1}_N$ and $\ket{3}_N$. A Critical point $\tilde{\Omega}_c$ emerges when transitions between reset states acquire nonzero probability (red arrows).
    For $\tilde\Omega<\tilde{\Omega}_c$, the states $\ket{1}_N$ and $\ket{3}_N$ are no longer absorbing.
    The associated stationary density $p^{\mathrm{NESS}}_j$ for state $j$ displays collective behavior at $\tilde{\Omega}_c$. This manifests in first-order jump discontinuities ($j=1$). For integer spin $S$, additionally, second-order discontinuities in the first derivative are present ($j=2$).}
    \label{fig: disc. jump in exden and reset protocol}
\end{figure}

In this manuscript, we provide an overarching classification of the collective behavior that emerges from the interplay between coherent dynamics and stochastic resetting in noninteracting many-body spin $S$ systems, as depicted in Fig.~\ref{fig: disc. jump in exden and reset protocol}(a). Our classification for the order of the emergent nonanalytic behavior is solely based on time-reversal symmetry, which distinguishes half-integer from integer spin values. In the former case only second-order nonanalytic points can be found, while on the latter one strikingly observes multicritical behavior where both first and second-order nonanalyticities coexist [cf. Fig.~\ref{fig: disc. jump in exden and reset protocol}(b)]. Such transitions fall within an extended region of nonergodicity, where the stationary state depends on the initial state of the dynamics. 

We explain these results by putting forward an exact mapping of the quantum dynamics in the presence of global measurements and resets to a Markov chain model defined on the space of reset states, which is shown in Fig.~\ref{fig: disc. jump in exden and reset protocol}(b). The nice aspect about this correspondence relies on the fact that the complicated intertwining between quantum dynamics and monitoring is reflected into the transition probabilities of the Markov chain. Nonanalytic points emerge when some transition probabilities become zero and microscopic reversibility in the Markov chain is broken. The associated collective behavior can be therefore classified as a genuinely nonequilibrium phenomenon, where irreversibility is caused by the global monitoring. As such, it is also remarkably different from equilibrium phase transitions, e.g., in the Potts model \cite{potts1,potts2,potts3,potts4}. In this case, collective behavior is necessarily rooted into interactions among the microscopic constituents and it is crucially affected by space dimensionality. In our case, on the contrary, interactions are not necessary for the emergence of nonanalytic behavior. The stationary phase diagram we find consequently does not depend on the space dimensionality of the system. Our analysis thereby identifies a novel mechanism to induce emergent nonlinearities, which is solely based on global monitoring of the system via resets and measurements. This approach is robust with respect to any choice of the underlying lattice geometry and to interactions and could therefore be implemented in state-of-the-art quantum simulators, where global measurements and resets have already proven to be realizable \cite{rqh1,rqh2,rqh3,rqh4}.    

The rest of the manuscript is organized as follows. In Sec.~\ref{sec:summary_results}, we briefly review our main results. In Sec.~\ref{sec: Dynamics and reset protocol}, we introduce the noninteracting spin Hamiltonian and the reset states. We also recall, for the sake of comparison, the resetting protocol to a prescribed reset state (no measurement prior to resetting is performed). In Sec.~\ref{sec:conditional}, we introduce the conditional resetting protocol for the spin Hamiltonian (a magnetization measurement prior to resetting is performed). This is the central nonequilibrium protocol addressed throughout the whole manuscript. We explain how the associated stationary state can be computed via a mapping to a Markov chain defined in the discrete set of reset states. In Sec.~\ref{sec:results}, we apply this method to the case of spin $S=1,3/2$ and $2$ spins. In Sec.~\ref{sec:conclusions}, we draw our conclusions, while technical aspects of the presentation and additional calculations are reported in the appendices.

\section{Summary of main results}
\label{sec:summary_results}
The dynamics follow a conditional reset protocol with $2S+1$ possible reset states corresponding to the magnetization eigenstates. In this protocol, a global measurement of the magnetization right before resetting is performed, and the number of spins in each magnetization eigenstate is determined. The system is then reinitialized to the reset state corresponding to the magnetization measured for the majority of spins, see Fig.~\ref{fig: disc. jump in exden and reset protocol}(a). In the thermodynamic limit, we find that the combination of measurements and resets leads to collective behavior for all the values of $S$ studied. This is quantified by considering the stationary density of spins $p_j^{\mathrm{NESS}}$ in a given magnetization eigenstate $\ket{j}_{\alpha}$ ($j=1,2,\dots 2S+1$) at lattice site $\alpha$, 
\begin{equation}
p_j^{\mathrm{NESS}}=\mbox{Tr}[\ket{j}_{\alpha}\tensor*[_{\alpha}]{\bra{j}}{} \rho^{\mathrm{NESS}}], 
\label{eq:stationary_density_summary}
\end{equation}
where $\rho^{\mathrm{NESS}}$ is the density matrix describing the stationary state. The density of spins $p_j^{\mathrm{NESS}}$ plays the role of an order parameter for the stationary state. The three main results concerning this quantity are here summarized.    
\begin{enumerate}
\item \textbf{Mapping to Markov chain model}: We map a sequence of measurements and resets events in the conditional protocol into a sequence of jumps on a Markov chain model (in discrete time) defined over the $2S+1$ reset states. In Fig.~\ref{fig: MC S=1}, we report, as an example, the Markov chain describing a system of spin $S=1$, with three possible reset states. 
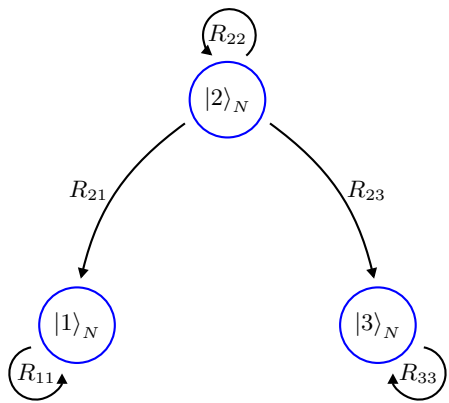
\begin{figure}[H]
\centering
\begin{tikzpicture}[myarrow/.style={-{Triangle[length=1.4mm, width=1.4mm]}, thick}]
  \def\circleRadius{0.5cm}
  \def\arrowdistance{0.05cm}
  \def\horizontalSpacing{2cm}
  \def\verticalSpacing{1.5cm}
  \def\selfarrowRadius{0.35cm}

  \node[draw=blue, circle, thick, minimum size=2*\circleRadius] (state1) at (-\horizontalSpacing, -\verticalSpacing) { $\ket{1}_N$};
  \node[draw=blue, circle, thick, minimum size=2*\circleRadius] (state2) at (0, \verticalSpacing) { $\ket{2}_N$};
  \node[draw=blue, circle, thick, minimum size=2*\circleRadius] (state3) at (\horizontalSpacing, -\verticalSpacing) { $\ket{3}_N$};
   \draw[myarrow] ($(state2.south west) + (-4*\arrowdistance, \arrowdistance)$) to[bend right=20] node[left]{ $R_{21}$} ($(state1.north) + (\arrowdistance, 2*\arrowdistance)$);

  \draw[myarrow] ($(state2.south east) + (4*\arrowdistance, \arrowdistance)$) to[bend left=20] node[right]{ $R_{23}$} ($(state3.north) + (-\arrowdistance, 2*\arrowdistance)$);

  \draw[myarrow] ($(state1.south west) + (-0.7*\selfarrowRadius, 0.2*\selfarrowRadius)$) arc (100:360:\selfarrowRadius);
  \draw[myarrow] ($(state2.north) + (0.5*\circleRadius, 0.2*\selfarrowRadius)$) arc (310:590:\selfarrowRadius);
  \draw[myarrow] ($(state3.south east) + (0.7*\selfarrowRadius, 0.2*\selfarrowRadius)$) arc (440:180:\selfarrowRadius);

  \node at ($(state1.south west) + (-0.5*\selfarrowRadius, -0.8*\selfarrowRadius)$) {$R_{11}$};
  \node at ($(state2.north) + (0, 1*\selfarrowRadius)$) {$R_{22}$};
  \node at ($(state3.south east) + (0.5*\selfarrowRadius, -0.8*\selfarrowRadius)$) {$R_{33}$};

\end{tikzpicture}
\caption{\textbf{Markov chain of reset states in a non-ergodic spin-$1$ system.} The reset states $\ket{1}_N$ and $\ket{3}_N$ are, in this case, absorbing. State $\ket{2}_N$ is transient.
}
\label{fig: MC S=1}
\end{figure}
The transition probabilities $R_{jk}$ from reset state $j$ to $k$ depend on the many-body quantum state just prior to the global measurement and therefore on the total number $N$ of spins. The spins are therefore effectively coupled by the global measurement in spite of the original Hamiltonian not being interacting. 
\item \textbf{Absorbing reset states}: As $N\to \infty$, some of the transition probabilities vanish and we find that \textit{absorbing reset states} appear in the Markov chain. An absorbing state $j$ is defined as 
\begin{equation}
R_{jk}=0  \,\, \mbox{when} \,\, k\neq j, \,\,\, R_{jj}=1.    
\end{equation}
As such these reset states violate microscopic reversibility of the Markov chain since once accessed, they cannot be any longer left. Mapping this back to the quantum dynamics, once a reset to one of those states takes place, all the subsequent resets of the many-body wavefunction take place towards the same state. Absorbing reset states are the reason for the emergence of nonanalytic behavior and nonergodicity in the stationary density \eqref{eq:stationary_density_summary}. 
\item \textbf{Nonanalytic behavior}: Each time system parameters are tuned so that transition probabilities out of an absorbing reset state become nonzero, nonanalytic behavior in the stationary state is obtained. An example is the activation of the transition probabilities $R_{12}$ and $R_{32}$ in Fig.~\ref{fig: MC S=1} and in the sketch in Fig.~\ref{fig: disc. jump in exden and reset protocol}(b). The order of the nonanalytic point is solely determined by time-reversal symmetry of the underlying unitary dynamics. It therefore depends on the spin being half integer or integer. In the former case, one has only first-order analytic points in the stationary density \eqref{eq:stationary_density_summary}, while in the latter, one finds both first and second order nonanalytic points. We name this physical phenomenon multicritical behavior.   
\end{enumerate}
The key technical tools at the basis of our analysis are the renewal structure \cite{evans2020review,nagar2023review} of stochastic resetting and Markov processes \cite{feller1957introduction}. Our results consequently establish a intriguing connection between ergodicity breaking, collective behavior in monitored quantum dynamics and irreversibility in stochastic processes. This hightlights a pathway towards the realization of complex stationary states, which can be analytically characterized and possess an high level of experimental control. As a matter of fact, the stationary behavior we unveil does not require to tune interactions and space geometries, but it is only based on global monitoring and resetting of a noninteracting dynamics. This might pave the way to the development of new paradigms in many-body physics, which are especially relevant in light of modern developments of quantum hardware. These platforms, indeed, provide a natural test bed for exploring and applying our results: resetting in these systems has already been demonstrated and shown to be a useful mechanism, e.g. in the context of quantum error correction \cite{rqh1,rqh2,rqh3,rqh4}. 

\section{Model and stochastic resetting dynamics}\label{sec: Dynamics and reset protocol}
In Subsec.~\ref{subsec:model_symm}, we define the noninteracting spin model that we consider in the whole text. In Subsec.~\ref{subsec:resetting_unc} we introduce stochastic resetting to a predefined state. This protocol will be used for comparison with the conditional resetting protocol that we develop in the next Sec.~\ref{sec:conditional}.
\subsection{Model and symmetries}
\label{subsec:model_symm}
We consider a system of $N$ noninteracting spins with Hamiltonian
\begin{equation}
H = \Omega \sum_{\alpha=1}^N \hat S^x_{\alpha} + \Delta \sum_{\alpha=1}^N \hat S^z_{\alpha}.
\label{eq:Hamiltonian}
\end{equation}
This Hamiltonian can be implemented with cold atoms on an optical lattice. The atoms are considered as distinguishable as they are located on different sites of the optical lattice, and the $2S+1$ spin states correspond, for instance, to $2S+1$ hyperfine electronic levels which are exposed to magnetic fields with longitudinal $\Delta$ and transverse $\Omega$ component \cite{S1cold1,S1cold2,S1cold3,S1cold4}. Other implementations via laser-excited Rydberg atoms are also possible \cite{S1Ryd1,S1Ryd2,S1Ryd3}. The operators $\hat S_\alpha^{x,y,z}$ are the spin matrices at lattice site $\alpha=1,2,\dots N$ for total spin $S$ with components $x,y,z$, and $N$ being the total number of spins. The basis states $\ket{j}_\alpha$ for each site $\alpha$ are chosen as the eigenstates of $\hat S^z_{\alpha}$. For total spin $S$, there are $2S+1$ eigenstates of $S^z_{\alpha}$ with $j \in \{1,2,\dots, 2S+1\}$. We will consider the total spins $S=1/2, 1, 3/2$ and $2$, meaning that we explore systems whose local Hilbert space has dimensions $2,3,4$ and $5$, respectively. Unitary dynamics dictated by the Hamiltonian \eqref{eq:Hamiltonian} starts from $N$-spin product states $\ket{j}_N$ built from the tensor product of single spin states\footnote{The relation between the spin eigenvalue in the $z$ direction $s_z\in [-S,S]$ and the index $j$ chosen to label the eigenstate is $s_z(j)=j-S-1$, so that $j=1$ is the label to eigenstate with spin $s_z(1)=-S$ and $j=2S+1$ to spin $s_z(2S+1)=S$.}: 
\begin{equation}
\ket{j}_N=\otimes_{\alpha=1}^N \ket{j}_\alpha, \quad j=1,2, \dots, 2S+1.
\label{eq:reset_states}
\end{equation} 
The motivation behind this choice is that such states can be straight-forwardly prepared experimentally, e.g., through optical pumping. Given that the states \eqref{eq:reset_states} are products of single-spin states, and the Hamiltonian \eqref{eq:Hamiltonian} is noninteracting, we have that the many-body density matrix $\rho_j^F(t)$ is a product state at all times $t$
\begin{equation}
\rho_j^F(t)= \otimes_{\alpha=1}^N \hat{U}_{\alpha}(t) \ket{j}_{\alpha}\tensor*[_{\alpha}]{\bra{j}}{}\hat{U}^{\dagger}_{\alpha},
\label{eq:product_state_t}
\end{equation}
with the single-spin unitary propagator $U_\alpha(t) =\mbox{exp}(-iH_\alpha t)$, where $H_\alpha =\Omega \hat S^x_\alpha + \Delta \hat S_\alpha^z$. Importantly, this propagator is invariant under time-reversal symmetry,
\begin{equation}
\Theta_{\alpha} \, \hat{U}_{\alpha}(t) \, \Theta_{\alpha}^{-1}=\hat{U}_{\alpha}(t),
\label{eq:time_reversal}
\end{equation}
which is implemented by the antiunitary operator $\Theta$ \cite{sakurai1994modern}
\begin{equation}
\Theta_{\alpha}= i \, e^{-i\pi \hat{S}^y_{\alpha}} K.
\label{eq:antiunitary}
\end{equation}
Here, $K$ is an operator which takes the complex conjugate of any coefficient multiplying the state on which $\Theta_{\alpha}$ is acting. The eigenvalue of the $\Theta_{\alpha}^2$ operator conveniently distinguishes half-integer from integer spins
\begin{equation}
\Theta_{\alpha}^2\ket{j}_{\alpha}=(-1)^{2S}\ket{j}_{\alpha}. 
\label{eq:t_reversal}
\end{equation}
The different behavior of half-integer and integer spins under time reversal will later underpin the different kind of collective behavior observed in the two cases. In particular, the symmetry \eqref{eq:time_reversal} has important effects on reset-free spin dynamics on the probability $p_{jk}^{F,\alpha}(t)$ that the spin at site $\alpha$ initialized in the state $\ket{j}_\alpha$ is measured in state $\ket{k}_{\alpha}$ at time $t$. This probability is expressible in terms of single-spin dynamics only
\begin{equation}
p^{F,\alpha}_{jk}(t)=\mbox{Tr}[\rho_j^F(t) \ket{k}_{\alpha}\tensor*[_{\alpha}]{\bra{k}}{}]=\vert \tensor*[_{\alpha}]{\braket{k | \hat U_\alpha(t) |j}}{_\alpha}\vert^2.
\label{eq:single_particle_propagator}
\end{equation}
It obeys the following symmetry relations from \eqref{eq:time_reversal}
\begin{equation}
p_{jk}^{F,\alpha}(t)\!=\!p_{kj}^{F,\alpha}(t)\!=\!p_{\tilde{j}\,\tilde{k}}^{F,\alpha}(t)\!=\!p_{\tilde{k}\,\tilde{j}}^{F,\alpha}(t),
\label{eq:symmetry_p}
\end{equation}
with the label $\tilde{x}=\tilde{j},\tilde{k}$ of the magnetization eigenstate being related to $x=j,k$ as
\begin{equation}
\tilde{x}=2S+2-x, \quad x=1,2,\dots 2S+1.
\label{eq:symmetry}    
\end{equation}
The proof of \eqref{eq:symmetry_p} is detailed in Appendix \ref{app:t_reversal}. The latter equation implies that, as a a consequence of time-reversal invariance, the probability of a coherent propagation from any state $\ket{j}_{\alpha}$ to $\ket{k}_{\alpha}$ is equal to the probability of propagation for the reverse transition (from $\ket{k}_{\alpha}$ to $\ket{j}_{\alpha}$). In addition, we also see that eigenstates of the magnetization $S^z_{\alpha}$ can be grouped in symmetry pairs, with the state $\ket{x}_{\alpha}$ ($\ket{x}_{N}$) being paired with $\ket{\tilde{x}}_{\alpha}$ ($\ket{\tilde{x}}_N$) and $\tilde{x}$ from \eqref{eq:symmetry}. In terms of the eigenvalue $s_z\in [-S,S]$ of $S_{\alpha}^z$, symmetric pairs are made of states with opposite magnetization in the $z$ direction $s_z(x)=-s_z(\tilde{x})$ (cf. the footnote after Eq.~\eqref{eq:antiunitary}). This follows from the fact that the quantum number $s_z\to -s_z$ is reversed under time-reversal symmetry. For integer spin $S$, the state $\ket{S+1}_{\alpha}$ ($\ket{S+1}_{N}$), corresponding to spin eigenvalue in the $z$ direction $s_z=0$, has no symmetric partner and will be called henceforth the center state. Note that, because of translational invariance, the single spin propagation probability $p^{F,\alpha}_{ij}(t)$ is independent of the site $\alpha$. We will therefore drop the symbol $\alpha$ henceforth to ease notation. The explicit analytical expressions of Eq.~\eqref{eq:single_particle_propagator} for the studied values of $S$ are reported in Appendix \ref{app:appendix_1} ($S=1$), \ref{app:S=1.5} ($S=3/2$) and \ref{app:S=2} ($S=2$). In Eqs.~\eqref{eq:product_state_t}-\eqref{eq:symmetry_p}, the superscript $F$ highlights that $\rho_i^F(t)$ and $p^F_{ij}(t)$ characterize the reset-free unitary dynamics. We now consider the addition of stochastic resets on top of such dynamics. 

\subsection{Unconditional resetting}
\label{subsec:resetting_unc}
The simplest way to implement a resetting protocol is by reinitializing all the $N$ spins simultaneously to a predefined reset state $\ket{r}_N$ at stochastically distributed times. We refer henceforth to this protocol as \textit{unconditional} resetting to distinguish it from the conditional resetting protocol of the next section. The time $\tau>0$ elapsing between two consecutive resets is distributed according to a waiting-time probability density function $f(\tau)$. In the whole manuscript, we consider for the sake of illustration purposes the case of resetting at a constant rate $\gamma$. This is called Poisson resetting, which is described by the waiting time distribution 
\begin{equation}
f(\tau)=\gamma e^{-\gamma \tau}.
\label{eq:waiting_time_distribution}
\end{equation}
The main results are, however, independent of the choice for the waiting time distribution. The many-body reset state $\ket{r}_N=\ket{j}_N$ is chosen as one of the states in Eq.~\eqref{eq:reset_states}. The density matrix $\rho_j(t)$ describing the many-body spin system in the presence of resets to state $\ket{j}_N$ can be obtained from the knowledge of the reset-free unitary via the \textit{last renewal equation} \cite{Mukherjee2018,evans2020review} 
\begin{equation}
    \rho_j(t)=e^{-\gamma t} e^{-iHt} \ket{\psi_0}\bra{\psi_0} e^{iHt}
    + \gamma \int _0 ^t dt' e^{\gamma t'} \rho^F_j(t').
\label{eq:last_renewal_time}
\end{equation}
The first term on the right-hand side describes realization of the process where no reset event takes place in the time interval $(0,t)$, and it is therefore given by the unitary dynamics from the initial state $\ket{\psi_0}$. This is weighted by the survival probability $\mbox{exp}(-\gamma t)$, i.e., the probability of not resetting after a time $t$ since the previous reset. The second term describes realizations where the last reset happened at time $t-t'$, with rate $\gamma$, and then coherent dynamics take place for a time duration $t'$.
From Eq.~\eqref{eq:last_renewal_time}, a NESS at long times $t\to \infty$ is attained
\begin{equation}
\label{eq:density matrix ness}
    \rho_j^{\mathrm{NESS}}=\gamma \int _0 ^\infty dt' e^{-\gamma t'} \rho^F_j(t'). 
\end{equation}
The NESS therefore, does not depend on the initial state $\ket{\psi_0}$, whose contribution in the renewal equation \eqref{eq:last_renewal_time} is transient in time. The stationary state density matrix \eqref{eq:density matrix ness} is a separable state since it is expressed as a convex combination of the product states \eqref{eq:product_state_t}. In classical systems, a similar structure in the joint probability density of $N$ independent Brownian walkers that are simultaneously reset to the same point has been found in Ref.~\cite{Biroli_extreme}. From Eq.~\eqref{eq:density matrix ness} expectation values of observables can be computed. In particular, we here compute from Eqs.~\eqref{eq:single_particle_propagator} and \eqref{eq:density matrix ness} the stationary probability $p_{jk}^{\mathrm{NESS},\mathrm{uc}}$ of finding a spin in state $\ket{k}_{\alpha}$ in the presence of resets to state $\ket{j}_N$:
\begin{equation}
    p_{jk}^{\mathrm{NESS},\mathrm{uc}}=\mathrm{Tr}\left[ \ket{k}_{\alpha}\tensor*[_{\alpha}]{\bra{k}}{} \rho_{j}^{\mathrm{NESS}}\right]=\gamma \int _0 ^\infty dt\, e^{-\gamma t} p_{jk}^F (t).
\label{eq:ness_pij_unconditional}
\end{equation}
In Appendix \ref{app:appendix_1}, \ref{app:S=1.5} and \ref{app:S=2}, we report the explicit analytic expressions for the excitation densities in Eq.~\eqref{eq:ness_pij_unconditional} for spin $S=1,3/2$ and $2$, respectively. Importantly, we find that, in all the cases, $p_{jk}^{\mathrm{NESS},\mathrm{uc}}(\Omega,\Delta)$ are analytic functions of the ratio $\Omega/\Delta$. Within the unconditional reset protocol, no collective behavior is therefore found in the stationary state. The physics motivation stems from the fact that within the unconditional reset protocol, $p_{jk}^{\mathrm{NESS},\mathrm{uc}}$ is completely determined by single spin dynamics and it is, indeed, independent of the number of spins $N$. In the next section, we show that for the \textit{conditional} resetting protocol, instead, the measurement of the many-body state prior to resetting introduces fluctuations explicitly dependent on $N$. As a consequence, collective behavior can be found in the thermodynamic limit.

\section{Conditional resetting}
\label{sec:conditional}
In Subsec.~\ref{subsec:measurement}, we define the global measurement protocol at the basis of the conditional choice of the reset reset in the conditional reset protocol. In Subsec.~\ref{subsec:stationary_ergodic}, we map the ensuing stationary state to a Markov chain defined in the space of reset states. In Subsecs.~\ref{subsec:nonergodic_emergence} and \ref{Subsec:weight_vec_nonergodic}, we show how the stationary state is computed when nonergodicity in the reset Markov chain arises as $N\to \infty$. 
\subsection{Measurement protocol}
\label{subsec:measurement}
In the \textit{conditional reset} protocol, the unitary dynamics according to Eq.~\eqref{eq:Hamiltonian} is interspersed with projective measurements and resets, the latter being performed right after the former. The reset state is chosen among the $2S+1$ states \eqref{eq:reset_states} on the basis of the outcome of a global projective measurement of the magnetization density $s^z$ in the $z$ direction
\begin{equation}
s^z=\frac{1}{N}\sum_{\alpha=1}^N S^z_{\alpha}.
\label{eq:measurement_global}
\end{equation}
This kind of measurement finds natural motivation and realization from quantum-gas microscope experiments \cite{Qmicroscope_review,Qmicroscope_1,Qmicroscope_2,Qmicroscope_3}, where site-resolved measurements are possible. Single-shot measurements of the global magnetization of the system can therefore also be performed. From this measurement, one then computes the fraction $n_j$ of spins measured in the state $j$ out of the total number $N$. Immediately after the measurement, the system is reinitialized to a state chosen from the set of $2S+1$ reset states \eqref{eq:reset_states}. The choice of the reset state is dictated by the majority rule, i.e., one picks the state $\ket{r}_N$ corresponding to the magnetization where the majority of spins were measured: $r=\mbox{argmax}_{j=1,2,\dots 2S+1} n_j$. Unitary dynamics then starts over for a time duration $\tau$ from the chosen reset state until the next measurement and reset are performed. The time $\tau$ elapsing between consecutive resets is a random variable chosen from the waiting time distribution $f(\tau)$ \eqref{eq:waiting_time_distribution}, similarly to the unconditional reset protocol. The main difference from the latter therefore lies in the fact that the choice of the reset state now depends on the instantaneous many-body state of the system via the measurement. This is the key mechanism leading to collective behavior even in the absence of interactions among the spins.  

The conditional resetting protocol represents a semi-Markov process, see, e.g., Ref.~\cite{janssen2006applied}, since it is entirely characterized by the probability $W_{jk}^{(N)}(\tau)$ of resetting to state $\ket{k}_N$, given a time $\tau$ elapsed since the previous reset to $\ket{j}_N$. This probability can be immediately written on the basis of the Born rule for the outcome of a measurement on a quantum system
\begin{equation}
W_{jk}^{(N)}(\tau)= \tensor*[_{N}]{\braket{\psi_j(\tau)|  \mathcal{P}_k|\psi_j(\tau)}}{_N},
\label{eq:W_general}
\end{equation}
with $\ket{\psi_j(\tau)}_N=\mbox{exp}(-i H t) \ket{j}_N$ the many-body unitary evolution from state $j$, and 
\begin{equation}
\mathcal{P}_k=\sum_{C_k} \ket{S^z}\bra{S^z}
\label{eq:projector}
\end{equation}
the projector over the sector of the Hilbert space $C_k$ spanned by the eigenstates $\ket{S^z}$ of $S^z=\sum_{\alpha=1}^N S^z_{\alpha}$ with a majority of spins in the state $k$. We therefore define $\{m_s\}_{s=1}^{2S+1}$ as the set of numbers $m_s$, with $\sum_{s=1}^{2S+1}m_s=N$, for which a single-spin eigenvalue $s$ of $S^z_{\alpha}$ is found in the eigenstate $\ket{S^z_{\alpha}}$. Exploiting the product state structure of the reset states \eqref{eq:reset_states} and of the Hamiltonian \eqref{eq:Hamiltonian}, one can simplify the expression of $W_{jk}^{(N)}(\tau)$ to 
\begin{equation}
W_{jk}^{(N)}(\tau)=\sum_{m_1, \dots m_s, \dots m_{2S+1}}^{k'} P_{j}(N,\{m_s\},\tau),
\label{eq:W_intermediate}
\end{equation}
where the superscript $k'$ denotes that the sum is restricted only to those configurations where $m_k=\mbox{max}_s{m_s}$, i.e., where the state $\ket{k}_{\alpha}$ is measured for the majority of spins. Here, $P_{j}(N,\{m_s\},\tau)$ denotes the probability of measuring a set of $\{m_s\}$ spins in the corresponding eigenstate $s$ at time $\tau$ after the prior reset to $j$. This is given by the multinomial distribution due to the noninteracting structure of the Hamiltonian
\begin{align}
\label{eq:multinomial_distribution}
   P_{j}(N, \{m_s\}, \tau)=& \binom{N}{m_1,\dots m_s, \dots m_{2S+1}} (p^F_{j1}(\tau))^{m_1} \dots \nonumber \\
   &\, \,\, \quad \quad (p^F_{js}(\tau))^{m_s}\dots (p^F_{j 2S+1}(\tau))^{m_{2S+1}}  .
\end{align}
Here $p_{jk}^F(\tau)$ is the single-spin probability \eqref{eq:single_particle_propagator}. From Eq.~\eqref{eq:multinomial_distribution} we see that for a finite number of spins $N$, the probability $p^F_{jk}(\tau)$ determines only the mean number of spins found in the state $\ket{k}_{\alpha}$, which is given by $N p^F_{jk}(\tau)$. Fluctuations from the average value are, however, possible and they reflect the projection noise characterizing measurements in quantum systems. In addition to this, in Eq.~\eqref{eq:multinomial_distribution}, the time $\tau$ elapsing between two consecutive measurements-resets is a random variable distributed according to the waiting time distribution $f(\tau)$. It is then convenient to combine the latter with Eq.~\eqref{eq:waiting_time_distribution} by defining the transition matrix $\hat{R}_{jk}^{(N)}(\tau)$:
\begin{equation}\label{eq:def_R}
    \hat{R}_{jk}^{(N)}(\tau)=f(\tau)\, W_{jk}^{(N)}(\tau)=\gamma e^{-\gamma \tau}\, W_{jk}^{(N)}(\tau),
\end{equation}
which gives the probability density of resetting to state $\ket{k}_{N}$ and in the time interval $(\tau,\tau+d\tau)$ from the previous reset to $\ket{j}_N$. The symmetry relations among the spins unitary probabilities from time-reversal invariance in Eq.~\eqref{eq:symmetry_p} impact the structure of the matrix $\hat{R}^{(N)}$. The latter inherits the following symmetry property
\begin{equation}
\hat{R}_{jk}^{(N)}(\tau)=\hat{R}_{\tilde{j}\,\tilde{k}}^{(N)}(\tau),
\label{eq:symmetry_R}
\end{equation}
from the analogous symmetry among the reset-free probabilities and $\tilde{j}$ ($\tilde{k}$) defined as in Eq.~\eqref{eq:symmetry}. In general, we can see that the transition probabilities are \textit{not} symmetric $\hat{R}_{j,k}^{(N)} \neq \hat{R}_{k,j}^{(N)}$ for any pair of states $\ket{j}_N$ and $\ket{k}_N$, differently from the unitary propagators \eqref{eq:symmetry_p}. This is physically a consequence of the monitoring of the system via projective measurements, which introduces irreversibility into the dynamics. It in general causes the asymmetry between the transition probabilities, which do connect a pair of states corresponding to opposite magnetization eigenvalues [as in Eq. \eqref{eq:symmetry_R}]. The relation \eqref{eq:symmetry_R} characterizes, specifically, the matrix $\hat{R}^{(N)}$ to be centrosymmetric.
\subsection{Stationary state in the ergodic regime}
\label{subsec:stationary_ergodic}
\begin{figure}[H]
\centering

\begin{tikzpicture}[myarrow/.style={-{Triangle[length=1.4mm, width=1.4mm]}, thick}]
  \def\circleRadius{0.5cm}
  \def\arrowdistance{0.1cm}
  \def\horizontalSpacing{3.5cm}
  \def\selfarrowRadius{0.35cm}

  \node[draw=blue, circle, thick, minimum size=2*\circleRadius] (state1) at (-\horizontalSpacing/2, 0) { $\ket{1}_N$};
  \node[draw=blue, circle, thick, minimum size=2*\circleRadius] (state2) at (\horizontalSpacing/2,0) { $\ket{2}_N$};

  \draw[myarrow] ($(state2.west) + (-\arrowdistance, 2*\arrowdistance)$) to[bend right=15,<-] node[above]{ $R_{21}$} ($(state1.east) + (\arrowdistance, 2*\arrowdistance)$);
  \draw[myarrow] ($(state1.east) + (\arrowdistance, -2*\arrowdistance)$) to[bend right=15,<-] node[below]{ $R_{12}$} ($(state2.west) + (-\arrowdistance, -2*\arrowdistance)$);

  \draw[myarrow] ($(state1.west) + (-0.35*\selfarrowRadius, 0.7071*\selfarrowRadius)$)  arc (60:320:\selfarrowRadius);
  \draw[myarrow] ($(state2.east) + (0.35*\selfarrowRadius, 0.7071*\selfarrowRadius)$)  arc (130:-130:\selfarrowRadius);

  \node at ($(state1.west) + (-0.9*\selfarrowRadius, 0)$) {$R_{11}$};
  \node at ($(state2.east) + (0.9*\selfarrowRadius, 0)$) { $R_{22}$};

\end{tikzpicture}

\caption{\textbf{Markov chain of reset states in ergodic spin-$1/2$ system.} In the thermodynamic limit $N\to \infty$ when $\tilde\Omega>\tilde \Omega_c$, sequences of resets from state $\ket{1}_N$ to $\ket{2}_N$ (and viceversa) become possible. Accordingly, $R_{12}=R_{21}\neq 0$ and the Markov chain becomes ergodic. The definition of the critical value $\tilde{\Omega}_c$ of the ratio $\tilde{\Omega}=\Omega/\Delta$ is reported in Eq.~\eqref{eq: def of om_c} below. Note that both transition probabilities become simultaneously nonzero as a consequence of the symmetry \eqref{eq:symmetry_R}.}
\label{fig: MC S=0.5}
\end{figure}
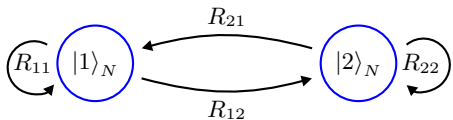

With the developed tools, one can eventually derive the expression for the stationary density matrix $\rho^{\mathrm{NESS}}$. This is obtained by averaging over all possible realizations of the reset dynamics. By considering an observation time $t$ and the density matrix $\rho(t)$, one namely averages over the possible number $n=0,1,2 \dots $ of resets taking place within the observation time and on the possible outcomes for the reset state $\ket{k}_N$ according to the matrix $\hat{R}_{jk}(\tau)$. This procedure has been detailed in Ref.~\cite{perfetto2021designing} and we thereby here provide only the final expression for the stationary state density matrix $\rho^{\mathrm{NESS}}=\lim_{t\to\infty}\rho(t)$. This reads for generic total spin $S$ 
\begin{equation}
\label{eq:ness}
    \rho^{\mathrm{NESS}}=\gamma \sum_{m=1}^{2S+1} c_m \int_0^\infty dt' e^{-\gamma t'} \rho^F_m(t'),
\end{equation}
with the stationary probability for a spin in state $\ket{j}_{\alpha}$ consequently written as
\begin{equation}
\label{eq:exciations density - general}
    p^{\mathrm{NESS}}_j=\gamma  \sum_{m=1}^{2S+1}  c_m \int_0^\infty dt' e^{-\gamma t'}\ p^F_{mj}(t') \, .
\end{equation}
The formulas \eqref{eq:ness} and \eqref{eq:exciations density - general} extend Eqs.~\eqref{eq:density matrix ness} and \eqref{eq:ness_pij_unconditional} to the conditional reset protocol. The coefficients $c_m$ ($m=1,2,\dots 2S+1$) give the stationary probability of resetting to the state $\ket{m}_{N}$ and non-trivially embody the combined effect of the monitoring via resetting and measurements. They can be computed from the knowledge of the matrix $\hat{R}_{jk}^{(N)}$. To do so, we define its time-integral:
\begin{equation}
R_{jk}^{(N)}=\int_0^{\infty} d \tau \hat{R}_{jk}^{(N)}(\tau),
\label{eq:R_matrix_final}
\end{equation}
which defines a Markovian matrix, since $\sum_k R_{jk}^{(N)}=1$. The vector of steady-state weights $\vec{c}^{\ T}=(c_1, c_2,...,c_{2S+1})$ in Eq.~\eqref{eq:ness} is given as the left eigenvector of the transition matrix $R^{(N)}$ \eqref{eq:R_matrix_final} with eigenvalue $1$:  
\begin{equation}
    \vec{c}^{\ T}=\vec{c}^{\ T} R^{(N)} \, .
\label{eq:c_eigenvalue}
\end{equation}
This formula allows us to develop a clear connection between the stationary state \eqref{eq:ness} in the conditional resetting protocol and a Markov chain, in discrete time, model defined in the space of $\ket{j}_N$ reset states, with $j=1,2\dots 2S+1$. A sequence of resetting events is understood as a sequence of jumps in a Markov chain whose transition probabilities \eqref{eq:R_matrix_final} are determined by the waiting time distribution and the measurement monitoring. In Fig.~\ref{fig: MC S=0.5}, we report the simplest example of such a Markov chain for the case of spin $1/2$. The stationary distribution $\vec{c}$ of the Markov chain can then be readily evaluated from the eigenvalue equation \eqref{eq:c_eigenvalue} and therefore the stationary-state density matrix $\rho^{\mathrm{NESS}}$ \eqref{eq:ness} is fully specified. 

At this point, it is, however, crucial to highlight that this method applies only if the Markov-chain model defined by $R_{jk}^{(N)}$ is \textit{ergodic} and therefore the stationary probability distribution $\vec{c}$ given by the eigenvalue equation \eqref{eq:c_eigenvalue} is unique. We refer to this regime as \textit{ergodic}. The derivation of Eq.~\eqref{eq:ness}, indeed, assumes that the system at a given time $t$ can be reset to any state $\ket{j}_N$. This translates in the Markov-chain model to the fact that any state $\ket{j}_N$ can be reached in a certain number of time-steps regardless of the initial condition. This is precisely the definition of an ergodic Markov chain, which is irreducible into smaller chains and aperiodic. The latter ensures that the stationary distribution \eqref{eq:c_eigenvalue} is always reached regardless of the initial condition and the system does not get trapped into persistent oscillations, see the discussion in Refs.~\cite{feller1957introduction,tutorial_Markov}. Aperiodicity of the Markov chain $R_{jk}^{(N)}$ is always ensured by the presence of self-transitions ($R_{jj}^{(N)}\neq 0$), while irreducibility fundamentally depends on the number $N$ of spins. For finite $N$, $R_{jk}^{(N)}>0$ for any pair of states $(j,k)$ according to Eqs.~\eqref{eq:W_intermediate}-\eqref{eq:symmetry_R} and therefore the stationary state \eqref{eq:ness} can be computed by inserting the unique solution of Eq.~\eqref{eq:c_eigenvalue} for weight vector $\vec{c}$ and it does not depend on the initial state.

\subsection{Emergence of nonergodicity as $N\to \infty$}
\label{subsec:nonergodic_emergence}
For $N\to \infty$, however, we find that some transition probabilities can become zero as a function of $\Omega$ and $\Delta$. The Markov chain then becomes reducible, and therefore nonergodic. This reflects the fact that the stationary densities of spins $p_j^{\mathrm{NESS}}$ depend on the initial state. We refer to this behavior as \textit{nonergodic} and we now explain how it emerges in the thermodynamic limit $N\to \infty$. 

When we consider the fraction of spins $\{n_s=m_s/N\}$, with $s=1,2,\dots 2S+1$, measured in a state $s$, the corresponding probability large $N$ can be obtained by approximating the multinomial distribution \eqref{eq:multinomial_distribution} with a Gaussian, see, e.g., Ref.~\cite{gnedenko2018theory}. This specifically reads
\begin{equation}
P_{j}(N\to\infty,\{n_s\},\tau) \propto \mbox{exp}\left(-\frac{N}{2} \sum_{s=1}^{2S+1} \frac{(n_s-p_{js}^F(\tau))^2}{p_{js}^F(\tau)}\right).
\label{eq:multinomial_approx}
\end{equation}
This form is a consequence of the central limit theorem, and it shows that for each spin state $s$ as $N\to \infty$ fluctuations in the outcome of measurement shrink and the measured fraction of spins $n_s$ approaches the mean value $p_{is}^F(\tau)$. Mathematically, we have that the probability $P_{j s}(N\to \infty,n_s,\tau)$ of measuring a fraction $n_s$ of spins in state $s$ concentrates around its mean value 
\begin{equation}
    P_{j s}(N\to \infty,n_s,\tau)=\delta(n_s-p_{js}^F(\tau)).
\label{eq:law_large_numbers}
\end{equation}
Accordingly, in the thermodynamic limit, the outcome of a measurement of the density of spins $n_s$ in a given single-particle state $\ket{s}_{\alpha}$ is deterministic and fully determined by the mean value $p_{js}^F(\tau)$. The projection noise from the measurement is thus suppressed in the thermodynamic limit. 

The concentration property \eqref{eq:law_large_numbers} is at the root of the collective behavior emerging in the stationary state and nonergodicity for the stationary occupation probabilities $p_j^{\mathrm{NESS}}$ in Eq.~\eqref{eq:exciations density - general}. In the Markov chain picture, in the limit $N\to \infty$, some transition probabilities \eqref{eq:W_intermediate} vanish, rendering the latter reducible. This can be readily realized by taking the limit $N\to \infty$ of Eq.~\eqref{eq:W_intermediate} and inserting the result \eqref{eq:law_large_numbers}:
\begin{equation}\label{eq:def_W}
    W^{(\infty)}_{jk}(t)=\bigg\{
    \begin{array}{ll}
    1 & p^F_{jk}(t)=\max_{m=1}^{2S+1} p^F_{jm}(t),\\
    0 & \,  \textrm{else} \\
    \end{array}.
\end{equation}
The reset state $\ket{k}_N$ chosen after the measurement is therefore fully determined by the mean densities $p_{js}^F$, as sketched in Fig.~\ref{fig: disc. jump in exden and reset protocol}(b). It then clear that, if $p_{jj}^F(\tau)>p_{js}^F(\tau)$, with $s\neq j$, reset to states $s\neq j$ become impossible. In terms of the Markov chain rates from Eq.~\eqref{eq:def_W} and \eqref{eq:R_matrix_final}, we see that 
\begin{equation}
p_{jj}^F(\tau)>p_{js}^F(\tau), \,\,\, \forall \tau \,\to\, R_{j*}=0, \, R_{jj}=1.
\label{eq:absorbing_state}
\end{equation}
This equation defines an \textit{absorbing-reset state}: if the system gets reset to $\ket{j}_{\alpha}$, then all subsequent resets necessarily take place to the same state $\ket{j}_{\alpha}$. Here and henceforth, for lightness of notation, we denote by $R_{jk}^{(\infty)}\equiv R_{jk}$ the transition probabilities obtained in the thermodynamic limit. The validity of Eq.~\eqref{eq:absorbing_state} depends on the ratio $\tilde{\Omega}=\Omega/\Delta$. Consequently, it is possible to define a critical value $\tilde \Omega_{c}$ at which resets with probability $R_{jk}$, with $k\neq j$, out of state $\ket{j}_N$ become nonzero. In general, the critical values $\tilde\Omega_{c}$ are numerically computed in this work, as the values of $\tilde\Omega$ for which a $R_{jk}$ becomes nonzero while being zero for all $\tilde\Omega<\tilde\Omega_{c}$ according to \eqref{eq:def_W} and \eqref{eq:R_matrix_final}. In formulas:
\begin{equation}\label{eq: def of om_c}
    \tilde{\Omega}_c = \min \{ \tilde{\Omega} \mid R_{jk} > 0 \}\, .
\end{equation}
Note that this condition includes both the case where the state $j$ was formerly absorbing or not. In the specific case where $j$ and/or $k$ are absorbing reset states, we find that the stationary spin densities \eqref{eq:density matrix ness} features nonanalytic behavior in the form of a jump discontinuity or a discontinuity in the first derivative at $\tilde{\Omega}_c$. These nonanalytic points identify collective behavior.

\begin{figure}[t!]
    \centering
    \def\svgwidth{0.48\textwidth} 
    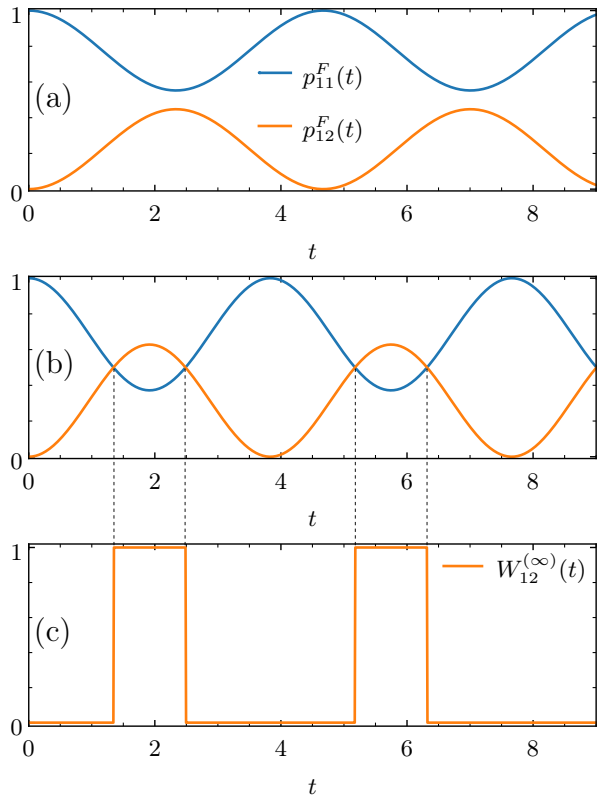
   \caption{\textbf{Reset-free evolution and Markov transition probabilities for spin $1/2$.}
   In panels (a) and (b), we plot reset-free single-particle occupation probabilities $p^F_{is}(t)$ as a function of the time $t$ after a reset to state $\ket{1}_N$, with $s=1,2$. In panel (a), one has $\tilde\Omega=0.9<\tilde{\Omega}_c=1$. It then holds at $p^F_{11}(t)>p^F_{12}(t)$. The state $\ket{1}_N$ is therefore an absorbing reset state and all subsequent resets will be to this state. In panel (b), it is $\tilde\Omega=1.3>\tilde{\Omega}_c$. Resets to the state $\ket{2}_N$ are now allowed.
   In panel (c), we plot the probability  $W^{(\infty)}_{12}(t)$ of resetting to state $2$ at time $t$ after the last reset to state $1$ for $\tilde\Omega=1.3$ as in panel (b). In the thermodynamic limit $N\to \infty$, $W^{(\infty)}_{12}(t)$ is fully determined by the mean occupation probabilities $p^F_{1s}(t)$. We have that, $W^{(\infty)}_{12}(t)=1$ for all the times $t$ such that $p^F_{12}(t)>p^F_{11}(t)$ holds. Otherwise, it is $W^{(\infty)}_{ij}(t)=0$. In the figure we set $\Delta =1$.}
    \label{fig: Free evolution s=0.5 + step function}
\end{figure}

As long as at least one absorbing reset state is present, then the Markov chain is reducible and therefore nonergodic. In general, we find that for small $\tilde{\Omega}$, the Markov chain model is reducible to $2S+1$ noncommunicating classes made of $2S+1$ absorbing states. As $\Omega$ is increased, new transitions have nonzero probability and the corresponding critical points are defined as in Eq.~\eqref{eq: def of om_c}. These new transitions reduce the number of noncommunicating subclasses in the Markov chain. For large $\tilde{\Omega}$, we eventually find resets towards all the reset states $\ket{s}_N$ take place and therefore the Markov chain is irreducible (made of a single class of states) and ergodic. Once a transition is, indeed, allowed for a certain $\tilde\Omega_{c}$, it persists for all values of $\tilde\Omega$ larger than $\tilde\Omega_{c}$. In Fig.~\ref{fig: Free evolution s=0.5 + step function}, we exemplify this concept for the simplest case of total spin $S=1/2$ and two reset states (see Subsec.~\ref{subsec:qubit} later for more details). Here we can identify a single critical value $\tilde\Omega_{c}=1$. For $\tilde{\Omega}<\tilde{\Omega}_c$ and the system being initialized in state $\ket{1}_N$, cf. Fig.~\ref{fig: Free evolution s=0.5 + step function}(a), equation \eqref{eq:absorbing_state} is satisfied for $j=1$. The latter is an absorbing reset state, and the two-state Markov chain reduces to two Markov chains made of a single state only. For $\tilde{\Omega}>\tilde{\Omega}_c$, instead, $R_{12}\neq 0$, since there are times $t$ for which $p^F_{12}(t)>p^F_{11}(t)$, as shown in Fig.~\ref{fig: Free evolution s=0.5 + step function}~(b)-(c). The transition probability simultaneously becomes $R_{21} \neq 0$ as a consequence of the symmetry \eqref{eq:symmetry_R}. The Markov chain then becomes irreducible and ergodic, as depicted in Fig.~\ref{fig: MC S=0.5}. This marks the transition from the nonergodic to the ergodic regime. 

Importantly, the eigenvalue equation \eqref{eq:c_eigenvalue} admits multiple solutions whenever the underlying Markov chain is reducible and nonergodic. One therefore needs additional analysis to uniquely fix the stationary-state weight $\vec{c}$ needed for \eqref{eq:ness}. We explain how to determine the stationary reset probabilities $\vec{c}$ for a nonergodic resetting Markov chain in the next section.

\subsection{Stationary state in the nonergodic regime as $N\to \infty$}
\label{Subsec:weight_vec_nonergodic}
In the nonergodic regime, we have at least two absorbing reset states. As a matter of fact, from the symmetry relation \eqref{eq:symmetry_R}, we find that if the state $\ket{j}_N$ is absorbing, then also $\ket{2S+2-j}_N$ is. The presence of absorbing states clearly makes the eigenvalue equation \eqref{eq:c_eigenvalue} degenerate. Indeed, if, for instance, $\ket{1}_N$ and $\ket{2S+1}_N$ are absorbing states, then both $c_1=(1,0,0,...,0)^T$ and $(0,0,0,\dots,1)^T$ are left eigenvector of $R$ with eigenvalue $1$. This property reflects the fact that at long times the system surely ends up in one of the absorbing states. All the other states are therefore named transient and have zero-stationary probability. The stationary occupation probabilities of the absorbing-states, and therefore the stationary vector $\vec{c}$, then depend on the chosen initial state.  

To compute $\vec{c}$ in this case, we follow the approach taken for Markov chains presenting absorbing states, see, e.g., chapter 15 of Ref.~\cite{feller1957introduction}. We divide reset states into two sets. The first set $\mathcal{A}$ is made of $u$ absorbing reset states, according to Eq.~\eqref{eq:absorbing_state}, while the second set $\mathcal{T}$ is made of $b$ transient states. We then rearrange the entries of the transition matrix $R$ into a new matrix $R'$ so that the rows and columns corresponding to absorbing states come first. This rearranging is equivalent to renumbering the $2S+1$ states in such a way that, in the case of $u$ absorbing states and $b$ non-absorbing states, they are indexed $1, 2,\dots u$ and the non-absorbing are indexed $u+1, u+2, ..., u+b$. The rows of the absorbing states form an identity matrix $I_u$ with dimension $u\times u$ in the upper left corner of the rearranged transition matrix $R'$. 
Then we construct a transition matrix $B$, with dimension $b \times u$, that accounts for the transitions from the non-absorbing states to the absorbing ones. In the bottom right corner, one eventually has the $b \times b$ matrix $Q$ that accounts for transitions among the transient states:
\begin{equation}
\label{eq: transition matrix form for formalism}
    R' = \left( \begin{array}{c c}
    I_u & 0 \\ 
    B & Q
\end{array} \right).
\end{equation}
In Fig.~\ref{fig: MC S=1}, we show as an example the case of total spin $S=1$, where the center state $\ket{2}_N$ is transient ($b=1$), while the state $\ket{1}_N$ and $\ket{3}_N$ are absorbing ($u=2$).
If now the system is initialized into a transient reset state $j\in \mathcal{T}$, the stationary probability $P^{a}_j$ that the system ends up in one absorbing state $a\in \mathcal{A}$ satisfies
\begin{equation}
P^{a}_j=\sum_{\nu \in \mathcal{T}}R'_{j\nu}P^{a}_\nu +  R'_{j a}.
\label{eq:absorbing_stationary_equation}
\end{equation}
The first term on the right-hand side describes transition ($R'_{j\nu}$) between the transient states $j$ and $\nu$, before eventual absorption ($P^a_{\nu}$) into the absorbing reset state $a$ from the initial transient state $\nu$. The second term describes transitions ($R'_{ja}$) from the transient state $j$ to the absorbing one $a$. The previous equation can be written in terms of the matrices defined in Eq.~\eqref{eq: transition matrix form for formalism} as
\begin{equation}
P^{a}_j=[(I_b-Q)^{-1} B]_{ja}.
\label{eq:matrix_stationary_prob}
\end{equation}
Here we introduced the matrix $(I_b-Q)^{-1}$, where $I_b$ is the $b\times b$ identity matrix. The latter is defined from the geometric series:
\begin{align}
    (I_b-Q)^{-1}=\sum_{n=0}^\infty Q^n.
\label{eq:geometric_sum_matrices}
\end{align}
The entries $(I_b-Q)^{-1}_{jk}$ can be interpreted as the expected number of times the system visits the transient state $k$ when initialized in the transient state $j$, before eventual absorption into one of the absorbing states. This statement clearly makes sense only when the geometric sum is convergent, which is true if and only if the spectral radius is smaller than one, see, e.g., page 618 of Ref.~\cite{meyer2023matrix}. We verified that in all the cases discussed later in Sec.~\ref{sec:results} this applies. As a consequence, we find that $\lim_{n\to\infty}Q^n=0$ and therefore at long times no transitions take place within the set of transient states. The entries of the stationary vector $\vec{c}$ of \eqref{eq:c_eigenvalue} are then given by Eq.~\eqref{eq:matrix_stationary_prob} for the absorbing reset states, and zero for the transient ones. This stationary vector is eventually plugged in Eqs.~\eqref{eq:ness} and \eqref{eq:exciations density - general} to determine the spin stationary state.

We can now take the case in Fig.~\ref{fig: MC S=1} for spin $S=1$ (see further below in Subsec.~\ref{subsec:qutrits}) as a first example to illustrate the aforementioned method to compute the stationary state weight for a nonergodic Markov chain in the space of reset states. Here, $\ket{1}_N$ and $\ket{3}_N$ are absorbing reset states, while $\ket{2}_N$ is transient. We relabel the absorbing states as $\ket{1'}_N$ and $\ket{2'}_N$, and the center state (former $\ket{2}_N$) as $\ket{3'}_N$, to obtain the transition matrix in the form \eqref{eq: transition matrix form for formalism}. The transition probabilities from the transient state $\ket{3'}_N$ to the absorbing ones $R'_{3'1'}=R'_{3'2'}=p$ are equal as a consequence of the symmetry \eqref{eq:symmetry_R}. We then have
\begin{align}
    R' = \left( \begin{array}{c c c}
    1 & 0 & 0 \\
    0 & 1 & 0\\
    p & p & 1-2p\\    
\end{array} \right).
\label{eq:RprimeS1}
\end{align}
Here, $B=(p,p)$ and $Q=1-2p<1$ (a scalar since only one transient state is present). Then from Eq.~\eqref{eq:matrix_stationary_prob} we immediately obtain $(P_{3'}^{1'},P_{3'}^{2'})=1/(2p) \cdot (p,p)=(1/2,1/2)$.
Rearranging the states to our previous labelling and inserting zeros for the transient states leads to the weight vector $\vec{c}^{\ T}=(\frac{1}{2},\ 0,\ \frac{1}{2})$. This uniquely identifies the stationary state \eqref{eq:ness} of the conditional reset protocol.

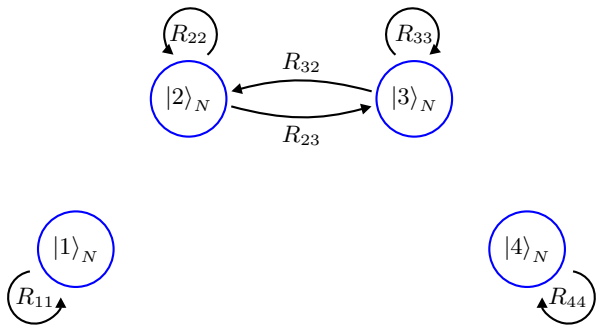
\begin{figure}[H]
\centering

\begin{tikzpicture}[myarrow/.style={-{Triangle[length=1.4mm, width=1.4mm]}, thick}]
  \def\circleRadius{0.5cm}
  \def\arrowdistance{0.05cm}
  \def\horizontalSpacing{3cm}
  \def\verticalSpacing{1cm}
  \def\selfarrowRadius{0.35cm}

  \node[draw=blue, circle, thick, minimum size=2*\circleRadius] (state1) at (-\horizontalSpacing, -\verticalSpacing) { $\ket{1}_N$};
  \node[draw=blue, circle, thick, minimum size=2*\circleRadius] (state2) at (-0.5*\horizontalSpacing, \verticalSpacing) { $\ket{2}_N$};
  \node[draw=blue, circle, thick, minimum size=2*\circleRadius] (state3) at (0.5*\horizontalSpacing, \verticalSpacing) { $\ket{3}_N$};
  \node[draw=blue, circle, thick, minimum size=2*\circleRadius] (state4) at (\horizontalSpacing, -\verticalSpacing) { $\ket{4}_N$};

    \draw[myarrow] ($(state3.west) + (-\arrowdistance, 2*\arrowdistance)$) to[bend right=15,<-] node[above]{ $R_{32}$} ($(state2.east) + (\arrowdistance, 2*\arrowdistance)$);
    \draw[myarrow] ($(state2.east) + (\arrowdistance, -2*\arrowdistance)$) to[bend right=15,<-] node[below]{ $R_{23}$} ($(state3.west) + (-\arrowdistance, -2*\arrowdistance)$);
    
    \draw[myarrow] ($(state1.south west) + (-0.7*\selfarrowRadius, 0.2*\selfarrowRadius)$) arc (100:360:\selfarrowRadius);
    \draw[myarrow] ($(state2.north) + (0.5*\circleRadius, 0.2*\selfarrowRadius)$) arc (310:590:\selfarrowRadius);
    \draw[myarrow] ($(state3.north) + (-0.5*\circleRadius, 0.2*\selfarrowRadius)$) arc (230:-50:\selfarrowRadius);
    \draw[myarrow] ($(state4.south east) + (0.7*\selfarrowRadius, 0.2*\selfarrowRadius)$) arc (440:180:\selfarrowRadius);

  \node at ($(state1.south west) + (-0.5*\selfarrowRadius, -0.8*\selfarrowRadius)$) {$R_{11}$};
  \node at ($(state2.north) + (0, 1*\selfarrowRadius)$) {$R_{22}$};
  \node at ($(state3.north) + (0, 1*\selfarrowRadius)$) {$R_{33}$};
  \node at ($(state4.south east) + (0.5*\selfarrowRadius, -0.8*\selfarrowRadius)$) {$R_{44}$};
\end{tikzpicture}
\caption{
\textbf{Markov chain of reset states in a non-ergodic spin-$3/2$ system.} The reset states $\ket{1}_N$ and $\ket{4}_N$ are absorbing. The Markov chain is reducible into three noncommunicating classes. The first one is formed by the reset state $\ket{1}_N$, the second by the states $\ket{2}_N$ and $\ket{3}_N$, and the third by $\ket{4}_N$. The reset states $\ket{2}_N$ and $\ket{3}_N$ form an ergodic subsystem. Due to the symmetry \eqref{eq:symmetry_R}, the transition probabilities obey $R_{11}=R_{44}$, $R_{22}=R_{33}$ and $R_{23}=R_{32}$.}
\label{fig: MC S=1.5}
\end{figure}

We also note that it is possible that the Markov chain model admits absorbing reset states, but these are not connected by any path to the initial reset state. An example is reported in Fig.~\ref{fig: MC S=1.5} for the example total spin $S=3/2$ and four reset states (see below in Subsec.~\ref{subsec:S32}). In this case, the distinction between absorbing and transient states does not have meaning since the stationary occupation probability still remains concentrated on the non-absorbing states if the system is initialized in one of them. To compute the stationary probability vector $\vec{c}$, here, we, instead, consider the non-absorbing states (in Fig.~\ref{fig: MC S=1.5} the states $\ket{2}_N$ and $\ket{3}_N$) as an ergodic subsystem and we derive the corresponding stationary probabilities by restricting the eigenvalue equation \eqref{eq:c_eigenvalue} to the Markovian subsystem of non-absorbing states.
\section{Stationary-state order parameter in the conditional resetting protocol}
\label{sec:results}
\begin{figure*}[t!]
    \centering
    \def\svgwidth{\textwidth} 
    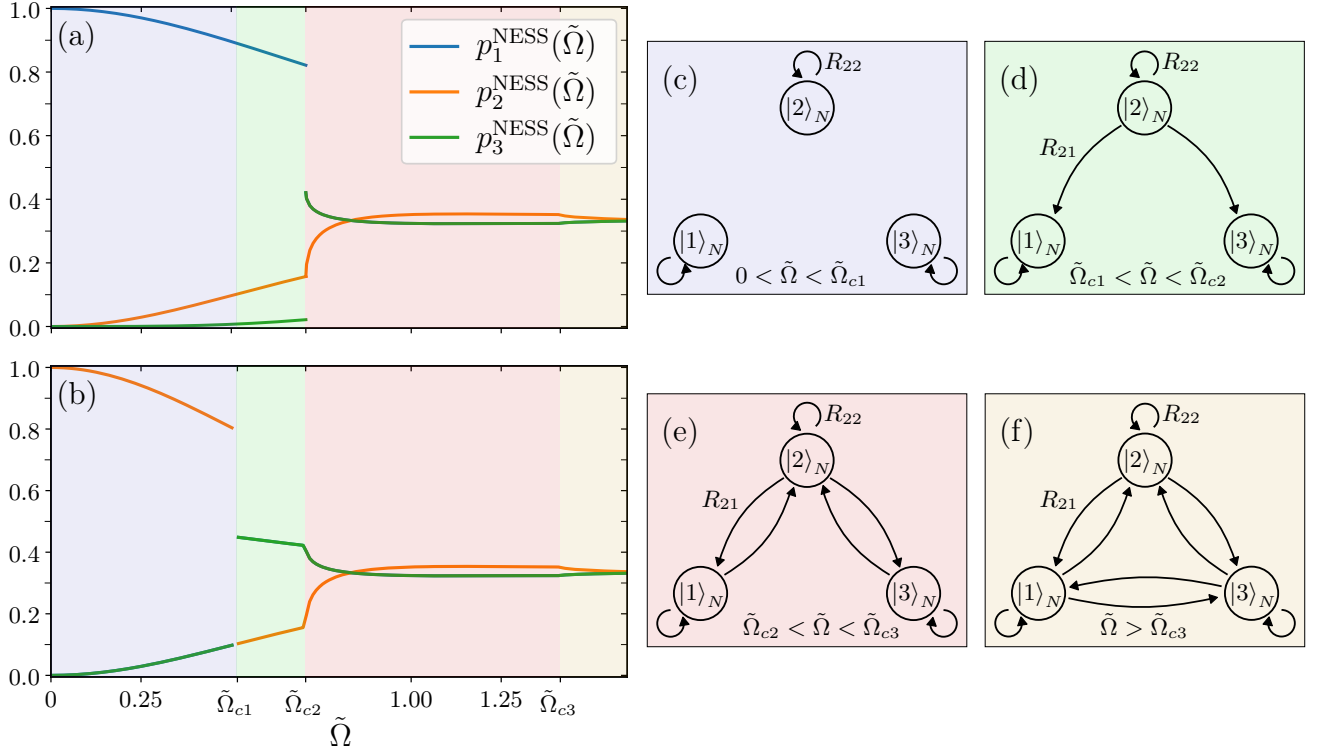
    \caption{\textbf{Stationary spin densities for a spin-$1$ system.}
    Plot of the stationary densities $p_j^{\mathrm{NESS}}$ as a function of $\tilde\Omega=\Omega/\Delta$. Four different regions (highlighted with different colors) are identified as a function of $\tilde{\Omega}$ and the three critical values \eqref{eq:S1_C1}-\eqref{eq:S1_C3}. In the panels (c)-(d), we report the reset Markov chain corresponding to each of the four regions. In panel (a), we consider the case where the system is initialized in $\ket{1}_N$. Here for $\tilde{\Omega}>\tilde{\Omega}_{c2}$ the curves for $p_1^{\mathrm{NESS}}=p_3^{\mathrm{NESS}}$. We can see that at the critical point $\tilde{\Omega}_{c2}$ the stationary density features both a first-order discontinuity (in $p_1^{\mathrm{NESS}}$) and a second-order one (in $p_2^{\mathrm{NESS}}$). In panel (b), we consider the case where the system is initialized in $\ket{2}_N$. In the whole plot one has $p_1^{\mathrm{NESS}}=p_3^{\mathrm{NESS}}$. The critical value $\tilde{\Omega}_{c2}$ separates the nonergodic regime ($\tilde{\Omega}<\tilde{\Omega}_{c2}$), from the ergodic one ($\tilde{\Omega}>\tilde{\Omega}_{c2}$). In the latter, the stationary state does not depend on the chosen initial state, and panels (a) and (b) coincide. 
    In the figure, we set $\gamma=\Delta=1$.
    }
    \label{fig: S1}
\end{figure*}

In this section, we apply the methods discussed before in Sec.~\ref{sec:conditional} to compute the stationary densities of spins $p_j^{\mathrm{NESS}}$ \eqref{eq:exciations density - general} in the conditional reset protocol. This quantity plays the role of the order parameter since it allows to classify collective behavior in the stationary state. We consider various values of the total spin $S$.
In Subsec.~\ref{subsec:qubit}, we discuss the spin $1/2$ model as a preliminary case. In Subsec.~\ref{subsec:qutrits}, we consider qutrits with $S=1$, with the main results summarized in Fig.~\ref{fig: S1}. In Subsec.~\ref{subsec:S32}, we move to the case of $S=3/2$ and four reset states, with the results in Fig.~\ref{fig: S1.5}. In Subsec.~\ref{subsec:S2}, we eventually study the case of spin $S=2$ and five reset states. The corresponding results are reported in Fig.~\ref{fig: S2}.

\subsection{Spin system $S=1/2$}
\label{subsec:qubit}
For total spin $S=1/2$, the spin matrices in Eq.~\eqref{eq:Hamiltonian} are $S^{x,y,z}_{\alpha}=\sigma^{x,y,z}_{\alpha}/2$, where $\sigma_{x,y,z}$ are Pauli matrices. The dynamics ensuing in the conditional reset protocol had been previously addressed in Ref.~\cite{Magoni_PRA_quantum} (note that therein Pauli matrices are used without the prefactor $1/2$). We therefore briefly summarize the main results for $S=1/2$ reformulated in terms of the Markov chain mapping of Sec.~\ref{sec: Dynamics and reset protocol}. This analysis will be useful to compare with the richer physics observed for the higher spin values  of the next subsections. In the conditional reset protocol, there are two possible states: $\ket{1}_N=\ket{\downarrow}_N$ (down spins in the $z$ direction) and $\ket{2}_N=\ket{\uparrow}_N$ (up spins in the $z$ direction). The majority condition in Eq.~\eqref{eq:def_W} in this case simply translates to 
\begin{equation}
 W^{(\infty)}_{ij}(t)=\bigg\{
    \begin{array}{ll}
    1 & p^F_{ij}(t)>1/2,\\
    0 & \,  \textrm{else}.
    \end{array}
\end{equation}
A majority of spins in the state $\ket{j}_{\alpha}$ is, indeed, reached when the corresponding density exceeds the threshold value $1/2$. The reset-free densities for the down and up spin state reads
\begin{align}
p^{F}_{11}(t)&= 1-\frac{\tilde{\Omega}^2}{1+\tilde{\Omega}^2}\sin^2\left(\frac{\Gamma t}{2}\right) = p^{F}_{22}(t), \label{eq:qubit1}\\
p^{F}_{12}(t)&= \frac{\tilde{\Omega}^2}{1+\tilde{\Omega}^2}\sin^2\left(\frac{\Gamma t}{2}\right)   =p^{F}_{21}(t),
\label{eq:qubit2}
\end{align}
with $\tilde{\Omega}=\Omega/\Delta$ and $\Gamma=\sqrt{\Omega^2+\Delta^2}$. The last equality in both equations follows from the symmetry \eqref{eq:symmetry_R}. The expression for the NESS density matrix in the conditional reset protocol is readily obtained as a particular case of Eq.~\eqref{eq:ness}
\begin{equation}
\label{eq:ness_s_1_2}
    \rho^{\mathrm{NESS}}=\gamma  \int_0^\infty dt' e^{-\gamma t'} \left[c_1 \rho^F_{1}(t') + c_2 \rho^F_{2}(t') \right]
\end{equation}
The stationary mean occupation probabilities consequently are
\begin{equation}
\label{eq: excitation density S=1/2}
    p^{\mathrm{NESS}}_j=\gamma \int_0^\infty dt' e^{-\gamma t'} \left[c_1 p^F_{1j}(t') + c_2 p^F_{2j}(t')\right].
\end{equation}
These formulas match those derived in Ref.~\cite{perfetto2021designing} for the Hamiltonian of the transverse field Ising chain. It then remains to determine the value of the stationary resetting probabilities $c_{1,2}$ to fully determine the NESS. Based on the discussion of Sec.~\ref{sec: Dynamics and reset protocol}, it is crucial to distinguish the case where the Markov chain over reset states is ergodic from the one where it is nonergodic. As discussed for Fig.~\ref{fig: Free evolution s=0.5 + step function}, this happens at the critical point $\tilde{\Omega}_c=1$, with $\tilde{\Omega}<\tilde{\Omega}_c$ being the nonergodic regime, while $\tilde{\Omega}>\tilde{\Omega}_c$ the ergodic.   

\begin{figure}[H]
    \centering
    \def\svgwidth{0.48\textwidth}
    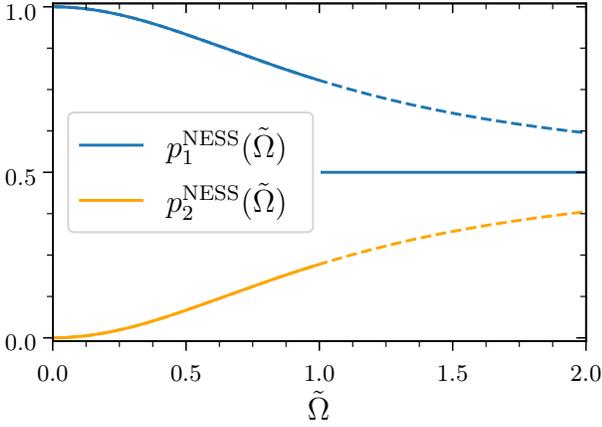
    \caption{\textbf{Stationary occupation probabilities for noninteracting spins $S=1/2$ under conditional resetting.} Plot of the stationary occupation probabilities $p_{1,2}^{\mathrm{NESS}}$ from Eq.~\eqref{eq: excitation density S=1/2} as a function of $\tilde{\Omega}=\Omega/\Delta$. The system is initialized in the state $\ket{1}_N$. For $\tilde \Omega<1$, the Markov chain of reset states is nonergodic and the system is always reset to $\ket{1}_N$. The curves of $p_{1,2}^{\mathrm{NESS}}$ therefore coincide with the ones obtained for unconditional resetting to state $\ket{1}_N$. At $\tilde\Omega=\tilde\Omega_c=1$, both curves have a jump discontinuity, since for $\tilde \Omega >1$, the resetting Markov chain becomes ergodic $R_{12}=R_{21}\neq 0$. In particular, $p_1^{\mathrm{NESS}}=p_2^{\mathrm{NESS}}=1/2$ for $\tilde\Omega>1$ independently on the initial state. The dashed curves are obtained by implementing the unconditional reset protocol to state $\ket{1}_N$ for comparison. No discontinuity is present within this protocol, and the stationary probabilities are analytic functions of $\tilde \Omega$. 
    Note that the same plot also holds for initializing the system in $\ket{2}_N$ simply swapping the roles of $\ket{1}$ and $\ket{2}$. In the figure, we set $\gamma=\Delta=1$.}
    \label{fig: Plot exden S=0.5}
\end{figure}

In particular, for $\tilde \Omega<1$,  if the system initialized in $\ket{1}_N$, one has $W_{11}(t)=1$, $W_{12}(t)=0$ and $R_{11}=1$, $R_{12}=0$. The Markov chain of reset states is made of two noncommunicating absorbing reset states. The weights in Eq.~\eqref{eq:ness_s_1_2} are then $c_1=1$ and $c_2=0$. The system will always be reset to the state $\ket{1}_N$ and the conditional reset protocol is equivalent to unconditional resetting (the NESS \eqref{eq:ness_s_1_2} reduces exactly to \eqref{eq:density matrix ness}). This result is shown in Fig.~\ref{fig: Plot exden S=0.5}, where the stationary probabilities $p_{1,2}^{\mathrm{NESS}}$ are plotted as a function of $\tilde{\Omega}$. If the system is initialized in $\ket{2}_N$, the dynamics analogously reduces to unconditional resetting to state $\ket{2}_N$, with $c_1=0$ and $c_2=1$. 

For $\tilde\Omega >1$, instead, for a system initialized in $\ket{1}_N$ resets to state $\ket{2}_N$ become possible. The opposite holds as well, as a consequence of the symmetry \eqref{eq:symmetry_R} obeyed by the transition probabilities. The Markov chain of reset states consequently becomes ergodic, as shown in Fig.~\ref{fig: MC S=0.5}. We then compute the weights $c_{1,2}$ from the eigenvalue equation \eqref{eq:c_eigenvalue}, which gives a unique solution
\begin{equation}\label{eq: weights S=0.5}
    c_1=\frac{R_{12}}{R_{12}+R_{21}}\  \text{and}\ c_2=\frac{R_{21}}{R_{12}+R_{21}}.
\end{equation} 
Since, one additionally has $R_{12}=R_{21}$ the coefficients \eqref{eq: weights S=0.5} simplify to $c_1=c_2=\frac{1}{2}$ and Eq.~\eqref{eq: excitation density S=1/2} gives $p_{1}^{\mathrm{NESS}}=p_{2}^{\mathrm{NESS}}=1/2$. This result is shown in Fig.~\ref{fig: Plot exden S=0.5}. 

We fundamentally note that the ergodic, and nonergodic regimes of the resetting Markov chains are separated in Fig.~\ref{fig: Plot exden S=0.5} by a first-order nonanalytic point at $\tilde{\Omega}_c$. This collective behavior is similar to that found in first-order phase transitions, and it follows from the monitoring of the system with global measurements. In the next sections, we show that an even richer behavior and multiple nonanalyticities, both of first and second-order type are, possible for higher spin values. 

\subsection{Spin system $S=1$}
\label{subsec:qutrits}
The reset-free probabilities for $S=1$ can be analytically derived by exploiting the result of Ref.~\cite{curtright2014compact}, where an expansion of the one-spin unitary time evolution operator $\hat U_{\alpha}(t)$ in a finite sum of powers of the generators of $S_{\alpha}^{x,y,z}$ of the $SU(2)$ Lie algebra is derived. For $S=1$, this expansion reads 
\begin{align}
\label{eq: eponential expansion S=1 - first}
    \hat{U}_{\alpha}(t)&=e^{-i\hat{H}_{\alpha}t}=e^{-i(\Omega \hat{S}_{\alpha}^x+\Delta \hat{S}^z_{\alpha})t} \nonumber \\
    &=\mathbbm{1}_{3\times 3} -i \frac{\tilde \Omega \hat{S}^x_{\alpha}+ \hat{S}^z_{\alpha}}{\sqrt{1+\tilde\Omega ^2}}\sin \left( \Gamma t\right) \nonumber \\
    &+\frac{(\tilde \Omega \hat{S}^x_{\alpha} + \hat{S}^z_{\alpha})^2}{1+\tilde\Omega ^2} \left (\cos\left(\Gamma t\right) -1\right)\, ,
\end{align}
with $\tilde\Omega=\Omega/\Delta$ and $\Gamma = \sqrt{\Omega ^2 + \Delta ^2}$. This expansion of the unitary evolution operator allows to explicitly calculate the reset-free single spin occupation probabilities $p^F_{ij}(\tilde\Omega, \Gamma, t)$. We report their expressions in the Appendix \ref{app:appendix_1} for the sake of brevity. The transition probabilities $W^{(\infty)}_{ij}(t)$ and $R_{ij}$ are then computed computed according to Eqs.~\eqref{eq:def_W} and \eqref{eq:R_matrix_final}. From the latter, we find the following three critical values for the Hamiltonian driving parameters $\tilde\Omega$ according to \eqref{eq: def of om_c}:
\begin{align}
    \tilde\Omega_{c1}\approx 0.52\ \ \  (R_{21}=R_{23}\neq 0), \label{eq:S1_C1} \\
    \tilde\Omega_{c2}\approx0.71\ \ \  (R_{12}=R_{32}\neq 0), \label{eq:S1_C2}   \\
    \tilde\Omega_{c3}\approx1.41\ \ \  (R_{13}=R_{31}\neq 0). \label{eq:S1_C3}
\end{align}
In Fig.~\ref{fig: S1}, we plot the stationary density of spins $p_j^{\mathrm{NESS}}$ in each magnetization eigenstate as a function of $\tilde\Omega$. The plot is divided into four regimes delimited by the three critical values \eqref{eq:S1_C1}-\eqref{eq:S1_C3}. For $S=1$, there are three eigenstates of the magnetization $\hat{S}^z$ and therefore three possible reset states. In Fig.~\ref{fig: S1}~(a), we show the curves obtained by initializing in $\ket{1}_N$, while in Fig.~\ref{fig: S1}~(b) the ones for the initial state $\ket{2}_N$. The reset state $\ket{3}_N$ is here the symmetric partner of $\ket{1}_N$ from \eqref{eq:symmetry_R}. Therefore the stationary state obtained by initializing the system in $\ket{3}_N$ is analogous to that found in Fig.~\ref{fig: S1}(a) simply swapping the role of $\ket{1}_N$ and $\ket{3}_N$. In each region of the stationary-state phase diagram the weight vector $\vec{c}$ in Eq.~\eqref{eq:exciations density - general} is eventually computed from the eigenvalue equation \eqref{eq:c_eigenvalue}, when the reset Markov chain is ergodic, or via the formalism in Eqs.~\eqref{eq: transition matrix form for formalism}-\eqref{eq:geometric_sum_matrices} for nonergodic chains featuring absorbing reset states. The reset Markov chain corresponding to each region of the stationary-state plot is reported in Fig.~\ref{fig: S1}(c)-(f).  

In Fig.~\ref{fig: S1}(a), we see that the system initialized in $\ket{1}_N$ is always reset to this same state for $\tilde\Omega<\tilde\Omega_{c2}$. For such $\tilde\Omega$ values, the reset state $\ket{1}_N$ is, indeed, absorbing as shown in Fig.~\ref{fig: S1}(c)-(d). This leads to a higher value of the density of spins $p_1^{\mathrm{NESS}}$ in the initial state. The state $\ket{2}_N$ and $\ket{3}_N$ acquire, instead, a smaller occupation due to the time evolution in between consecutive resets. In this regime, the dynamics is thus equivalent to the unconditional reset protocol to state $\ket{1}_N$, since the weights of the stationary state are $c_1=1$, $c_2=c_3=0$. The curve obtained for $p_1^{\mathrm{NESS}}$ in Fig.~\ref{fig: S1}~(a) for $\tilde\Omega<\tilde\Omega_{c2}$, indeed, coincides with the one obtained within the unconditional reset protocol, which is reported in Appendix \ref{app:appendix_1}. In general,  as we also saw previously for the case $S=1/2$ in Fig.~\ref{fig: Plot exden S=0.5}, whenever a single component $c_i=1$ of the stationary weight is non zero, the dynamics from the initial state $\ket{j}_N$ reduces to that of the unconditional reset protocol with $\ket{j}_N$ as reset state. The qualitative difference from the unconditional resetting protocol emerges, however, for $\tilde{\Omega}>\tilde{\Omega}_{c2}$. Here, as one can see from Fig.~\ref{fig: S1}(e), the reset state $\ket{1}_N$ ceases to be absorbing (and the same for $\ket{3}_N$) and the reset Markov chain becomes irreducible and ergodic. Consequently, the system can now be reset also to the state $\ket{2}_N$ and $\ket{3}_N$, which lead to a first-order discontinuity in the stationary density of spins $p_1^{\mathrm{NESS}}$. The mechanism behind this collective behavior is akin to that observed in Fig.~\ref{fig: Plot exden S=0.5}, and it relies on the presence of a single absorbing reset state. Different from the spin $1/2$ case, we can see that at the same critical value $\tilde{\Omega}_{c2}$, the density $p_2^{\mathrm{NESS}}$ features a second-order nonanalytic point. At $\Omega_{c2}$, one therefore simultaneously observes collective behavior akin to a first and second order phase transition. This is possible only for integer total spin values, where a center state with spin eigenvalue $s=0$ is present. The associated transition probabilities are different compared to the outer states, which are grouped into symmetric pairs according to time-reversal symmetry, and determine the observed multicritical behavior.

In Fig.~\ref{fig: S1}~(b), we plot the stationary state obtained when the system is initialized in the center state $\ket{2}_N$. Due to the symmetry relation \eqref{eq:symmetry_R} and the initialization in the center state $\ket{2}_N$, it holds that  $p^{\mathrm{NESS}}_1=p^{\mathrm{NESS}}_3$ in the whole plot. For $\tilde\Omega<\tilde\Omega_{c1}$, the system is always reset to $\ket{2}_N$, which is absorbing. The ensuing behavior is thus qualitatively similar to that obtained for initializing in $\ket{1}_N$ in the same $\tilde{\Omega}$ regime, and it is analogous to the unconditional reset protocol to state $\ket{2}_N$. At $\tilde\Omega_{c1}$, cf Fig.~\ref{fig: S1}(d), the transition from the state $\ket{2}_N$ to $\ket{1}_N$ and $\ket{3}_N$ becomes possible with probability $R_{21}=R_{23}$ (they simultaneously become nonzero as a consequence of Eq.~\eqref{eq:symmetry_R}). The state $\ket{2}_N$ ceases to be absorbing and the associated density $p_2^{\mathrm{NESS}}$ features a jump discontinuity. The reset Markov chain is, however, still nonergodic for $\tilde\Omega_{c1}<\tilde{\Omega}<\tilde{\Omega}_{c2}$ as a consequence of the states $\ket{1}_N$ and $\ket{3}_N$, which are still absorbing. The structure in Fig.~\ref{fig: S1}(d) is, in particular, identical to that we discussed before in Fig.~\ref{fig: MC S=1}, where the state $\ket{2}_N$ is classified as transient. The steady state weight $\vec{c}$ is then computed according to absorbing-Markov chain formalism in Eq.~\eqref{eq:RprimeS1} and it is equal to $\vec{c}^{\ T}=(\frac{1}{2},0,\frac{1}{2})$. In this regime, moreover, the stationary density in $\ket{2}_N$ does not depend on the chosen initial state as it is identical in Fig.~\ref{fig: S1}~(a) and (b). This clearly comes from the fact that in Fig.~\ref{fig: S1}(d) the center state is transient, the system asymptotically resets always to $\ket{1}_N$ or $\ket{3}_N$, and both these states symmetrically contribute to the density of spins in state $\ket{2}_N$. As $\tilde\Omega>\tilde\Omega_{c2}$, the transitions from $\ket{1}_N$ and $\ket{3}_N$ to $\ket{2}_N$ become also possible. All the reset states cease to be absorbing, and the resetting Markov chain becomes ergodic. The density $p_2^{\mathrm{NESS}}$ features at $\tilde{\Omega}=\tilde\Omega_{c2}$ a second-order nonanalytic point since it is directly coupled to $\ket{1}_N$ and $\ket{3}_N$, which stops being absorbing. 

For $\tilde\Omega>\tilde\Omega_{c2}$, the system is in the ergodic regime and can be reset to each of the three states regardless of the initial condition. The curves in Fig.~\ref{fig: S1}~(a) and (b) consequently coincide for $\tilde\Omega>\tilde\Omega_{c2}$. Interestingly, one can identify a third critical point \eqref{eq:S1_C3} within the ergodic phase at $\tilde\Omega_{c3}$. Here, the two transitions between $\ket{1}_N$ and $\ket{3}_N$ acquire nonzero transition probability, which render the reset Markov chain fully connected. The stationary occupation probability in the center state $p_2^{\mathrm{NESS}}$ accordingly decreases because now the system after a reset $\ket{1}_N$ can be directly reset to $\ket{3}_N$ (and vice versa). In this case, we again find a second-order discontinuity in the stationary spins probabilities, albeit such a discontinuity is tiny on the scale of the figure. In general, we observe that the inclusion of new transitions in an already ergodic Markov chain at most results in small discontinuities of the first-derivative of the order parameter $p^{\mathrm{NESS}}$. First-order discontinuities of the latter are necessarily related to the inclusion of new transitions connecting formerly absorbing reset states. For large $\tilde{\Omega}$, we find the stationary vector $\vec{c}^{\ T}=(0.32213, 0.35573, 0.32213)$ and accordingly from \eqref{eq:exciations density - general}  the stationary densities $p_1^{\mathrm{NESS}}(\tilde{\Omega}\to \infty)=p_3^{\mathrm{NESS}}(\tilde{\Omega}\to \infty)\approx 0.33053$ and $p_2^{\mathrm{NESS}}(\tilde{\Omega}\to \infty)\approx 0.33893$. A large Rabi frequency $\Omega$, indeed, enforces fast oscillations among the different magnetization eigenstates, making the stationary occupation probabilities almost equal. The center state $\ket{2}_N$, however, is not symmetric with respect to the pair of states $\ket{1}_N$ and $\ket{3}_N$, according to Eq.~\eqref{eq:symmetry_R}, and therefore it still keeps a slightly different occupation probability.   

\begin{figure*}[t!]
    \centering
    \def\svgwidth{0.95\textwidth} 
    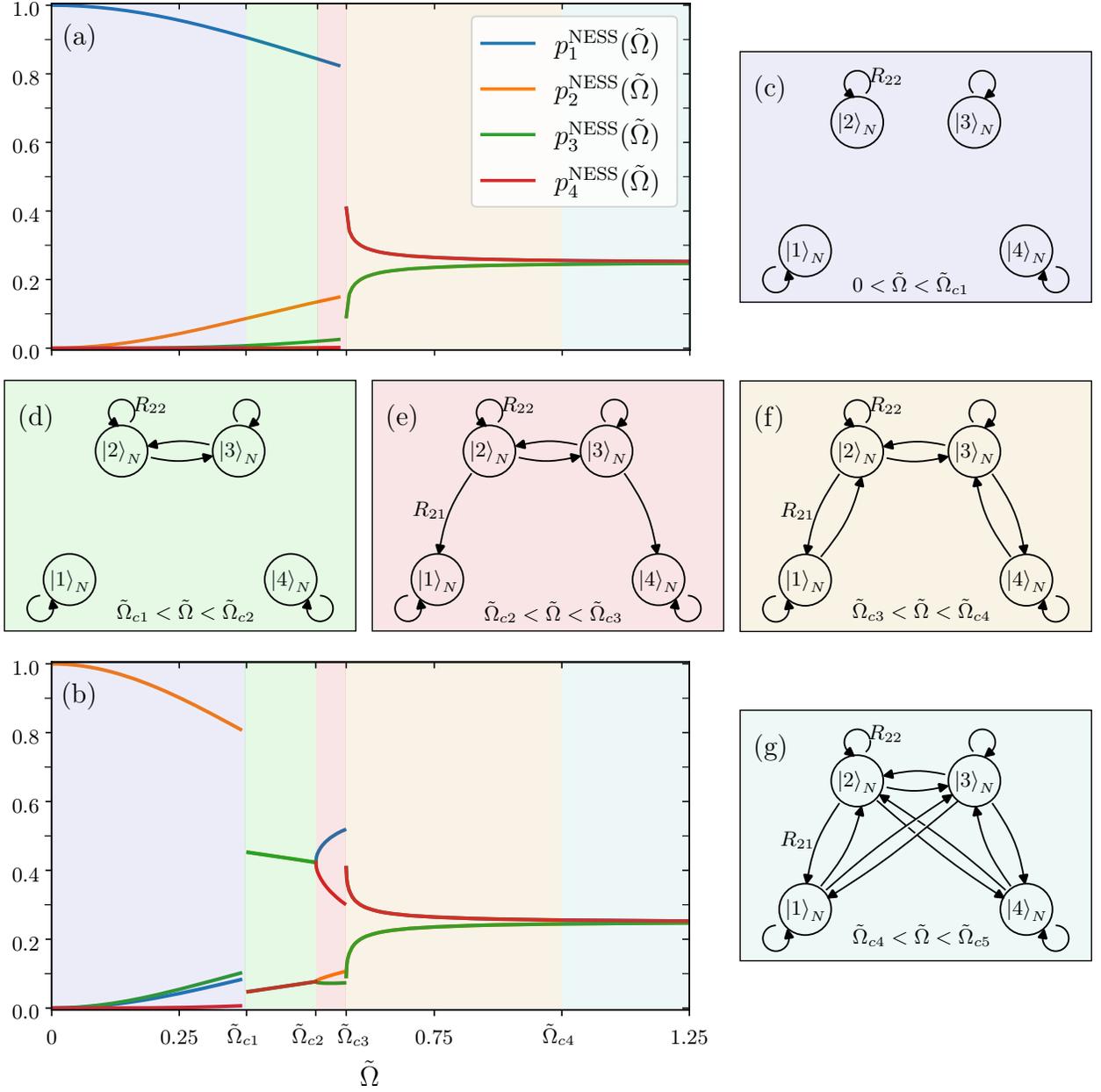
    \caption{\textbf{Stationary spin densities for a spin-$3/2$ system.} Plot of the stationary densities $p_{j}^{\mathrm{NESS}}$ as a function of $\tilde{\Omega}=\Omega/\Delta$. Five different regions (highlighted with different colors) are identified as a function of $\tilde{\Omega}$ and the five critical values \eqref{eq:S1.5_C1}-\eqref{eq:S1.5_C4}. In the panels (c)-(g), we report the resetting Markov chain corresponding to each region in $\tilde{\Omega}$. In panel (a), the initial state is chosen as $\ket{1}_N$. Here for $\tilde{\Omega}>\tilde{\Omega}_{c3}$, one has $p_1^{\mathrm{NESS}}=p_4^{\mathrm{NESS}}$ and $p_2^{\mathrm{NESS}}=p_3^{\mathrm{NESS}}$. In panel (b), the initial state is taken as $\ket{2}_N$. For $\tilde{\Omega}_{c1}<\tilde{\Omega}<\tilde{\Omega}_{c2}$ as well as for $\tilde{\Omega}>\tilde{\Omega}_{c3}$, one has $p_1^{\mathrm{NESS}}=p_4^{\mathrm{NESS}}$ and $p_2^{\mathrm{NESS}}=p_3^{\mathrm{NESS}}$. The critical value $\tilde{\Omega}_{c3}$ separates the nonergodic regime ($\tilde{\Omega}<\tilde{\Omega}_{c3}$), from the ergodic one ($\tilde{\Omega}>\tilde{\Omega}_{c3}$). All the nonanalytic points in the nonergodic regime are of first-order (discontinuous) type for half-integer spin. In the latter, the stationary state does not depend on the chosen initial state, and panels (a) and (b) coincide. In the figure, we set $\gamma=\Delta=1$.
    }
    \label{fig: S1.5}
\end{figure*}

\subsection{Spin system $S=3/2$} 
\label{subsec:S32}
The reset-free dynamics for noninteracting spins of spin $3/2$ can be worked out from the expansions of the single-spin propagator according to Ref.~\cite{curtright2014compact}:
\begin{align}
\label{eq:exp_expansion_S32}
    \hat{U}_{\alpha}(t)&=e^{-i\hat{H}_{\alpha}t}=e^{-i(\Omega \hat{S}^x_{\alpha}+\Delta \hat{S}^z_{\alpha})t} \nonumber \\
    &=\mathbbm{1}_{4\times 4} \cos\left(\frac{\Gamma t}{2}\right) \left[ 1 + \frac{1}{2} \sin^2\left(\frac{\Gamma t}{2}\right) \right] \nonumber \\ 
    &- 2i \left[\frac{\tilde \Omega \hat S^x_{\alpha} + \hat S^z_{\alpha}}{\sqrt{1+\tilde\Omega^2}}\sin\left(\frac{\Gamma t}{2}\right)\right] \left[1 + \frac{1}{6} \sin^2\left(\frac{\Gamma t}{2}\right) \right] \nonumber \\
    &- 2 \left[\frac{\tilde \Omega \hat S^x_{\alpha} + \hat S^z_{\alpha}}{\sqrt{1+\tilde\Omega^2}}\sin\left(\frac{\Gamma t}{2}\right)\right]^2 \cos\left(\frac{\Gamma t}{2}\right) \nonumber \\ 
    &+ \frac{4i}{3} \left[\frac{\tilde \Omega \hat S^x_{\alpha} + \hat S^z_{\alpha}}{\sqrt{1+\tilde\Omega^2}}\sin\left(\frac{\Gamma t}{2}\right)\right]^3.
\end{align}
The explicit expressions of the occupation probabilities $p^F_{ij}(\tilde\Omega, \Gamma, t)$ are reported in Appendix \ref{app:S=1.5}. From the calculation of the transition matrices \eqref{eq:def_W} and \eqref{eq:R_matrix_final}, we eventually find the following five critical values of $\tilde\Omega$ according to \eqref{eq: def of om_c}:
\begin{align}
    \tilde\Omega_{c1}&\approx 0.39\quad  (R_{23}=R_{32}\neq 0), \label{eq:S1.5_C1} \\
    \tilde\Omega_{c2}&\approx0.52\quad  (R_{21}=R_{34}\neq 0), \\
    \tilde\Omega_{c3}&\approx0.58\quad (R_{12}=R_{43}\neq 0), \label{eq:S1.5_C3}\quad \\
    \tilde\Omega_{c4}&\approx1.00\quad  \,\,(R_{13}=R_{42}=R_{31}=R_{24}\neq 0),\label{eq:S1.5_C4}\\
    \tilde\Omega_{c5}&\approx 1.73  \quad(R_{14}=R_{41} \neq 0). \label{eq:S1.5_C5}
\end{align}
For total spin $S=3/2$, there are $4$ eigenstates of $\hat S^z$ and four reset states. They can be grouped into two pairs of symmetric states:
the two outer states $\ket{1}_N$ and $\ket{4}_N$, and the center states $\ket{2}_N$ and $\ket{3}_N$. In Fig.~\ref{fig: S1.5}(a), we plot the stationary occupation probabilities \eqref{eq:exciations density - general} when the initial state is chosen as $\ket{1}_N$. The same stationary state is found from the initial state $\ket{4}_N$ simply swapping the curves of the symmetric states $p_1^{\mathrm{NESS}}$, $p_4^{\mathrm{NESS}}$ and $p_2^{\mathrm{NESS}}$, $p_3^{\mathrm{NESS}}$. In Fig.~\ref{fig: S1.5}(b), we consider the stationary state from the initial state $\ket{2}_N$. The case of $\ket{3}_N$ as the initial state is also found from the latter panel by swapping the curves associated with the symmetric states. In the Fig.~\ref{fig: S1.5}(c)-(g), we plot the reset Markov chain in the five regions as a function of $\tilde{\Omega}$ identified by the critical values \eqref{eq:S1.5_C1}-\eqref{eq:S1.5_C4}. In the Appendix \ref{app:S32_ergodic}, we report additional details concerning the interval $\tilde{\Omega}>\tilde{\Omega}_{c5}$ which are not reported here for brevity.

In Fig.~\ref{fig: S1.5}(a) with the initial state $\ket{1}_N$, this state is absorbing for the whole nonergodic regime, i.e., for $\tilde\Omega<\tilde\Omega_{c3}$. The weight vector is therefore $\vec{c}^{\ T}=(1,0,0,0)$ and we have the very same curves as those observed in the unconditional reset protocol to state $\ket{1}_N$, which are reported in Appendix \ref{app:S=1.5}. When the initial state is chosen as one of the center states, say $\ket{2}_N$ in Fig.~\ref{fig: S1.5}(b), we find, instead, additional nonanalytic behavior across the nonergodic regime. As $\tilde\Omega<\tilde\Omega_{c1}$, state $\ket{2}_N$, in Fig.~\ref{fig: S1.5}(c), is absorbing and all resets take place towards this state. The stationary densities therefore, coincide with the unconditional reset protocol. For $\tilde\Omega_{c1}<\tilde\Omega<\tilde\Omega_{c2}$, in Fig.~\ref{fig: S1.5}(d), the center state $\ket{2}_N$ stops being absorbing since it connects to its symmetric counterpart $\ket{3}_N$ with transition probabilities $R_{23}=R_{32}$. In this way, the resetting Markov chain is now reducible into three noncommunicating sets, where $\ket{2}_N$ and $\ket{3}_N$ form an ergodic subsystem. This is the structure we discussed earlier in Fig.~\ref{fig: MC S=1.5} and the weight vector is $(0,1/2,1/2,0)^T$, similarly to the spin $1/2$ Markov chain in Fig.~\ref{fig: MC S=0.5}. The difference compared to the latter is that albeit resetting to the outer states is impossible here, there is still a nonzero density of spins in states $\ket{1}_{\alpha}$ and $\ket{4}_{\alpha}$. The excitation densities of the respective symmetric states are equal, $p_1^{\mathrm{NESS}}=p_4^{\mathrm{NESS}}$ and $p_2^{\mathrm{NESS}}=p_3^{\mathrm{NESS}}$, resulting in the overlapping of the corresponding curves. For $\tilde\Omega_{c2}<\tilde\Omega<\tilde\Omega_{c3}$, in Fig.~\ref{fig: S1.5}~(e), the transitions to the outer states, which are still absorbing, become possible with probability $R_{21}=R_{34}\neq 0$. The resetting Markov chain now presents two absorbing states (outer states) and two transient states (center states). All the resets will then eventually take place either to $\ket{1}_N$ or $\ket{4}_N$. In order to properly compute the associated stationary state vector, we again exploit the formalism of absorbing Markov chains, and we get (see Appendix \ref{app:steady-state_weights_S32} for the details of the calculations) 
\begin{equation}
\label{eq: weight vec S=1.5}
    \vec{c}^{\ T}=\left(\frac{R_{21}+R_{23}}{R_{21}+2R_{23}},0, 0, \frac{R_{23}}{R_{21}+2R_{23}}\right)\, .
\end{equation}
The stationary resetting probability to state $\ket{1}_N$ is therefore larger than that to $\ket{4}_N$ since $R_{21}\neq 0$. This follows from the choice of the initial state $\ket{2}_N$, which favours resets to eventually take place to the state $\ket{1}_N$. The choice of the initial state is therefore responsible for the breaking of the symmetry between the reset states and, in this way, also for the splitting of the stationary probabilities for both symmetric pairs. We also note that all the nonanalyticities taking place in the nonergodic regime, both in Fig.~\ref{fig: S1.5}(a) and \ref{fig: S1.5}(b), are of first-order type. This follows from the absence of a center state, which is present only for integer total spin. 

For $\tilde\Omega>\tilde\Omega_{c3}$, cf. Fig.~\ref{fig: S1.5}(f), the system eventually enters the ergodic regime. The weight vector reads as 
\begin{equation}
    \vec{c}^{\ T}=\frac{1}{2(R_{12}+R_{21})}(R_{21}, R_{12}, R_{12}, R_{21}),
\label{eq:absorbing_c_S32}
\end{equation}
independently of the initial state. In addition, the ergodic regime hosts other critical values for $\tilde{\Omega}$. At $\tilde{\Omega}_{c4}$, cf. Fig.~\ref{fig: S1.5}(g), the additional transitions acquire nonzero probability $R_{13}=R_{42}=R_{31}=R_{24}$. All these four probabilities coincide, not only those referring to symmetric pairs ($R_{13}=R_{42}$ and $R_{31}=R_{24}$). No discontinuity of any kind is therefore observed. The resetting Markov chain becomes eventually fully connected only as $\tilde{\Omega}\geq \tilde{\Omega}_{c5}$ when the direct transitions between the outer reset states $\ket{1}_N$ and $\ket{4}_N$ are enabled. This leads to a tiny second-order discontinuity in the spin densities (see Appendix \ref{app:S32_ergodic}). Similarly to the spin $S=1$ case, we find that also in this case the inclusion of transitions in an already ergodic chain does not lead to any singularity (as in $\tilde{\Omega}_{c4}$) or at most to tiny second-order discontinuities (in $\tilde{\Omega}_{c5}$). 

\begin{figure*}[t!]
    \centering
    \def\svgwidth{1\textwidth} 
    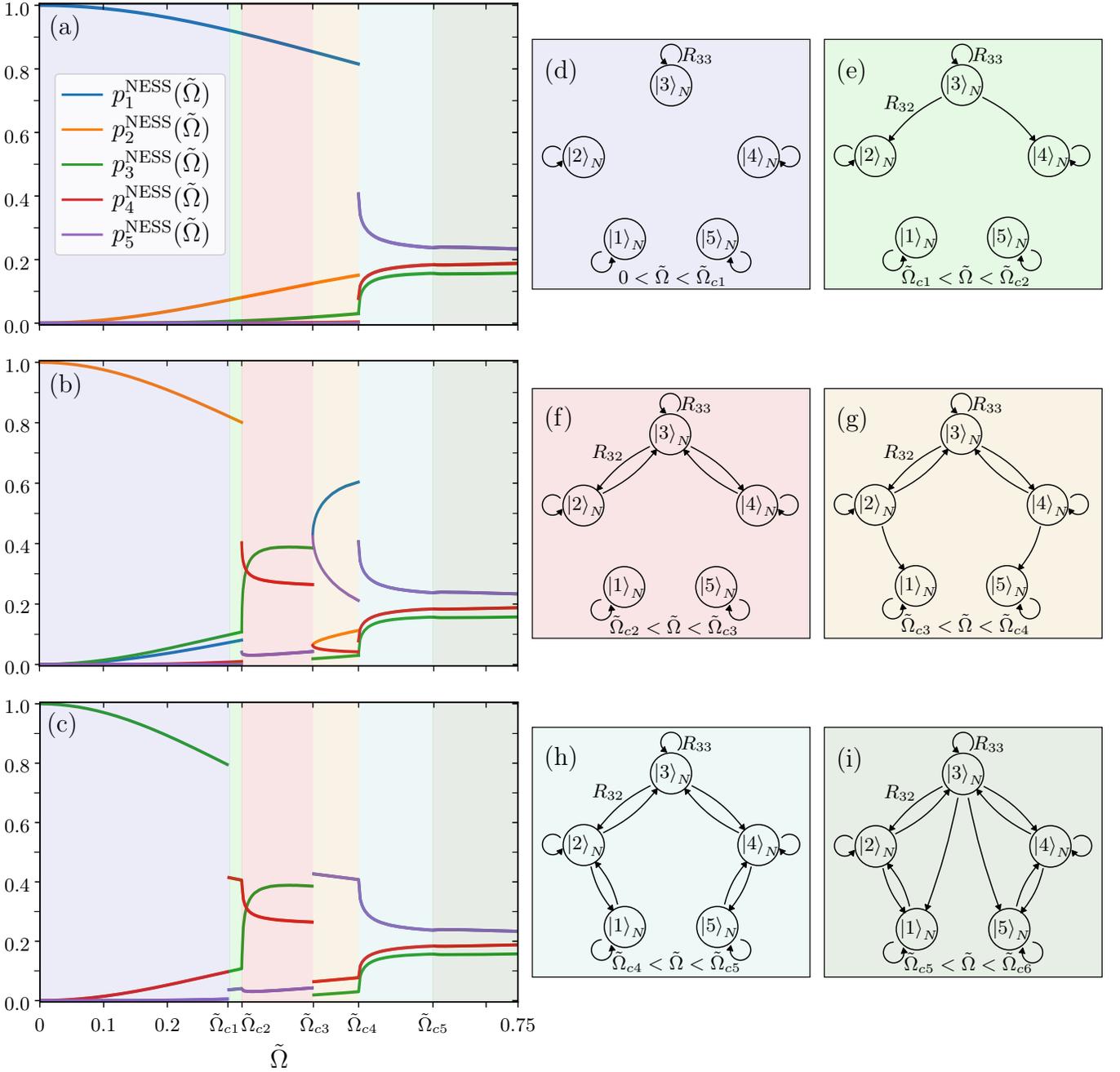
    \caption{\textbf{Stationary spin densities for a spin-$2$ system.}
    Plot of the stationary densities $p_j^{\mathrm{NESS}}$ as a function of $\tilde\Omega=\Omega/\Delta$. Six different regions (highlighted with different colors) are identified as a function of $\tilde{\Omega}$ and the critical values \eqref{eq:S2_C1}-\eqref{eq:S2_C5}. In the panels (d)-(i) we report the reset Markov chain corresponding to each of the six regions. In panel (a), we consider the case where the system is initialized in $\ket{1}_N$. The value $\tilde{\Omega}_{c4}$ marks multicritical behavior where the center state density $p_2^{\mathrm{NESS}}$ shows a second-order continuous transition, while the densities for all the other states display a first-order discontinuous jump. In panel (b), we take $\ket{2}_N$ as the initial state. Here at the value $\tilde{\Omega}_{c2}$, one again finds the simultaneous presence of first and second-order collective behavior. In panel (c), we take $\ket{3}_N$ as the initial state. In this case, all the spin densities display first-order singularities at $\tilde{\Omega}_{c1}$ and $\tilde{\Omega}_{c3}$, and second-order ones at $\tilde{\Omega}_{c2}$ and $\tilde{\Omega}_{c4}$ in the nonergodic regime. The critical value $\tilde{\Omega}_{c4}$ separates the nonergodic regime ($\tilde{\Omega}<\tilde{\Omega}_{c4}$), from the ergodic one ($\tilde{\Omega}>\tilde{\Omega}_{c4}$). In the latter, the stationary state does not depend on the chosen initial state, and panels (a), (b) and (c) coincide. 
    In the figure, we set $\gamma=\Delta=1$.}
    \label{fig: S2}
\end{figure*}

\subsection{Spin system $S=2$}
\label{subsec:S2}
The reset-free probabilities $p^F_{ij}(\tilde\Omega, \Gamma, t) $ can be obtained also in this case from the expansion of the single spin propagator $\hat{U}_{\alpha}$ in a sum over a finite number of terms containing the generators of the $5$-dimensional representation of the $SU(2)$ algebra. The expression for $\hat{U}_{\alpha}$ and the reset-free probabilities are reported in Appendix \ref{app:S=2}. From those expressions, we get the following set of critical values of $\tilde\Omega$ according to \eqref{eq: def of om_c}:
\begin{align}
    \tilde\Omega_{c1}&\approx 0.29\ \ \  (R_{32}=R_{34}\neq 0), \label{eq:S2_C1} \\
    \tilde\Omega_{c2}&\approx0.32\ \ \  (R_{23}=R_{43}\neq 0), \\
    \tilde\Omega_{c3}&\approx 0.43\ \ \  (R_{21}=R_{45}\neq 0), \\
    \tilde\Omega_{c4}&\approx0.50\ \ \  (R_{12}=R_{54}\neq 0), \label{eq:S2_C4}\\
    \tilde\Omega_{c5}&\approx0.62\ \ \  (R_{31}=R_{35}\neq 0), \label{eq:S2_C5} \\
    \tilde\Omega_{c6}&\approx0.82\ \ \  (R_{13}=R_{53}\neq 0), \label{eq:S2_C6}\\
    \tilde\Omega_{c7}&\approx1.00\ \ \  (R_{25}=R_{41}\neq 0), \\
    \tilde\Omega_{c8}&\approx1.22\ \ \  (R_{52}=R_{14}\neq 0), \\
    \tilde\Omega_{c9}&\approx2.00\ \ \  (R_{15}=R_{51}\neq 0), \\
    \tilde\Omega_{c10}&\approx3.15\ \ \  (R_{24}=R_{42}\neq 0) \,  .\label{eq:S2_C10}
\end{align}
In Fig.~\ref{fig: S2}, we plot the stationary density of spins $p_j^{\mathrm{NESS}}$ in each magnetization eigenstates for the intervals of $\tilde{\Omega}$ marked by the critical values \eqref{eq:S2_C1}-\eqref{eq:S2_C6}.  For $S=2$ there are $2S+1=5$ eigenstates of $\hat S^z$. There is one center state $\ket{3}_N$, two symmetric, henceforth named intermediate, states $\ket{2}_N$ and $\ket{4}_N$, as well as two symmetric outer states $\ket{1}_N$ and $\ket{5}_N$. 

In Fig.~\ref{fig: S2}~(a), we consider the stationary state obtained when the initial state is $\ket{1}_N$. The same plot holds for the initial state $\ket{5}_N$ upon swapping the roles of the two symmetric outer and intermediate states. In this case, the system is always reset to state $\ket{1}_N$ across the whole nonergodic regime, i.e., for $\tilde\Omega<\tilde\Omega_{c4}$.
For this regime and initial state, the curves are equal to those obtained in the unconditional reset protocol with $\ket{1}_N$ as reset state, which are reported in Appendix \ref{app:S=2}. At $\tilde{\Omega}_{c4}$, we again observe multicritical collective behavior characteristic of integer spin values, where one simultaneously has a first-order singularity in $p_{j}^{\mathrm{NESS}}$, $j\neq 3$, and a second-order singularity in the center state $j=3$.   

In Fig.~\ref{fig: S2}(b), we initialize the system in $\ket{2}_N$. For $\tilde\Omega<\tilde\Omega_{c2}$, see Fig.~\ref{fig: S2}(d)-(e), the state $\ket{2}_N$ is absorbing and the stationary state is analogous to that of the unconditional reset protocol with the same reset state. For $\tilde\Omega_{c2}<\tilde\Omega<\tilde\Omega_{c3}$, cf. Fig.~\ref{fig: S2}(f), the intermediate states $\ket{2}_N$ and $\ket{4}_N$ together with the center state $\ket{3}_N$ form an ergodic subsystem. The structure of the resetting Markov chain is similar to that of total spin $S=1$ for $\tilde\Omega_{c2}<\tilde\Omega<\tilde\Omega_{c3}$ (compare Fig.~\ref{fig: S1}~(e) and Fig.~\ref{fig: S2}~(f)). Accordingly, the discontinuities at $\tilde\Omega_{c2}$ show the same features as those discussed in the spin $S=1$ case initialized in $\ket{1}_N$. In particular, we observe also in this case multicritical behavior at $\tilde{\Omega}_{c2}$, with simultaneous presence of first-order collective behavior for the density of spins in the center state, and second-order one for the densities in the other states. For $\tilde\Omega_{c3}<\tilde\Omega<\tilde\Omega_{c4}$, the absorbing reset states $\ket{1}_N$ or $\ket{5}_N$ can be eventually reached. The other three reset states, therefore, become transient. The symmetry between the absorbing states $\ket{1}_N$ and $\ket{5}_N$ is again broken by the choice of the initial state, with $\ket{2}_N$ leading to a larger stationary resetting probability to $\ket{1}_N$ (see Appendix \ref{app:steady-state_weights_S2} for the calculations): 
\begin{equation}
 \vec{c}^{\ T}=\left(\frac{2R_{21}+R_{23}}{2(R_{21}+R_{23})}, 0, 0, 0, \frac{R_{23}}{2(R_{21}+R_{23})}\right).
\label{eq:absorbing_c_S2}
\end{equation}
The curves corresponding to each spin state accordingly split up, and first-order collective behavior is found. 

In Fig.~\ref{fig: S2}(c), we eventually initialize the system in the center state $\ket{3}_N$. As $\tilde\Omega<\tilde\Omega_{c1}$, the dynamics reduces to the unconditional reset protocol to state $\ket{3}_N$. At $\tilde\Omega_{c1}$, resets to the intermediate absorbing states $\ket{2}_{N}$ and $\ket{4}_N$ are possible. This leads to eventual absorption in one of the two with equal probability (analogously to the discussion for Fig.~\ref{fig: S1}(d) in the spin $S=1$ case) and to a first-order discontinuity at $\tilde\Omega_{c1}$ of all curves. As $\tilde\Omega_{c2}<\tilde\Omega<\tilde\Omega_{c3}$, the states $\ket{2}_N$ and $\ket{4}_N$ are no longer absorbing and the states $\ket{2}_N$, $\ket{3}_N$ and $\ket{4}_N$ form an ergodic subsystem. The stationary state is therefore independent of the initial state, as long as the latter is picked within the ergodic subsystem.
This is why all the spin stationary densities Fig.~\ref{fig: S2}~(b) and Fig.~\ref{fig: S2}(c) coincide for $\tilde\Omega_{c2}<\tilde\Omega<\tilde\Omega_{c3}$. For $\tilde\Omega_{c3}<\tilde{\Omega}<\tilde{\Omega}_{c4}$, one has eventual absorption to one of the absorbing outer states $\ket{1}_N$ and $\ket{5}_N$ similarly to the case we discussed for the initial state $\ket{2}_N$. The fundamental difference from the latter case is that, here, on the contrary, the symmetry between the absorbing reset states is preserved and they are equally weighted in the stationary state according to the vector $\vec{c}^{\ T}=(1/2,0,0,0,1/2)$. There is consequently no splitting of the curves associated with the reset states that are symmetrically paired. This follows from the initialization in the center state $\ket{3}_N$ in the same way as in Fig.~\ref{fig: S1}~(b) for spin $S=1$, where the curves of $p_1^{\mathrm{NESS}}$ and $p^{\mathrm{NESS}}_3$ coincide for all values of $\tilde\Omega$ as well.

As $\tilde{\Omega}>\tilde{\Omega}_{c4}$, the ergodic regime is met, when the stationary state does not depend on the chosen initial state. The outer states $\ket{1}_N$ and $\ket{5}_N$ cease to be absorbing since the outwards transitions towards the states $\ket{2}_N$ and $\ket{4}_N$, respectively, get nonzero probability. The resetting Markov chain is, however, not fully connected. Based on the definition \eqref{eq: def of om_c}, we, indeed, find five additional critical points when direct connections among reset states start having nonzero transition probabilities. The first, for example, is $\tilde{\Omega}_{c5}$ in Eq.~\eqref{eq:S2_C5} when direct transitions from the center state $\ket{3}_N$ to the outer states $\ket{1}_N$ and $\ket{5}_N$ develop. We refer to Appendix \ref{app:S2_ergodic} for a discussion of these additional critical values appearing in the ergodic phase. As a general feature, we, however, observe that the stationary densities $p_j^{\mathrm{NESS}}$ feature in correspondence to these critical values only tiny, on the scale of Fig.~\ref{fig: S2}, second-order nonanalytic points (similarly to what we observed for spin $3/2$). 

\section{Conclusions and outlook} 
\label{sec:conclusions}
We have shown that collective behavior, similar to the nonanalyticities appearing in the presence of phase transitions, emerges in a noninteracting system of spins subject to a conditional reset protocol. For given total spin $S$ we considered $2S+1$ reset states, Eq. \eqref{eq:reset_states}, constructed as a tensor product of the single-spin magnetization eigenstates. Which one of the reset states is selected depends on the outcome of a global projective measurement of the magnetization \eqref{eq:measurement_global} and a majority rule. Applying this protocol, the spins acquire correlations since they are \emph{simultaneously} reset to a common state, which depends on the \emph{entire} many-body wavefunction just prior to the reset. 

Nonanalytic points in the stationary density of spins appears when the thermodynamic limit $N\to \infty$ is taken. For all the spin length $S$ considered, we find a rich stationary state for the density of spins as $N\to \infty$ that can be tuned by varying the Hamiltonian parameters. We developed a classification of these nonanaliticities on the basis of time-reversal symmetry of the unitary dynamics \eqref{eq:product_state_t}. The latter distinguishes half-integer from integer total spin. In the former case, exemplified by the values $S=1/2$ and $S=3/2$, one finds solely first-order discontinuous jumps in the nonergodic regime, see Fig.~\ref{fig: Plot exden S=0.5} and \ref{fig: S1.5}. For integer total spin $S=1$ and $S=2$, we find multicritical behavior since both first-order nonanalytic points and second-order ones are present in the nonergodic phase, cf. Fig.~\ref{fig: S1} and \ref{fig: S2}, respectively. 

To rationalize these findings, we developed an exact mapping of the stationary state expression for the density matrix to a Markov chain model defined on the space of reset states. By virtue of this construction a sequence of reset events is mapped onto a sequence of jumps on the Markov chain. Collective behavior in the stationary state accordingly arises from an a sudden change of the stationary vector of the Markov chain. This happens due to the presence of absorbing reset states: once the system is reset to one of those states, all the subsequent resets take place towards the same state. For critical values of the Hamiltonian driving parameters, transitions out of an absorbing-reset state become nonzero. Here the system shows nonanalytic behavior. 

From the structure of the Markov chain it becomes apparent that the difference between half-integer and integer spins follows from the fact that only in the latter case there is a reset state with zero magnetization eigenvalue. This state is not connected by time-reversal to any other state (with opposite magnetization) and therefore obeys different transition probabilities in the Markov chain model compared to the other pairs of states. This feature might become relevant when engineering collective many-body states, e.g., for sensing protocols (see  Refs.~\cite{sensing_1,sensing_2,sensing_3,sensing_4,sensing_5}), whose performance depends on the order of the associated nonanalytic emergent behavior.

In light of such applications, it will certainly be relevant to study the reset dynamics both for a finite number $N$ of spins and for finite times. Such small space-time scales are necessarily considered in experimental implementations of related noninteracting models subject to measurements, e.g., on quantum hardware \cite{tornow2023,yin2025resonance,yin2025restart}. A finite-size analysis of that kind will also shed light on how collective stationary behavior emerges in experimental setups as larger space-time scales are probed. Here, it will be further interesting to characterize the stationary state beyond the mean spin density, e.g., via the study of dynamical correlation functions of the spin densities at different times. 

In addition, one may ask other questions related to reset protocols. For example, one can investigate the mean time necessary to reset for the first time to a given reset state and how this changes as a function of the Hamiltonian parameters. Addressing such research question connects to current research on quantum first detection time protocols \cite{dhar2015,dhar2015b,grunbaum2013recurrence} and may pave the way for the formulation of hitherto unexplored persistence exponents \cite{persistence_exp_classical} in quantum many-body monitored dynamics. 

\section*{Acknowledgements}
We acknowledge funding from the Deutsche Forschungsgemeinschaft (DFG, German Research Foundation) under Project No. 435696605 and through the Research Units FOR 5413/1, Grant No. 465199066 and FOR 5522/1, Grant No. 499180199. 
Furthermore, we acknowledge the financial support by the German Federal Ministry of Research, Technology and Space (BMFTR) within the project ``Neuronale Quantennetzwerke auf NISQ-Quantencomputern (NeuQuant)'' under grant 13N17065. This work was supported by the QuantERA II programme (project CoQuaDis, DFG Grant No. 532763411) that has received funding from the EU H2020 research and innovation programme under GA No. 101017733.
This work is also supported by the ERC grant OPEN-2QS (Grant No. 101164443, https://doi.org/10.3030/101164443).

\onecolumngrid

\appendix
 
\section{time-reversal symmetry and reset-free probabilities}
\label{app:t_reversal}
In this Appendix, we discuss the time-reversal symmetry of the noninteracting spin Hamiltonian \eqref{eq:Hamiltonian} for generic total spin $S$ and we show how it leads to the symmetry \eqref{eq:symmetry_p} among the reset-free probabilities. We focus on the Hamiltonian at a single site $\alpha$
\begin{equation}
H_{\alpha}=\Omega \hat{S}^x_{\alpha}+\Delta \hat{S}^z_{\alpha}.
\label{app:single_qudit_Hamiltonian}
\end{equation}
The extension to the many spins case is trivial since the Hamiltonian \eqref{eq:Hamiltonian} is noninteracting. We therefore drop henceforth in this Appendix the lattice site subscript $\alpha$ for lightness of notation. Time-reversal symmetry is implemented by the antiunitary operator $\Theta$ in Eq.~\eqref{eq:antiunitary}, which we here rewrite as
\begin{equation}
\theta= \hat{U}^{\pi}_y K, \quad \mbox{with}, \quad \hat{U}^{\pi}_y= i\, \mbox{exp}(-i\pi \hat{S}^y). \label{app:antiunitary_decomposition}   
\end{equation}
This expression highlights the decomposition of the antiunitary time-reversal operator into a unitary operator $\hat{U}^{\pi}_y$ and a complex conjugate operator $K$, which takes the complex conjugate of the state it acts onto. The appearance therein of $\hat{S}^y$ is a consequence of the choice for the representation of the spin matrices where $\hat{S}^z$ is diagonal. The phase factor $i$ is a convention and here we follow that adopted in Ref.~\cite{sakurai1994modern} (cf. Sec.~4.4 therein). The Hamiltonian in Eq.~\eqref{app:single_qudit_Hamiltonian} is, in fact, odd under time-reversal 
\begin{equation}
\Theta H\Theta^{-1}=-H,
\label{app:symmetry_odd_H}
\end{equation}
as a consequence of the analogous property obeyed by the spin operator $\Theta S^{x,y,z} \Theta^{-1}=-S^{x,y,z}$ \cite{sakurai1994modern}. The unitary evolution operator is consequently even under time-reversal (note the extra $i$ factor in the exponent)
\begin{equation}
\Theta \, \hat{U}(t) \, \Theta^{-1}=\hat{U}(t)=\mbox{exp}(-i H t),
\label{app:time_reversal}
\end{equation}
Let us now consider two eigenstates $\ket{j}$ and $\ket{k}$ of the magnetization $\hat{S}^z$, with $j,k=1,2\dots 2S+1$. The relation between the index $j$ ($k$) and the eigenvalue $s_z(j)\in [-S,S]$ ($s_z(k)$) is $s_z(j)=j-S-1$ . We wish to compute matrix matrix elements of the unitary propagator $\hat{U}$ with respect to such basis. To this aim, we define the state $\ket{\gamma}=\hat{U}^{\dagger}\ket{j}$ and we get, due to antiunitary nature of the time-reversal operator $\Theta$, that
\begin{equation}
\braket{j|\hat{U}|k}=\braket{\gamma|k}=\braket{\tilde{\gamma}|\tilde{k}}^{\ast}=\braket{\tilde{k}|\tilde{\gamma}}=\braket{\tilde{k}|\Theta \hat{U}^{\dagger}|j}=\braket{\tilde{k}|\Theta \hat{U}^{\dagger} \Theta^{-1}|\tilde{j}}=\braket{\tilde{k}| \hat{U}^{\dagger}|\tilde{j}}.
\label{eq:chain_t_reversal}
\end{equation}
In the second equality, we exploited that antiunitary transformations take the scalar product of two states into their complex conjugates. We also defined the time-reversal transformed state $\ket{\tilde{j}}$ (and analogously for $\ket{\tilde{k}}$)
\begin{equation}
\ket{\tilde{j}}=\Theta \ket{j}=i^{2(j-S-1)}\ket{2S+2-j},
\label{eq:time_reversal_state}
\end{equation}
which expresses that under time-reversal the quantum number associated with $\hat{S}^z$ is flipped $s_z \to -s_z$ and therefore $j\to 2S+2-j$. In the last equality of Eq.~\eqref{eq:chain_t_reversal}, we eventually used the invariance \eqref{app:time_reversal} of the unitary operator under time-reversal symmetry. In a similar way, we can consider the action of the unitary operator $\hat{U}^{\pi}_y$, defined in \eqref{app:antiunitary_decomposition}, which implements a rotation around the $y$ axis by an angle $\pi$. The unitary evolution operator of the noninteracting spins Hamiltonian accordingly changes as 
\begin{equation}
\hat{U}_y^{\pi} \, \hat{U} \, (\hat{U}_y^{\pi})^{\dagger}= \mbox{exp}(i H t) = \hat{U}^{\dagger},
\label{eq:unitary_pi_rotation}
\end{equation}
and its matrix elements transform as 
\begin{equation}
\braket{j|\hat{U}|k}=\braket{\tilde{j}|\hat{U}^{\dagger}|\tilde{k}}.    
\label{eq:rotation_pi_unitary_intermediate}
\end{equation}
Comparing now Eq.~\eqref{eq:chain_t_reversal} and \eqref{eq:rotation_pi_unitary_intermediate}, we eventually conclude that
\begin{equation}
\braket{\tilde{k}|\hat{U}^{\dagger}|\tilde{j}}=\braket{\tilde{j}|\hat{U}^{\dagger}|\tilde{k}},\quad \Longleftrightarrow \quad \braket{\tilde{k}|\hat{U}|\tilde{j}}=\braket{\tilde{j}|\hat{U}|\tilde{k}}.
\label{eq:time_reversal_proved}
\end{equation}
By comparing Eq.~\eqref{eq:chain_t_reversal} and \eqref{eq:time_reversal_proved}, remembering the definition of the reset-free probabilities $p^F_{jk}(t)$ in Eq.~\eqref{eq:single_particle_propagator}, we immediately get
\begin{equation}
p^F_{jk}(t)=p^F_{kj}(t)=\braket{k|\hat{U}|j}\braket{j|U^{\dagger}|k}=\braket{\tilde{j}|\hat{U}|\tilde{k}}\braket{\tilde{k}|\hat{U}^{\dagger}|\tilde{j}}=p^F_{\tilde{k}\tilde{j}}(t)= p^F_{\tilde{j}\tilde{k}}(t),
\label{eq:symmetry_t_proved_final}
\end{equation}
where in the third equality we used \eqref{eq:rotation_pi_unitary_intermediate}. Equation \eqref{eq:symmetry_t_proved_final} is eventually recognized to be identical to \eqref{eq:symmetry_p} of the main text with the identification of the time-reversal transformed states $\ket{\tilde{j}}$ and $\ket{\tilde{k}}$ according to Eq.~\eqref{eq:time_reversal_state}. Importantly, we note that in the reset-free dynamics the unitary propagation from any pair of states $\ket{j}$ and $\ket{k}$ is symmetrical, i.e., it takes place with the same probability both for the transition from $\ket{j}$ to $\ket{k}$ and vice versa.

\section{Reset-free probabilities for total spin $S=1$ and unconditional resetting}
\label{app:appendix_1}
In this Appendix, we report the explicit form of the spin matrices and the reset-free single spin probabilities \eqref{eq:single_particle_propagator} for total spin $S=1$, which have been used for the calculations in the main text. In the following, we denote as 
\begin{equation}
 \tilde\Omega= \frac{\Omega}{\Delta}, \quad \mbox{and} \quad  \Gamma = \sqrt{\Omega^2 + \Delta^2},   
\label{app:app_notation}
\end{equation}
consistently with the notation of the main text. The same notation as \eqref{app:app_notation} for the ratio $\tilde{\Omega}$ between the Rabi frequency $\Omega$ and the detuning $\Delta$, together with the frequency oscillation $\Gamma$, will be used for the other spin values $S=3/2$ and $S=2$ that will be presented in the next sections of the Appendix.  

For $S=1$, the spin matrices $\hat S^x_{\alpha}$, $S^y_{\alpha}$ $\hat{S^z_{\alpha}}$ correspond to the generators of the Lie algebra for the three-dimensional representation of the $SU(2)$ group and they are given by
\begin{equation}
\label{eq: Sx_S1}
    \hat{S}^x_{\alpha}=\frac{1}{\sqrt{2}}    
     \begin{pmatrix}
        0 & 1 & 0\\
        1 & 0 & 1\\
        0 & 1 & 0\\
    \end{pmatrix},
\quad
\hat{S}^y_{\alpha}=
\frac{1}{\sqrt{2}}
     \begin{pmatrix}
        0 & -i & 0\\
        i & 0 & -i\\
        0 & i & 0\\
    \end{pmatrix},    
\quad
    \hat{S}^z_{\alpha}=   
     \begin{pmatrix}
        1 & 0 & 0\\
        0 & 0 & 0\\
        0 & 0 & -1\\
    \end{pmatrix}.
\end{equation}
The subscript $\alpha$ here denotes the lattice site. The reset-free single-spin probabilities clearly do not depend on $\alpha$ since the Hamiltonian \eqref{eq:Hamiltonian} and the reset states are translational invariant and we therefore omit this subscript henceforth for lightness of notation. We also note that $S^x_{\alpha}$ is a centrosymmetric matrix:
\begin{equation}
S^x_{jk}=S^x_{2S-j+2,2S-k+2}, \quad j,k=1,2,\dots 2S+1,    
\label{eq:centrosymm}
\end{equation}
while $S^y_{\alpha}$ and $S^z_{\alpha}$ are skew-centrosymmetric  
\begin{equation}
S^{y,z}_{jk}=-S^{y,z}_{2S-j+2,2S-k+2}, \quad j,k=1,2,\dots 2S+1.
\label{eq:skew-centrosymm}
\end{equation}
\begin{figure*}[t]
    \centering
    \def\svgwidth{0.5\textwidth} 
    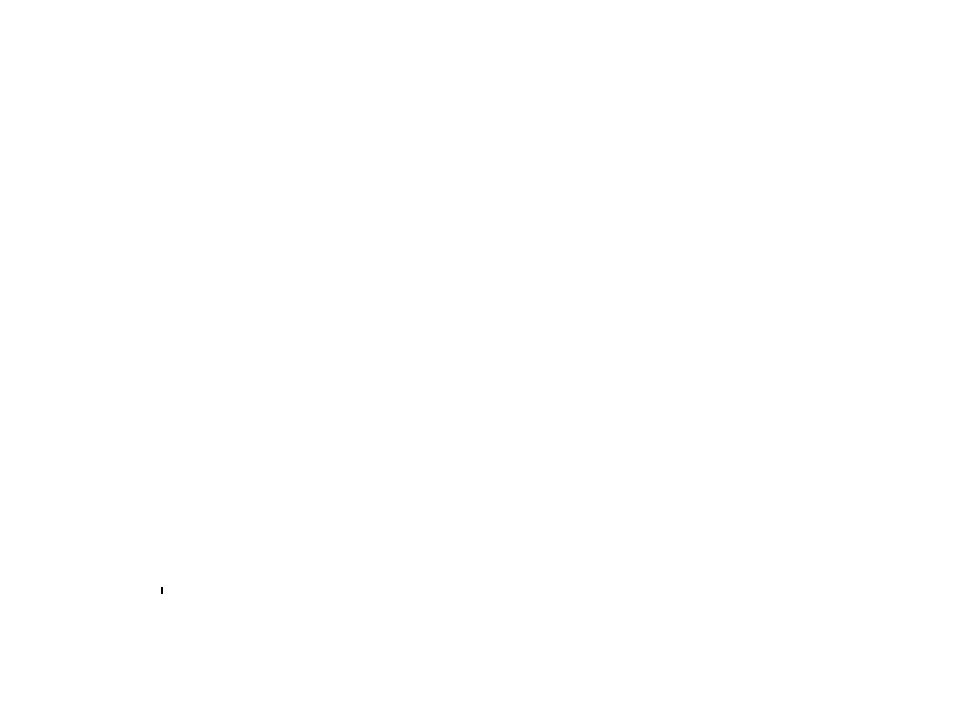
    \caption{\textbf{Stationary density of spins in the unconditional reset protocol for a noninteracting spin-$1$ system.} We plot the stationary density of spins $p_{11}^{\mathrm{NESS},uc}(\tilde \Omega)$ in the state $\ket{1}_{\alpha}$ for unconditional resetting towards the state $\ket{1}_N$ and total spin $S=1$. The curve for resetting rate $\gamma=1$ is the same as that for $p^{\mathrm{NESS}}_{1}$ for $\tilde\Omega<\tilde\Omega_{c2}$ in Fig.~\ref{fig: S1}~(a). In the figure, we set $\Delta=1$ and therefore it is $\Gamma=\sqrt{\Omega ^2 +1}$ and we take various resetting values $\gamma=1,0.5,0.1$ (from top to bottom).}
    \label{fig: exden-uncondreset-S1}
\end{figure*}
As a consequence, we find that the single spin unitary operator $(\hat{U}_{\alpha})_{j,k}$, is neither centrosymmetric nor skew-centrosymmetric. Its modulus squared, which determines the reset-free probabilities according to \eqref{eq:single_particle_propagator}, however, is still a centrosymmetric matrix. This eventually causes the matrix $p_{jk}^{F}(t)$ of reset-free single-spin propagators to be centrosymmetric in agreement with the relation \eqref{eq:symmetry_t_proved_final} derived based on the time-reversal symmetry of the unitary operator. The entries of the unitary propagator can be evaluated by a direct calculation starting from the expansion in Eq.~\eqref{eq: eponential expansion S=1 - first} of the propagator $\hat{U}_{\alpha}$ in terms of the spin matrices \eqref{eq: Sx_S1}. They read as
\begin{align}
\label{eq: p(t) explicit, S=1, first}
p^F_{11}=p^F_{33}&=\frac{ \left(1+ \tilde \Omega ^2\cos^2{\left(\frac{\Gamma t}{2}\right)}\right)^2}{\left(1+\tilde \Omega^2\right)^2}\, ,\\
p^F_{12}=p^F_{21}=p^F_{32}=p^F_{23}&= \frac{2\tilde{\Omega}^2 \sin^2\left(\frac{\Gamma t}{2}\right) \left(1 + \tilde{\Omega}^2 \cos^2\left(\frac{\Gamma t}{2}\right)  \right)}{\left(1 + \tilde{\Omega}^2\right)^2},
\\
p^F_{13}=p^F_{31}&=\frac{\tilde\Omega^4 \sin^4\left(\frac{\Gamma t}{2}\right)}{\left( 1+ \tilde\Omega^2\right)^2}, \\ 
p^F_{22}&=\frac{ \left(1+ \tilde \Omega ^2\cos{\left(\Gamma t \right)}\right)^2}{\left(1+\tilde \Omega^2\right)^2} \, .
\end{align}
The parameter $\Gamma$ sets the frequency of oscillations in time of the occupation of the different magnetization eigenstates, while $\tilde{\Omega}$ dictates the amplitude of such oscillations (similarly to the spin $1/2$ case in Eqs.~\eqref{eq:qubit1} and \eqref{eq:qubit2}). We note the symmetric structure of the reset-free probabilities according to Eq.~\eqref{eq:symmetry_t_proved_final} determine $p_{12}^F=p^F_{32}$ and $p^F_{21}=p^F_{23}$, which is source of the symmetry \eqref{eq:symmetry_R} between the transition probabilities in the Markov chain of reset states. In addition to this, we also have that $p_{12}^F=p^F_{21}$ and $p_{23}^F=p_{32}^F$, consistently with Eq.~\eqref{eq:symmetry_t_proved_final}. It is, however, important to note that the latter equalities do not imply that $R_{12}=R_{21}$ and $R_{23}=R_{32}$. As a matter of fact, for example, if $p_{11}^F>p^F_{12}>p^F_{13}$, one has that $R_{12}=0$, while $R_{21}\neq 0$. This is precisely the case in Fig.~\ref{fig: S1}(d) of the main text for $\tilde{\Omega}_{c1}<\tilde{\Omega}<\tilde{\Omega}_{c2}$. As soon as resets from to the state $\ket{2}_N$ after the previous reset to $\ket{1}_N$, and vice versa, become possible, the associated transition probability $R_{12} \neq 0$, with $R_{12} \neq R_{21}$. The Markov chain between reset states is consequently directional when transitions between pair of states not having opposite magnetization eigenvalues are considered. This is the case, for instance, of Fig.~\ref{fig: S1}(e) for $\tilde{\Omega}_{c2}<\tilde{\Omega}<\tilde{\Omega}_{c3}$.      

From the expressions in Eq.~\eqref{eq: p(t) explicit, S=1, first}, it is immediate to compute the stationary density of spins in each $S^z_{\alpha}$ magnetization eigenstate in the unconditional reset protocol. Considering, for instance, the case of unconditional resetting to the state $\ket{1}_N$, the NESS density of spins $p_{11}^{\mathrm{NESS},uc}$ in the same state is obtained via Eq.~\eqref{eq:ness_pij_unconditional}:
\begin{equation}
\label{app:unconditional_S1}
     p^{\mathrm{NESS},uc}_{11}(\tilde \Omega, \gamma, \Gamma)
    =\frac{2 \gamma^4 \tilde\Omega^4+4 \gamma^4 \tilde\Omega^2+2 \gamma^4+8 \gamma^2 \Gamma^2 \tilde\Omega^4+18 \gamma^2 \Gamma^2 \tilde\Omega^2+10 \gamma^2 \Gamma^2+3 \Gamma^4 \tilde\Omega^4+8 \Gamma^4 \tilde\Omega^2+8 \Gamma^4}{2 \left(\tilde\Omega^2+1\right)^2 \left(\gamma^2+\Gamma^2\right) \left(\gamma^2+4 \Gamma^2\right)}.
\end{equation}
This function is plotted in Fig.~\ref{fig: exden-uncondreset-S1} for different values of the reset rate $\gamma$ with $\Delta=1$ and thus $\Gamma=\sqrt{\Omega ^2 +1}$. The expression in Eq.~\eqref{app:unconditional_S1} coincides with the one obtained in the conditional reset protocol starting from $\ket{1}_N$ for $\tilde{\Omega}<\tilde{\Omega}_{c1}$ in Fig.~\ref{fig: S1} when only resets to state $\ket{1}_N$ are possible. In the expression for the stationary state densities, this amounts to setting $c_1=1$ and $c_j=0$, for $j\neq 1$, in Eq.~\eqref{eq:ness}.

\section{Reset-free probabilities for total spin $S=3/2$ and unconditional resetting}
\label{app:S=1.5}
For $S=3/2$, the spin matrices $\hat S^{x,y,z}_{\alpha}$ for each site $\alpha$ are four dimensional and they are given by  
\begin{align}
\label{eq: Sx_S1.5}
    \hat{S}^x_{\alpha}=\frac{1}{2}    
     \begin{pmatrix}
        0 & \sqrt{3} & 0 & 0\\
        \sqrt{3} & 0 & 2 & 0\\
        0 & 2 & 0 & \sqrt{3}\\
        0 & 0 & \sqrt{3} & 0\\
    \end{pmatrix}, \quad
    \hat{S}^y_{\alpha}=\frac{i}{2}    
     \begin{pmatrix}
        0 & -\sqrt{3} & 0 & 0\\
        \sqrt{3} & 0 & -2 & 0\\
        0 & 2 & 0 & -\sqrt{3}\\
        0 & 0 & \sqrt{3} & 0\\
    \end{pmatrix}, \quad
    \hat{S}^z_{\alpha}=\frac{1}{2}    
     \begin{pmatrix}
        3 & 0 & 0 & 0\\
        0 & 1 & 0 & 0\\
        0 & 0 & -1 & 0\\
        0 & 0 & 0 & -3\\
    \end{pmatrix}\, .
\end{align}
As in the spin $1$ case, we observe that the matrix $S^x_{\alpha}$ is centrosymmetric as in \eqref{eq:centrosymm}, while $S^{y,z}_{\alpha}$ are skew-symmetric according to \eqref{eq:skew-centrosymm}. Consequentely we find that the modulus square of the single-spin propagator $\hat{U}_{\alpha}$ is centrosymmetric and so the single spin probabilities $p^F_{ij}(\tilde\Omega, \Gamma,t)$ are. For the latter we obtain
\begin{align}
\label{eq: p(t) explicit, S=1.5, first}
p^F_{11}=p^F_{44}&=
\frac{ \left(\left(7 \tilde\Omega^{2} + 9\right) \sin^{2}{\left(\frac{\Gamma t}{2} \right)} - 6 \left(\tilde\Omega^{2} + 1\right) \cdot \left(\frac{1}{6} \sin^{2}{\left(\frac{\Gamma t}{2} \right)} + 1\right)\right)^{2} \sin^{2}{\left(\frac{\Gamma t}{2} \right)}}{4 \left(\tilde\Omega^{2} + 1\right)^{3}}  \nonumber \\
&+\frac{16 \left(\left(\frac{1}{4} \tilde\Omega^{2} +1 \right) \cos^{2}{\left(\frac{\Gamma t}{2} \right)} - \frac{3}{4}\right)^{2} \cos^{2}{\left(\frac{\Gamma t}{2} \right)}}{\left(\tilde\Omega^{2} + 1\right)^{2}}\, ,\\ 
p^F_{12}= p^F_{43}=p^F_{21}=p^F_{34}&= \frac{3 \tilde\Omega^{2} \left(\tilde\Omega^{2} \cos^{2}{\left(\frac{\Gamma t}{2} \right)} + 1\right)^{2} \sin^{2}{\left(\frac{\Gamma t}{2} \right)}}{\left(\tilde\Omega^{2} + 1\right)^{3}}, \label{eq:reset_free_S32_4th} \\
p^F_{13}=p^F_{42}= p^F_{31}=p^F_{24}&= \frac{3 \tilde\Omega^{4} \sin^{4}{\left(\frac{\Gamma t}{2} \right)}
\left(\sin^{2}{\left(\frac{\Gamma t}{2}\right)}+ \left(\tilde\Omega^{2} + 1\right) \cos^{2}{\left(\frac{\Gamma t}{2} \right)}\right)}{\left(\tilde\Omega^{2} + 1\right)^{3}},
\label{eq:reset_free_S32_5th}
\end{align}
\begin{figure*}[t!]
    \centering
    \def\svgwidth{0.5\textwidth} 
    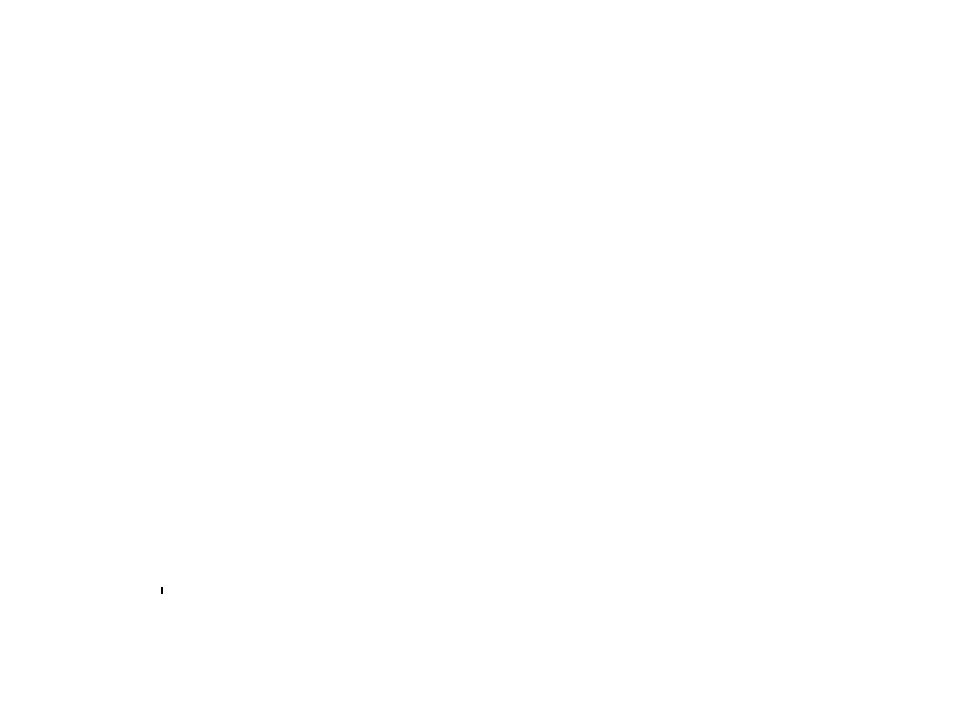
    \caption{\textbf{Stationary density of spins in the unconditional reset protocol for a noninteracting spin-$3/2$ system.} We plot the stationary density of spins $p_{11}^{\mathrm{NESS},uc}(\tilde \Omega)$ in the state $\ket{1}_{\alpha}$ for unconditional resetting towards the state $\ket{1}_N$ and total spin $S=3/2$. The curve for resetting rate $\gamma=1$ is the same as that for $p^{\mathrm{NESS}}_{1}$ for $\tilde\Omega<\tilde\Omega_{c3}$ in Fig.~\ref{fig: S1.5}~(a). In the figure, we set $\Delta=1$ and therefore it is $\Gamma=\sqrt{\Omega ^2 +1}$ and we take various resetting values $\gamma=1,0.5,0.1$ (from top to bottom).}
    \label{fig: exden-uncondreset-S1.5}
\end{figure*}
\begin{align}
p^F_{14}= p^F_{41}&=\frac{\tilde\Omega^{6} \sin^{6}{\left(\frac{\Gamma t}{2} \right)}}{\left(\tilde\Omega^{2} + 1\right)^{3}},\\
p^F_{22}=p^F_{33}&=
\frac{\left(\left(19 \tilde\Omega^{2} + 1\right) \sin^{2}{\left(\frac{\Gamma t}{2} \right)} - 6 \left(\tilde\Omega^{2} + 1\right) \cdot \left(\frac{1}{6} \sin^{2}{\left(\frac{\Gamma t}{2} \right)} + 1\right)\right)^{2} \sin^{2}{\left(\frac{\Gamma t}{2} \right)} }{36 \left(\tilde\Omega^{2} + 1\right)^{3}} \nonumber \\
 &+\frac{9 \left(- \tilde\Omega^{2} \sin^{2}{\left(\frac{\Gamma t}{2} \right)} + \frac{1}{3} \tilde\Omega^{2} + \frac{1}{3}\right)^{2} \cos^{2}{\left(\frac{\Gamma t}{2} \right)}}{ \left(\tilde\Omega^{2} + 1\right)^{2}}, \\     
p^F_{23}=p^F_{32}&=
\frac{\tilde\Omega^{2} \left(-\left(10 \tilde\Omega^{2} + 1\right) \sin^{2}{\left(\frac{\Gamma t}{2} \right)} + 6 \left(\tilde\Omega^{2} + 1\right) \cdot \left(\frac{1}{6} \sin^{2}{\left(\frac{\Gamma t}{2} \right)} + 1\right)\right)^{2} \sin^{2}{\left(\frac{\Gamma t}{2} \right)}}{9 \left(\tilde\Omega^{2} + 1\right)^{3}}.
\end{align}
Also in this case, the centrosymmetric structure \eqref{eq:symmetry_t_proved_final} of the reset-free probabilities from time-reversal invariance of the unitary propagator is manifest. This is implemented in the equalities $p^F_{12}=p^F_{43}$ and $p^F_{21}=p^F_{34}$ in Eq.~\eqref{eq:reset_free_S32_4th}, and in $p^F_{13}=p^F_{42}$, $p^F_{31}=p^F_{42}$ in Eq.~\eqref{eq:reset_free_S32_5th}. Also in this case, the additional equalities $p^F_{12}=p^F_{21}$ and $p^F_{43}=p^F_{34}$ in \eqref{eq:reset_free_S32_4th}, and $p^F_{13}=p^F_{31}$ and $p^F_{42}=p^F_{24}$ in \eqref{eq:reset_free_S32_4th}, do not imply equalities between the associated transition probabilities in the reset states after global monitoring is introduced. For example, in Fig.~\ref{fig: S1.5}(f), one has that $R_{12}\neq R_{21}$. 

For the unconditional resetting protocol, the evaluation of the spins stationary densities in each magnetization eigenstate in the $z$ direction is then straightforward from Eq.~\eqref{eq: p(t) explicit, S=1.5, first}. As an example, the NESS density of spins $p_{11}^{\mathrm{NESS},uc}$ in the state $\ket{1}_{\alpha}$ when uncondtional resetting towards the same state is performed reads from Eq.~\eqref{eq:ness_pij_unconditional} as
\begin{align} 
\label{eq: uncond exden explicit S1.5_1}
    p^{\mathrm{NESS},uc}_{11}(\tilde \Omega, \gamma, \Gamma)
    &=\frac{4 \gamma^6 \tilde \Omega^6+12 \gamma^6 \tilde \Omega^4+12 \gamma^6 \tilde \Omega^2+4 \gamma^6+50 \gamma^4 \Gamma^2 \tilde \Omega^6+156 \gamma^4 \Gamma^2 \tilde \Omega^4+162 \gamma^4 \Gamma^2 \tilde \Omega^2+56 \gamma^4 \Gamma^2}{4 \left(\tilde \Omega^2+1\right)^3 \left(\gamma^2+\Gamma^2\right) \left(\gamma^2+4 \Gamma^2\right) \left(\gamma^2+9 \Gamma^2\right)} \nonumber \\
    &+\frac{136 \gamma^2 \Gamma^4 \tilde \Omega^6+450 \gamma^2 \Gamma^4 \tilde \Omega^4+510 \gamma^2 \Gamma^4 \tilde \Omega^2+196 \gamma^2 \Gamma^4+45 \Gamma^6 \tilde \Omega^6+162 \Gamma^6 \tilde \Omega^4+216 \Gamma^6 \tilde \Omega^2+144 \Gamma^6}{4 \left(\tilde \Omega^2+1\right)^3 \left(\gamma^2+\Gamma^2\right) \left(\gamma^2+4 \Gamma^2\right) \left(\gamma^2+9 \Gamma^2\right)}\, .
\end{align}
This result is reported in Fig.~\ref{fig: exden-uncondreset-S1.5} for various values of the resetting rate $\gamma$. It coincides with the result obtained in the unconditional resetting protocol in Fig.~\ref{fig: S1.5}(a) for $\tilde{\Omega}<\tilde{\Omega}_{c3}$, i.e., the reset state $\ket{1}_N$ is absorbing and therefore all reset events take place towards this state. From Eq.~\eqref{eq: p(t) explicit, S=1.5, first} and \eqref{eq:ness_pij_unconditional} one can similarly work out the stationary expression for the density of spins in the other magnetization eigenstates $\ket{j}_{\alpha}$ in the $z$ direction. In general, such expressions coincide with the results from the conditional reset protocol, when, in the latter, only resets to the state $\ket{j}_N$ corresponding to the same magnetization eigenvalue are possible. 

\section{Reset-free probabilities for total spin $S=2$ and unconditional resetting}
\label{app:S=2}
For $S=2$, the spin matrices $\hat S^{x,y,z}_{\alpha}$ are five dimensional and they are given by 
\begin{equation}
\label{eq: Sx_S2}
    \hat{S}^x_{\alpha}=\frac{1}{2}    
     \begin{pmatrix}
        0 & 2 & 0 & 0 & 0\\
        2 & 0 & \sqrt{6} & 0 & 0\\
        0 & \sqrt{6} & 0 & \sqrt{6} & 0\\
        0 & 0 & \sqrt{6} & 0 & 2\\
        0 & 0 & 0 & 2 & 0\\
    \end{pmatrix}, \quad
    \hat{S}^y_{\alpha}=\frac{1}{2}    
     \begin{pmatrix}
        0 & -2 & 0 & 0 & 0\\
        2 & 0 & -\sqrt{6} & 0 & 0\\
        0 & \sqrt{6} & 0 & -\sqrt{6} & 0\\
        0 & 0 & \sqrt{6} & 0 & -2\\
        0 & 0 & 0 & 2 & 0\\
    \end{pmatrix}, \quad
    \hat{S}^z_{\alpha}=   
     \begin{pmatrix}
        2 & 0 & 0 & 0 & 0\\
        0 & 1 & 0 & 0 & 0\\
        0 & 0 & 0 & 0 & 0\\
        0 & 0 & 0 & -1 & 0\\
        0 & 0 & 0 & 0 & -2\\
    \end{pmatrix}\, .
\end{equation}
As well as in the previous cases of lower spin $S=1,3/2$, the matrix $S^x$ is centrosymmetric \eqref{eq:centrosymm}, while $S^{y,z}$ are skew-centrosymmetric \eqref{eq:skew-centrosymm}. In order to compute the reset-free propagators according to \eqref{eq:single_particle_propagator}, we again expand the single-spin operator in terms of a sum of a finite number of powers of $S^{x,y,z}_{\alpha}$ exploiting the result of Ref.~\cite{curtright2014compact}:
\begin{align}
    \hat{U}_{\alpha}(t)=e^{-i\hat{H}_{\alpha}t}=e^{-i(\Omega \hat{S}^x_{\alpha}+\Delta \hat{S}^z_{\alpha})t}&=\mathbbm{1}_{5\times 5} - 2i \left(\frac{\tilde \Omega \hat S^x_{\alpha} + \hat S^z_{\alpha}}{\sqrt{1+\tilde\Omega^2}}\sin\left(\frac{\Gamma t}{2}\right)\right) \cos\left(\frac{\Gamma t}{2}\right) \left(1 + \frac{2}{3} \sin^2\left(\frac{\Gamma t}{2}\right) \right) \nonumber \\
    &-2 \left(\frac{\tilde \Omega \hat S^x_{\alpha} + \hat S^z_{\alpha}}{\sqrt{1+\tilde\Omega^2}}\sin\left(\frac{\Gamma t}{2}\right)\right)^2\left(1 + \frac{1}{3} \sin^2\left(\frac{\Gamma t}{2}\right) \right) \nonumber \\
    &+ \frac{4i}{3} \left(\frac{\tilde \Omega \hat S^x_{\alpha} + \hat S^z_{\alpha}}{\sqrt{1+\tilde\Omega^2}}\sin\left(\frac{\Gamma t}{2}\right)\right)^3 \cos\left(\frac{\Gamma t}{2}\right) \nonumber + \frac{2}{3} \left(\frac{\tilde \Omega \hat S_x + \hat S_z}{\sqrt{1+\tilde\Omega^2}}\sin\left(\frac{\Gamma t}{2}\right)\right)^4,\\ \label{eq:U_exp_S2}.
\end{align}
The expression for the reset-free single-spin probabilities $p^F_{ij}(\tilde\Omega, \Gamma,t)$ can then be obtained from a lengthy, but otherwise simple, calculation and are given by
\begin{align}
\label{eq: p(t) explicit, S=2, first}
    p^F_{11}\cdot \left(\tilde\Omega^{2} + 1\right)^{4}&=p^F_{55}\cdot \left(\tilde\Omega^{2} + 1\right)^{4}=\Bigg (-\left(2 \tilde\Omega^{2} + 8\right) \left(\tilde\Omega^{2} + 1\right) \cdot \left(\frac{1}{3} \sin^{2}{\left(\frac{\Gamma t}{2} \right)} + 1\right) \sin^{2}{\left(\frac{\Gamma t}{2} \right)} \nonumber \\ &+\frac{2}{3}\left(\tilde\Omega^{2} \cdot \left(\frac{5}{2} \tilde\Omega^{2} + 7\right) + 10 \tilde\Omega^{2} + 16\right)
    \sin^{4}{\left(\frac{\Gamma t}{2} \right)} +
    \left(\tilde\Omega^{2} + 1\right)^{2} \Bigg )^{2}\nonumber \\
    &+ 4\left(\tilde\Omega^{2} + 1\right) \left(- 2 \cdot \left(\frac{5}{6} \tilde\Omega^{2} + \frac{4}{3}\right) \sin^{2}{\left(\frac{\Gamma t}{2} \right)} + \left(\tilde\Omega^{2} + 1\right) \cdot \left(\frac{2}{3} \sin^{2}{\left(\frac{\Gamma t}{2} \right)} + 1\right)\right)^{2}\sin^2(\Gamma t),
\end{align}
\begin{align}
    p^F_{12}\cdot \left(\tilde\Omega^{2} + 1\right)^{4}&=
    p^F_{54}\cdot \left(\tilde\Omega^{2} + 1\right)^{4}=
    p^F_{21}\cdot \left(\tilde\Omega^{2} + 1\right)^{4}=
    p^F_{45}\cdot \left(\tilde\Omega^{2} + 1\right)^{4} \nonumber \\
    &=\tilde\Omega^{2} \sin^{2}{\left(\frac{\Gamma t}{2} \right)} 
    \cdot \Bigg [\left(- \left(8 \tilde\Omega^{2} + 10\right) \sin^{2}{\left(\frac{\Gamma t}{2} \right)} + 6\left(\tilde\Omega^{2} + 1\right) \cdot \left(\frac{1}{3} \sin^{2}{\left(\frac{\Gamma t}{2} \right)} + 1\right)\right)^{2}\cdot \sin^{2}{\left(\frac{\Gamma t}{2} \right)}\nonumber \\
    &+ \left(\tilde\Omega^{2} + 1\right) \left(-\frac{2}{3} \left(5 \tilde\Omega^{2} + 14\right) \sin^{2}{\left(\frac{\Gamma t}{2} \right)} + 2 \left(\tilde\Omega^{2} + 1\right) \cdot \left(\frac{2}{3} \sin^{2}{\left(\frac{\Gamma t}{2} \right)} + 1\right)\right)^{2}\cos^{2}{\left(\frac{\Gamma t}{2} \right)} \Bigg ]\, ,\\
    p^F_{13} \cdot \left(\tilde\Omega^{2} + 1\right)^{4}&=
    p^F_{53}\cdot \left(\tilde\Omega^{2} + 1\right)^{4}=
    p^F_{31}\cdot \left(\tilde\Omega^{2} + 1\right)^{4}=
    p^F_{35}\cdot \left(\tilde\Omega^{2} + 1\right)^{4} \nonumber \\
    &=\tilde\Omega^{4}  \sin^{4}{\left(\frac{\Gamma t}{2} \right)} \cdot \Bigg[ 24 \cdot \left(\tilde\Omega^{2} + 1\right) \sin^{2}{\left(\frac{\Gamma t}{2} \right)}\cos^{2}{\left(\frac{\Gamma t}{2} \right)} \nonumber \\
    &+ \frac{2}{3}\left(- \left(4 \tilde\Omega^{2} + 7\right) \sin^{2}{\left(\frac{\Gamma t}{2} \right)} + 3 \left(\tilde\Omega^{2} + 1\right) \cdot \left(\frac{1}{3} \sin^{2}{\left(\frac{\Gamma t}{2} \right)} + 1\right)\right)^{2}\Bigg ]\, ,\\
    p^F_{14}&=p^F_{52}=p^F_{41}=p^F_{25}=\frac{4 \tilde\Omega^{6} \sin^{6}{\left(\frac{\Gamma t}{2} \right)} \left(\sin^2{\left(\frac{\Gamma t}{2} \right)} + \left(\tilde\Omega^{2} + 1\right)\cos^{2}{\left(\frac{\Gamma t}{2} \right)} \right)}{\left(\tilde\Omega^{2} + 1\right)^{4}}\, ,\\
     p^F_{15}&= p^F_{51}=\frac{\tilde\Omega^{8} \sin^{8}{\left(\frac{\Gamma t}{2} \right)}}{\left(\tilde\Omega^{2} + 1\right)^{4}}\, ,
\end{align}
\begin{align}
    p^F_{22}\cdot  \left(\tilde\Omega^{2} + 1\right)^{4} &=
    p^F_{44}\cdot \left(\tilde\Omega^{2} + 1\right)^{4} =\Bigg [ \left( \frac{1}{3}  \tilde\Omega^{2}\left(\left(5 \tilde\Omega^{2} + 14\right) + 3 \left(4 \tilde\Omega^{2} +1\right) + 14 \right) + \frac{2}{3}\right) \sin^{4}{\left(\frac{\Gamma t}{2} \right)} \nonumber \\
    &- \left(\tilde\Omega^{2} + 1\right) \cdot \left(5 \tilde\Omega^{2} + 2\right) \left(\frac{1}{3} \sin^{2}{\left(\frac{\Gamma t}{2} \right)} + 1\right) \sin^{2}{\left(\frac{\Gamma t}{2} \right)} + \left(\tilde\Omega^{2} + 1\right)^{2}\Bigg ]^{2}\nonumber \\
    &+ \left(\tilde\Omega^{2} + 1\right)\left(- \frac{2}{3} \cdot \left( 7\tilde\Omega^{2} + 1\right) \sin^{2}{\left(\frac{\Gamma t}{2} \right)} + \left(\tilde\Omega^{2} + 1\right) \cdot \left(\frac{2}{3} \sin^{2}{\left(\frac{\Gamma t}{2} \right)} + 1\right)\right)^{2} \sin^{2}(\Gamma t),\\ 
    p^F_{23}\cdot \left(\tilde\Omega^{2} + 1\right)^{4}&=
    p^F_{43}\cdot\left(\tilde\Omega^{2} + 1\right)^{4}=
    p^F_{32}\cdot\left(\tilde\Omega^{2} + 1\right)^{4}=
    p^F_{34}\cdot\left(\tilde\Omega^{2} + 1\right)^{4}= \nonumber \\
    &\tilde\Omega^{2} \sin^{2}{\left(\frac{\Gamma t}{2} \right)} \Bigg ( \frac{2}{3}\left(\tilde\Omega^{2} + 1\right) \left(-\left(8 \tilde\Omega^{2} + 2\right) \sin^{2}{\left(\frac{\Gamma t}{2} \right)} + 3\left(\tilde\Omega^{2} + 1\right) \left(\frac{2}{3} \sin^{2}{\left(\frac{\Gamma t}{2} \right)} + 1\right)\right)^{2} \cos^{2}{\left(\frac{\Gamma t}{2} \right)}\nonumber \\
    &+ 6 \left(- \frac{1}{3}\left(7 \tilde\Omega^{2}+1\right) \sin^{2}{\left(\frac{\Gamma t}{2} \right)} + \left(\tilde\Omega^{2} + 1\right) \left(\frac{1}{3} \sin^{2}{\left(\frac{\Gamma t}{2} \right)} + 1\right)\right)^{2} \sin^{2}{\left(\frac{\Gamma t}{2} \right)}\Bigg ) \,,
\end{align}
\begin{align}
    p^F_{24}\cdot \left(\tilde\Omega^{2} + 1\right)^{4}&=
    p^F_{42}\cdot \left(\tilde\Omega^{2} + 1\right)^{4}=\tilde\Omega^{4} \left(-\left(5 \tilde\Omega^{2} + 1\right) \sin^{2}{\left(\frac{\Gamma t}{2} \right)} + 3 \left(\tilde\Omega^{2} + 1\right) \cdot \left(\frac{1}{3} \sin^{2}{\left(\frac{\Gamma t}{2} \right)} + 1\right)\right)^{2} \sin^{4}{\left(\frac{\Gamma t}{2} \right)}\, , \\
   p^F_{33} \cdot  \left(\tilde\Omega^{2} + 1\right)^{4}&=\left(2 \tilde\Omega^{2}\left(4 \tilde\Omega^{2} + 1\right) \sin^{4}{\left(\frac{\Gamma t}{2} \right)} - 6 \tilde\Omega^{2} \left(\tilde\Omega^{2} + 1\right) \cdot \left(\frac{1}{3} \sin^{2}{\left(\frac{\Gamma t}{2} \right)} + 1\right) \sin^{2}{\left(\frac{\Gamma t}{2} \right)} + \left(\tilde\Omega^{2} + 1\right)^{2}\right)^{2}\, .
\end{align}
In all the cases, $\Gamma$ sets the frequency of the oscillations, while $\tilde{\Omega}$ the amplitude. The centrosymmetric structure \eqref{eq:time_reversal_proved} of the modulus squared of \eqref{eq:U_exp_S2} coming from time-reversal symmetry is again reflected into the equalities relating the reset-free probabilities of the two outer states ($\ket{1}_{\alpha}$ and $\ket{5}_{\alpha}$), and into the equalities between the reset-free probabilities of the two  intermediate states ($\ket{2}_{\alpha}$ and $\ket{4}_{\alpha}$). The state $\ket{3}_N$ is the so-called center state and has no symmetric counterpart.

\begin{figure*}[t!]
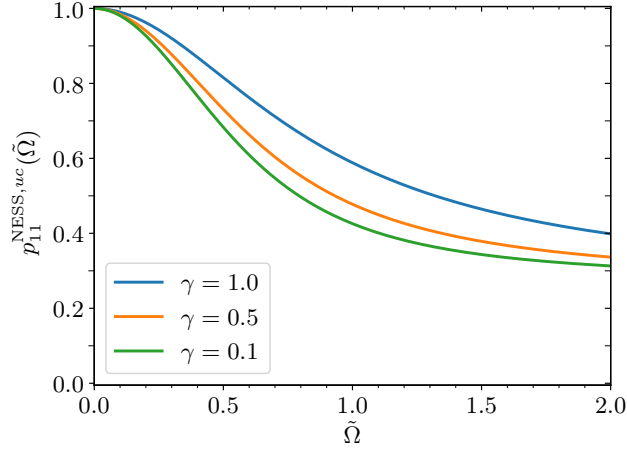

    \centering
    \def\svgwidth{0.5\textwidth} 
    
    \caption{\textbf{Stationary density of spins in the unconditional reset protocol for a noninteracting spin-$2$ system.} We plot the stationary density of spins $p_{11}^{\mathrm{NESS},uc}(\tilde \Omega)$ in the state $\ket{1}_{\alpha}$ for unconditional resetting towards the state $\ket{1}_N$ and total spin $S=2$. The curve for resetting rate $\gamma=1$ is the same as that for $p^{\mathrm{NESS}}_{1}$ for $\tilde\Omega<\tilde\Omega_{c4}$ in Fig.~\ref{fig: S2}~(a). In the figure, we set $\Delta=1$ and therefore it is $\Gamma=\sqrt{\Omega ^2 +1}$ and we take various resetting values $\gamma=1,0.5,0.1$ (from top to bottom).}
    \label{fig: exden-uncondreset-S2}
\end{figure*}

For the calculation of the stationary density of spins in a magnetization eigenstate in the $z$ direction we proceed as in the cases with lower spin from the last renewal equation \eqref{eq:ness_pij_unconditional}. Focusing on the paradigmatic case of the NESS density $p_{11}^{\mathrm{NESS},uc}$ of spins in the state $\ket{1}_{\alpha}$ with resetting towards the state $\ket{1}_N$ with the same magnetization in the $z$ direction for each site, we find
\begin{align}
\label{eq: uncond exden explicit S2_1}
    &p^{\mathrm{NESS},uc}_{11}(\tilde \Omega, \gamma, \Gamma) =\frac{35 \tilde\Omega^8+160 \tilde\Omega^6+288 \tilde\Omega^4+256 \tilde\Omega^2+128}{128 \left(\tilde\Omega^2+1\right)^4}+ \nonumber \\
&\frac{1}{128 \left(\tilde\Omega^2+1\right)^4}
    \left(\frac{\gamma^2 \tilde\Omega^8}{\gamma^2+16 \Gamma^2}
    \!+\!\frac{8 \gamma^2 \left(\tilde\Omega^2+2\right) \tilde\Omega^6}{\gamma^2+9 \Gamma^2}
    \!+\!\frac{4 \gamma^2 \left(7 \tilde\Omega^4+24 \tilde\Omega^2+24\right) \tilde\Omega^4}{\gamma^2+4 \Gamma^2}
    \!+\!\frac{8 \gamma^2 \left(\tilde\Omega^2+2\right) \left(7 \tilde\Omega^4+16 \tilde\Omega^2+16\right) \tilde\Omega^2}{\gamma^2+\Gamma^2}\right).
\end{align}
This function is reported in Fig.~\ref{fig: exden-uncondreset-S2}. In particular, it coincides with the result for the conditional resetting protocol in Fig.~\ref{fig: S2}(a) when the system is initialized in $\ket{1}_N$ and one has $\tilde{\Omega}<\tilde{\Omega}_{c4}$. In this case, the reset state $\ket{1}_N$ is absorbing and all the resets take place towards this state (cf. Figs.~\ref{fig: S2}(c)-(g)) rendering the protocol equivalent to unconditional resetting to the same state.

\section{Additional analysis of the ergodic regime}
In this Appendix, we provide additional details and plots for the stationary spin densities in the ergodic regime. In Appendix \ref{app:S32_ergodic}, we discuss the case of spin $S=3/2$, while spin $S=2$ is discussed in appendix \ref{app:S2_ergodic}.

\subsection{Spin-$3/2$ system}
\label{app:S32_ergodic}
\begin{figure*}[t!]
    \centering
    \def\svgwidth{\textwidth} 
    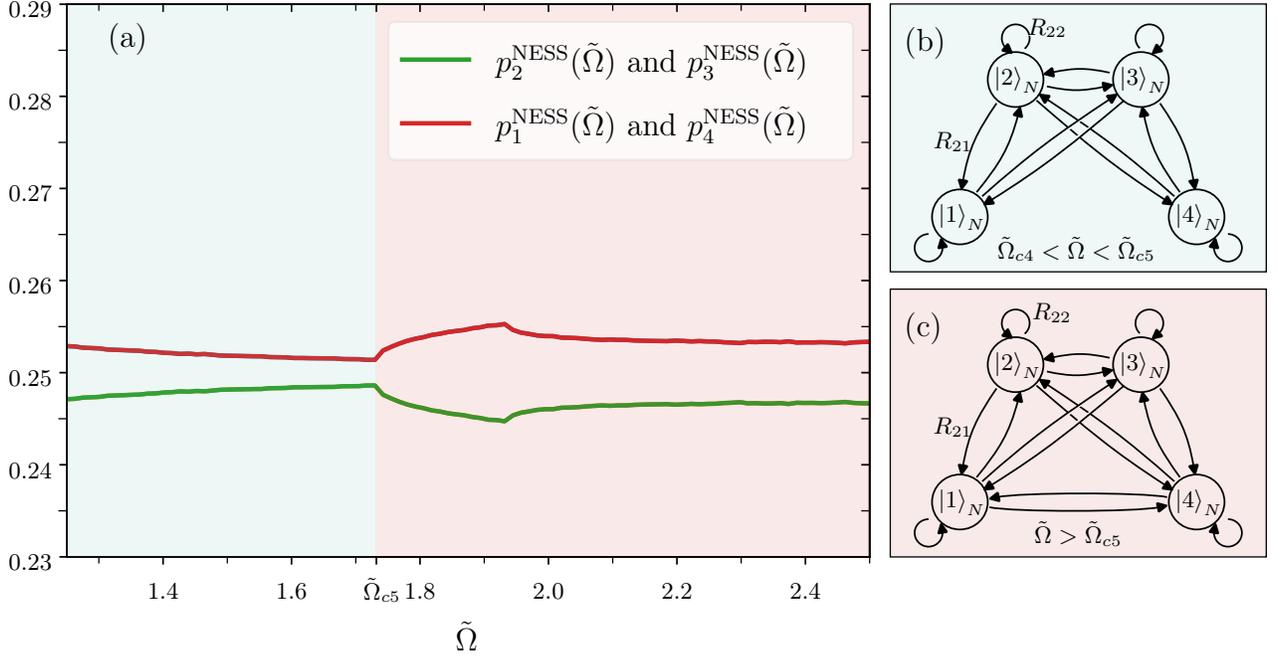
    \caption{\textbf{Stationary spin densities for a spin-$3/2$ system.}
    Plot of the stationary densities $p_j^{\mathrm{NESS}}$ as a function of $\tilde\Omega=\Omega/\Delta$ in the ergodic regime for $\tilde{\Omega}>1.25$. Two different regions (highlighted with different colors) are identified as a function of $\tilde{\Omega}$ and the critical value $\tilde\Omega_{c5}$ \eqref{eq:S1.5_C5}. In the panels (b) and (c), we report the reset Markov chain corresponding to the two regions. In panel (a), one observes a second-order continuous transition in all four densities at $\tilde\Omega_{c5}$.
    A similar transition appears at $\tilde\Omega\approx 1.93$, which, however, is not associated with any new connection in the Markov chain diagram. For $\tilde\Omega>\tilde\Omega_{c5}$ the Markov chain is, indeed, already fully connected, as shown in panel (c). In the figure, we set $\gamma=\Delta=1$.}
    \label{fig: S1.5-ergodic}
\end{figure*}

In the spin-$3/2$ system, the ergodic regime is realized for $\tilde{\Omega}>\tilde{\Omega}_{c3}$, with $\tilde{\Omega}_{c3}$ given in Eq.~\eqref{eq:S1.5_C3} of the main text. In Fig.~\ref{fig: S1.5-ergodic}~(a), we find two additional second-order discontinuities for $\tilde{\Omega}>1.25$. Both these discontinuities appear to be quantitatively small (note the zoomed-in scale of the vertical axis compared to that of Fig.~\ref{fig: S1.5}). The discontinuity at $\tilde{\Omega}=\tilde\Omega_{c5}$, with $\tilde{\Omega}_{c5}$ in \eqref{eq:S1.5_C5}, is found as a consequence of the formation of the formerly impossible transition between the two outer states $\ket{1}_N$ and $\ket{4}_N$ (cf. Figs.~\ref{fig: S1.5-ergodic}(b)-(c)).
Additionally, at  $\tilde\Omega\approx 1.93$ there is a discontinuity in the first derivative of all four curves that cannot be attributed to a new Markov chain transition, as our system is already fully connected, as can seen in Fig.~\ref{fig: S1.5-ergodic}(c). Instead, this can be traced back to a crossing in a small time window of the reset-free probabilities $p^F_{22}$ and $p^F_{24}$ (and analogously for $p^F_{33}$ and $p^F_{31}$), such that $p^F_{22}(t)>p^F_{24}(t)$ ($p^F_{33}(t)>p^F_{31}(t)$) for $\tilde{\Omega}$ slightly greater than 1.93, while $p^F_{22}(t)<p^F_{24}(t)$ ($p^F_{33}(t)<p^F_{31}(t)$) for $\tilde{\Omega}$ slightly smaller than $1.93$. According to the majority rule, this leads to an increase for $\tilde{\Omega}>1.93$ of $R_{22}$ ($R_{33}$) compared to $R_{24}$ ($R_{31}$) and therefore to an eventual growth of $p_2^{\mathrm{NESS}}=p_3^{\mathrm{NESS}}$, while $p_1^{\mathrm{NESS}}=p_4^{\mathrm{NESS}}$ decrease. This example shows that the majority rule in the conditional reset protocol can possibly lead to discontinuities in the first derivative of the curves in the ergodic regime, even without causing a change in the Markov chain diagram. Such discontinuities arise every time a crossing of the reset-free probabilities takes place in the ergodic regime as $\tilde{\Omega}$ is tuned. In general we, however, find that such second-order discontinuities are always tiny on the scale of the singularities arising in the nonergodic regime discussed in the main text.  For large values of $\tilde\Omega$ the stationary density of spins in the respective magnetization eigenstates approach the asymptotic values $p_1^{\mathrm{NESS}}(\infty)=p_4^{\mathrm{NESS}}(\infty)\approx 0.255$ and $p_2^{\mathrm{NESS}}(\infty)=p_3^{\mathrm{NESS}}(\infty)\approx0.245$. Similarly to the case of spin $S=1$, the stationary densities corresponding to opposite magnetization eigenvalues, and therefore connected by the time-reversal symmetry \eqref{eq:time_reversal_state}, coincide.

\subsection{Spin-$2$ system}
\label{app:S2_ergodic}
\begin{figure*}[t!]
    \centering
    \def\svgwidth{\textwidth} 
    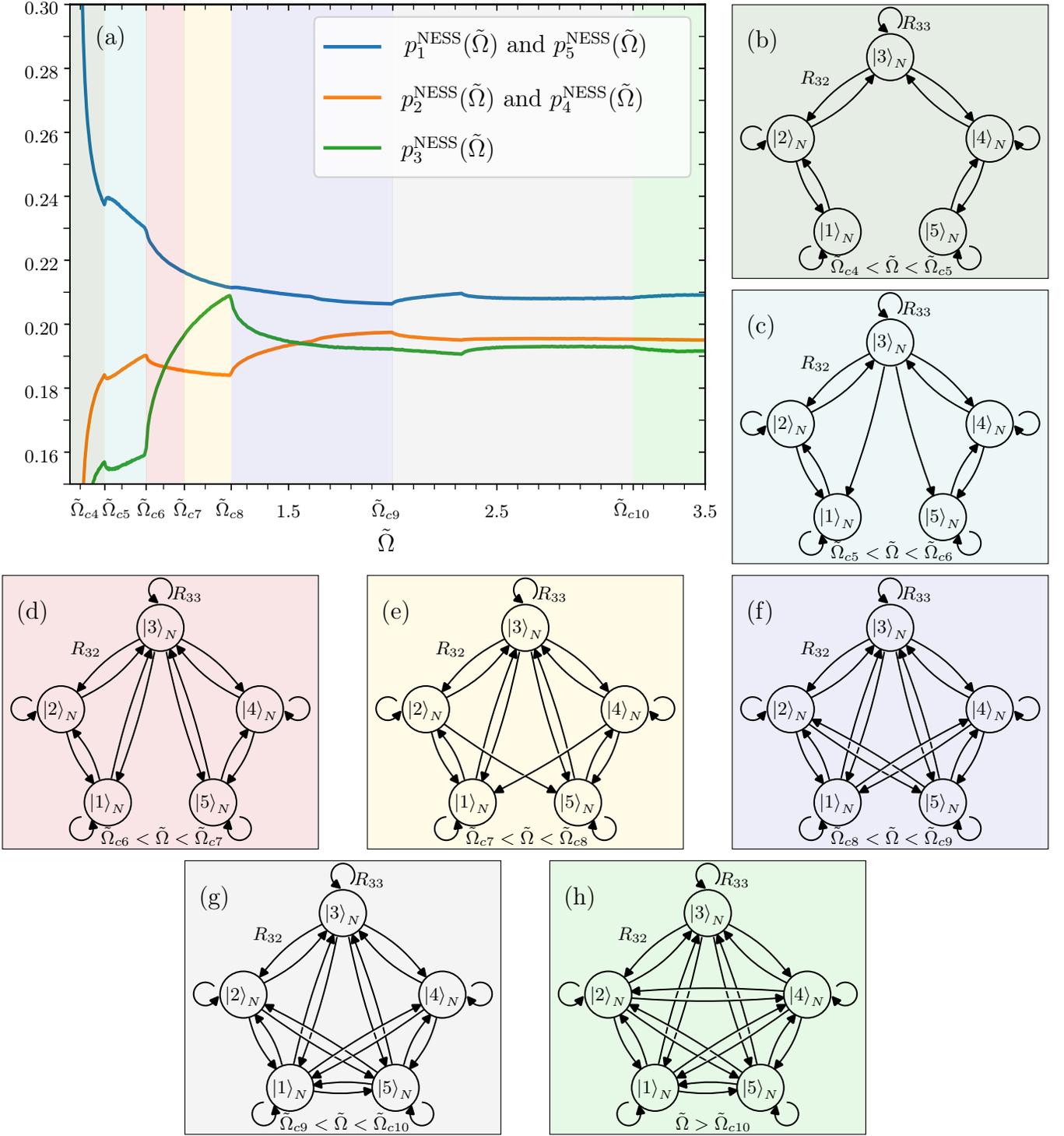
    \caption{\textbf{Stationary spin densities for a spin-$2$ system.}
    Plot of the stationary densities $p_j^{\mathrm{NESS}}$ as a function of $\tilde\Omega=\Omega/\Delta$ in the ergodic regime for $\tilde{\Omega}>\tilde{\Omega}_{c4}$. Seven different regions (highlighted with different colors) are identified as a function of $\tilde{\Omega}$ and the critical values \eqref{eq:S2_C4}-\eqref{eq:S2_C10}. In the panels (b)-(h), we report the reset Markov chain corresponding to the seven regions. In panel (a), one observes a second-order continuous transition in all five densities at all critical values except $\tilde\Omega_{c7}$.
    A discontinuity in the first derivative appears at $\tilde\Omega\approx 2.33$, which is not associated with any of the critical $\tilde{\Omega}$ values since no new connection in the Markov chain diagram is formed. In the figure, we set $\gamma=\Delta=1$.}
    \label{fig: S2-ergodic}
\end{figure*}

In the spin-$2$ system, the ergodic regime is obtained for $\tilde{\Omega}>\tilde{\Omega}_{c4}$, with $\tilde{\Omega}_{c4}$ given in Eq.~\eqref{eq:S2_C4}. In Fig.~\ref{fig: S2-ergodic}~(a), we observe a set of second-order discontinuities at the values $\tilde{\Omega}_{c5}-\tilde{\Omega}_{c10}$ in Eqs.~\eqref{eq:S2_C5}-\eqref{eq:S2_C10}, which are also in this case tiny on the scale of the sigularities taking place in the nonergodic regime in Fig.~\ref{fig: S2}. These transitions are observed due to formation of new transitions between two states in an already ergodic Markov graph (cf. Figs.~\ref{fig: S2-ergodic}(b)-(h)).
At the critical value $\tilde\Omega_{c7}$, specifically, the transitions from $\ket{2}_N$ to $\ket{5}_N$, with probability $R_{25}$, and that from $\ket{4}_N$ to $\ket{1}_N$, with probability $R_{41}$), are established (cf. Fig.~\ref{fig: S2-ergodic}(e)). In this case, we, however, find that the stationary weight-vector $\vec{c}$ changes continuously and all the spin densities in the respective magnetization eigenstates are continuous in $\tilde{\Omega}_{c7}$. This follows from the fact that the transition $R_{25}$ is activated because there are times $t$ where $p^{F}_{25}(t)>p^{F}_{21}(t)$. For all the other times, the ordering among the reset-free probabilities does not change across $\tilde{\Omega}_{c7}$. The change of the stationary resetting probability to $\ket{1}_N$ due to the decrease of $R_{21}$ is then exactly compensated by the growth of $R_{41}$. This applies since the state $\ket{4}_N$ is the symmetric partner of $\ket{1}_N$ under time-reversal. The same applies for the stationary resetting probability to $\ket{5}_N$, with analogous role played by the transitions $R_{45}$ and $R_{25}$. At $\tilde\Omega\approx 2.33$ in Fig.~\ref{fig: S2-ergodic}~(a) a discontinuity in the first derivative is visible. Similar to the discussion of Appendix~\ref{app:S32_ergodic} for spin $S=3/2$ at $\tilde\Omega\approx 1.93$, this kink is not linked to a new connection in the Markov chain. It is, instead, related to a crossing of the reset-free probabilities, which determines a sudden increase of the resetting probabilities $R_{ij}$ according to the majority rule. For large values of $\tilde\Omega$ we find that the stationary density of spins in the respective magnetization eigenstates approach the stationary values $p_1^{\mathrm{NESS}}(\infty)=p_5^{\mathrm{NESS}}(\infty)\approx 0.213$,  $p_2^{\mathrm{NESS}}(\infty)=p_4^{\mathrm{NESS}}(\infty)\approx0.192$
and $p_3^{\mathrm{NESS}}(\infty)\approx0.190$.

\section{Details about the calculations of the steady-state weights in the nonergodic regime}
\label{app:steady-state_weights}
In this section, we detail the calculation of the steady-state weight vector $\vec{c}$ for absorbing Markov chains using the formalism introduced in Sec.~\ref{Subsec:weight_vec_nonergodic}. In Subsec.~\ref{app:steady-state_weights_S32}, we detail the derivation of Eq.~\eqref{eq:absorbing_c_S32} for spin $3/2$, while in \ref{app:steady-state_weights_S2} we present the calculations leading to Eq.~\eqref{eq:absorbing_c_S2} for spin $2$.

\subsection{Steady-state weights for $S=3/2$ with absorbing reset states}
\label{app:steady-state_weights_S32}
In the spin-$3/2$ system, as discussed in the main text in Sec.~\ref{subsec:S32}, for $\tilde\Omega_{c2}<\tilde\Omega<\tilde\Omega_{c3}$ the corresponding Markov chain is absorbing (cf. Fig.~\ref{fig: S1.5}(e)). 
There are two absorbing states $\ket{1}_N$, $\ket{4}_N$ and two transient states $\ket{2}_N$, $\ket{3}_N$. The two absorbing (transient) states form a symmetric pair according to the time-reversal symmetry \eqref{eq:time_reversal_state}. Since both the absorbing states are reachable, all resets will eventually happen towards one of them. The choice of the initial state causes the splitting of the stationary resetting probabilities $c_1$ and $c_4$ and therefore of the associated stationary densities $p_1^{\mathrm{NESS}}$ and $p_4^{\mathrm{NESS}}$. Intuitively, if the initial state is $\ket{2}_N$, one would expect that the system is more likely to reset to $\ket{1}_N$ than in $\ket{4}_N$, since one extra transition is needed to get from the initial state $\ket{2}_N$ to $\ket{4}_N$ than in the case from $\ket{2}_N$ to $\ket{1}_N$.
We now quantitatively compute the asymmetry between the resetting probabilities in the absorbing reset states. 

Applying the formalism for absorbing Markov chain of Sec.~\ref{Subsec:weight_vec_nonergodic}, we first note that the symmetry in Eq.~\eqref{eq:symmetry_R} holds and set $R_{1k}=0$ respectively $R_{4k'}=0$ for $k\neq1$ respectively $k'\neq 4$. The transition matrix therefore is
\begin{align}
    R = \left( \begin{array}{c c c c}
    R_{11} & R_{12} & R_{13} & R_{14} \\
    R_{21} & R_{22} & R_{23} & R_{23}\\
    R_{31} & R_{32} & R_{33} & R_{34}\\ 
    R_{41} & R_{42} & R_{43} & R_{44}
    \end{array} \right)
    =\left( \begin{array}{c c c c}
    1 & 0 & 0 & 0 \\
    R_{21} & R_{22} & R_{23} & 0\\
    0 & R_{23} & R_{22} & R_{21}\\ 
    0 & 0 & 0 & 1
\end{array} \right)\, .
\end{align}
Rearranging the transition matrix into the form introduced in Eq.~\eqref{eq: transition matrix form for formalism} we obtain
\begin{align}
    R' = \left( \begin{array}{c c c c}
    1 & 0 & 0 & 0 \\
    0 & 1 & 0 & 0 \\
    R_{21} & 0 & R_{22} & R_{23}\\
    0 & R_{21} & R_{23} & R_{22}
\end{array} \right) 
=\left( \begin{array}{c c}
    I_2 & 0_{2\times2}\\
    B & Q\\
\end{array} \right),
\label{app:R_prime_S32}
\end{align}
with
\begin{align}
    B=\left( \begin{array}{c c}
    R_{21} & 0\\
    0 & R_{21}
\end{array} \right)\,\, \text{and} \,\,
    Q=\left( \begin{array}{c c}
    R_{22} & R_{23}\\
    R_{23} & R_{22}
\end{array} \right).
\end{align}
The structure of $R'$ in Eq.~\eqref{app:R_prime_S32} amounts to relabelling the reset states as $\ket{1}_N\leftrightarrow \ket{1'}_N$, $\ket{4}_N \leftrightarrow \ket{2'}_N$, $\ket{2}_N\leftrightarrow \ket{3'}_N$ and $\ket{3}_N\leftrightarrow \ket{4'}_N$. According to Eq.~\eqref{eq:matrix_stationary_prob} and using that the rows of the transition matrix sum up to $1$ since it is Markovian, we calculate the stationary probability as
\begin{align}
    P_j^a=[(I_b-Q)^{-1} B]_{ja}=\frac{1}{R_{21}+2R_{23}}
    \left[ \left( \begin{array}{c c}
    R_{21}+R_{23} & R_{23}\\
    R_{23} & R_{21}+R_{23}
    \end{array} \right) \right]_{ja}= \left( \begin{array}{c c}
    P_{3'}^{1'} & P_{3'}^{2'} \\
    P_{4'}^{1'} & P_{4'}^{2'}
    \end{array} \right)_{ja}.
\label{app:S32_absorbing_matrix}
\end{align}
In the last equality, we made explicit that the row index $j$ running over the transient states takes values $j=3',4'$ according to the relabelling of reset states. In the same way, the column index running over the absorbing states takes values $a=1',2'$. Going back to the original labelling of the reset states, for the initial transient state $\ket{2}^N$, the weight vector is obtained as the $j=3'$ row of the matrix \eqref{app:S32_absorbing_matrix}. For the initial state $\ket{3}^N$, the corresponding weight vector is given by the row $j=4'$ of \eqref{app:S32_absorbing_matrix}. Inserting zeros for the stationary probability of the transient states and rearranging the states according to the original labelling in the matrix $R$, we write down the weight vector for initializing in $\ket{2}^N$ as
\begin{align}\label{eq: weight vector S1.5 - c2}
    \vec{c}^{\ T}=\left(\frac{R_{21}+R_{23}}{R_{21}+2R_{23}}, 0, 0, \frac{R_{23}}{R_{21}+2R_{23}}\right).
\end{align}
This matches Eq.~\eqref{eq:absorbing_c_S32} of the main text. Since $R_{21}>0$, in the regime $\tilde\Omega_{c2}<\tilde\Omega<\tilde\Omega_{c3}$, one has $c_1>c_4$, as expected. This, in turn, gives $p_1^{\mathrm{NESS}}(\tilde\Omega)>p_4^{\mathrm{NESS}}(\tilde\Omega)$ and the stationary spin densities for the state $\ket{1}_N$ and $\ket{4}_N$ split for $\tilde\Omega_{c2}<\tilde\Omega<\tilde\Omega_{c3}$, as one can see from Fig.~\ref{fig: S1.5}. Since the probability for the spins to be found in $\ket{2}$ is higher than that to be found in $\ket{3}$ after a reset in $\ket{1}_N$, this analysis also explains the splitting of the two curves associated with the transient states, where $p^{\mathrm{NESS}}_2$ increases and $p^{\mathrm{NESS}}_3$ decreases in the interval $\tilde\Omega_{c2}<\tilde\Omega<\tilde\Omega_{c3}$.

\subsection{Steady-state weights for $S=2$ with absorbing reset states}
\label{app:steady-state_weights_S2}
In the spin-$2$ system, discussed in the main text in Sec.~\ref{subsec:S2},  for $\tilde\Omega_{c3}<\tilde\Omega<\tilde\Omega_{c4}$ the corresponding Markov chain in Fig.~\ref{fig: S2}(g) is absorbing. There are two absorbing states $\ket{1}_N$ and $\ket{5}_N$, related by time-reversal symmetry, and three transient states $\ket{2},\ket{3}_N,\ket{4}_N$.
As well as in the previous section, we determine the stationary state weight vector $\vec{c}$ upon initializing in a transient state exploiting the formalism of absorbing Markov chains. The transition matrix $R$ reads as
\begin{align}
    R = \left( \begin{array}{c c c c c}
    R_{11} & R_{12} & R_{13} & R_{14} & R_{15}\\
    R_{21} & R_{22} & R_{23} & R_{23} & R_{25}\\
    R_{31} & R_{32} & R_{33} & R_{34} & R_{35}\\ 
    R_{41} & R_{42} & R_{43} & R_{44} & R_{45}\\
    R_{51} & R_{52} & R_{53} & R_{54} & R_{55}
    \end{array} \right)
    =\left( \begin{array}{c c c c c}
    1 & 0 & 0 & 0 & 0 \\
    R_{21} & R_{22} & R_{23} & 0 & 0\\
    0 & R_{32} & R_{33} & R_{32} & 0\\ 
    0 & 0& R_{23} & R_{22} & R_{21}\\
    0 & 0 & 0 & 0 & 1
\end{array} \right),
\end{align}
where $R_{1k}=0$ for $k\neq1$, and $R_{5k'}=0$ for $k'\neq 5$ since the outer states are absorbing. We also used the centrosymmetric structure \eqref{eq:symmetry_R} of the matrix $R$. Rerranging the transition matrix into the form in Eq.~\eqref{eq: transition matrix form for formalism}, we obtain
\begin{align}
    R' = \left( \begin{array}{c c c c c}
    1      & 0      & 0      & 0      & 0\\
    0      & 1      & 0      & 0      & 0 \\
    R_{21} & 0      & R_{22} & R_{23} & 0\\
    0      & 0      & R_{32} & R_{33} & R_{32}\\
    0      & R_{21} & 0      & R_{23} & R_{22}
\end{array} \right) 
=\left( \begin{array}{c c}
    I_2 & 0_{2\times3}\\
    B & Q\\
\end{array} \right),
\label{app:R_prime_2}
\end{align}
with
\begin{align}
    B=\left( \begin{array}{c c}
    R_{21} & 0\\
    0 & 0\\
    0 & R_{21}
\end{array} \right)=R_{21}\left( \begin{array}{c c}
    1 & 0\\
    0 & 0\\
    0 & 1
\end{array} \right) \text{    and    } 
    Q=\left( \begin{array}{c c c}
    R_{22} & R_{23} & 0\\
    R_{32} & R_{33} & R_{32}\\
    0      & R_{23} & R_{22}
\end{array} \right).
\end{align}
The structure of $R'$ in \eqref{app:R_prime_2} amounts to relabelling of the reset states as $\ket{1}_N\leftrightarrow \ket{1'}_N$, $\ket{5}_N\leftrightarrow \ket{2'}_N$, $\ket{2}_N\leftrightarrow \ket{3'}_N$, $\ket{3}_N\leftrightarrow\ket{4'}_N$ and $\ket{4}_N \leftrightarrow \ket{5'}_N$. From Eq.~\eqref{eq:matrix_stationary_prob} the stationary probabilities of absorption in the absorbing states are given by
\begin{align}
    P_j^a=[(I_b-Q)^{-1} B]_{ja}=
    \frac{1}{2}\left[ 
    \left( \begin{array}{c c}
    \frac{2R_{21}+R_{23}}{R_{21}+R_{23}} & \frac{R_{23}}{R_{21}+R_{23}}\\
    1 & 1\\
    \frac{R_{23}}{R_{21}+R_{23}} & \frac{2R_{21}+R_{23}}{R_{21}+R_{23}}
    \end{array} \right) \right]_{ja}=\left( \begin{array}{c c}
   P_{3'}^{1'} & P_{3'}^{2'} \\
    P_{4'}^{1'} & P_{4'}^{2'}\\
    P_{5'}^{1'} & P_{5'}^{2'}
    \end{array} \right)_{ja}.
\label{app:matrix_absorbing_S2}
\end{align}
Turning back to the original labelling of the reset states, one has that stationary weight vectors for the initial transient states $\ket{2}_N$, $\ket{3}_N$ and $\ket{4}_N$ correspond to the $j=3',4',5'$ rows, respectively, of the matrix in Eq.~\eqref{app:matrix_absorbing_S2}. Inserting zeros for the transient states and rearranging the states according to the original labelling yields the correct weight vector for initializing in $\ket{2}_N$ in $\tilde\Omega_{c3}<\tilde\Omega<\tilde\Omega_{c4}$ as
\begin{align}
    \vec{c}^{\ T}=\left(\frac{2R_{21}+R_{23}}{2(R_{21}+R_{23})}, 0, 0, 0, \frac{R_{23}}{2(R_{21}+R_{23})}\right).
\end{align}
This expression matches \eqref{eq:absorbing_c_S2} of the main text. Since $R_{21}>0$, one has that $c_1>c_5$, as expected since the initial state $\ket{2}_N$ is closer to $\ket{1}_N$ in the Markov chain picture than to $\ket{5}_N$. The stationary spin densities accordingly split with $p_1^{\mathrm{NESS}}>p_5^{\mathrm{NESS}}$ and $p_2^{\mathrm{NESS}}>p_4^{\mathrm{NESS}}$. This splitting of the spin densities associated with the time-reversal symmetric states is observed in Fig.~\ref{fig: S2}(b) for $\tilde\Omega_{c3}<\tilde\Omega<\tilde\Omega_{c4}$.
\twocolumngrid
\bibliography{bibliography}
\end{document}